\documentclass[12pt,tightenlines,eqsecnum,floats,shownopacs,nofootinbib,amsmath,amssymb,aps,prd]{revtex4}

\usepackage{color}
\usepackage{graphicx}
\usepackage{amsmath,amssymb}

\def\be{\nopagebreak[3]\begin{equation}}
\def\ee{\end{equation}}
\def\ba{\nopagebreak[3]\begin{eqnarray}}
\def\ea{\end{eqnarray}}

\def\lp{\ell_{\rm Pl}}

\def\f{\frac}
\def\mpl{m_{\rm Pl}}
\def\rp{\rho_{\rm Pl}}
\def\rcr{\rho_{\rm max}}
\def\hom{\rm hom}

\def\dyn{\rm Dyn}
\def\BD{\rm BD}
\def\LQC{\rm LQC}
\def\b{{\rm b}}
\def\B{{\rm B}}%used for the bounce point

\def\bu{$\bullet\,\,$}

\def\t{\tilde}
\def\h{\hat}

\def\tr{\rm Trun}

\def\pphi{p_{(\phi)}}

\def\x{\vec{x}}
\def\vp{\varphi}
\def\vk{\vec{k}}
\def\R{\mathcal{R}}

\def\qzero{\mathring{q}}

\def\q{\mathfrak{q}}

\def\p{\mathfrak{p}}

\def\r{\mathfrak{f}}
\def\g{\mathfrak{A}}

\def\T{\mathcal{T}}
\def\ps{\Gamma}

\def\H{\mathcal{H}}

\def\R{\mathcal{R}}
\def\Q{\mathcal{Q}}

\begin{document}

%\title{A Quantum Gravity Extension of the Inflationary Scenario}

%\title{An Extension of the Quantum Theory of Cosmological %Perturbations to the
%Planck Era}

\title{The pre-inflationary dynamics of loop quantum cosmology:\\
Confronting quantum gravity with observations}

\author{Ivan Agullo$^{1,2}$}
\email{I.AgulloRodenas@damtp.cam.ac.uk}
\author{Abhay Ashtekar$^{1}$}
\email{ashtekar@gravity.psu.edu}
\author{William Nelson$^{1,3}$}
\email{nelson@gravity.psu.edu}
\affiliation{$^{1}$ Institute for
Gravitation and the Cosmos \& Physics
  Department, Penn State, University Park, PA 16802, U.S.A.}
\affiliation{$^{2}$ Center for Theoretical Cosmology, DAMTP, Wilberforce Road,
University of Cambridge, Cambridge CB3 OWA, U.K.}
\affiliation{$^{3}$ Radboud University Nijmegen, Institute for Mathematics, Astrophysics and Particle Physics,Ê Heyendaalsweg 135,Ê6525-AJ Nijmegen, Netherlands }

\begin{abstract}

Using techniques from loop quantum gravity, the standard theory of
cosmological perturbations was recently generalized to encompass the
Planck era. We now apply this framework to explore pre-inflationary
dynamics. The framework enables us to isolate and resolve the true
trans-Planckian difficulties, with interesting lessons both for
theory and observations. Specifically, for a large class of initial
conditions at the bounce, we are led to a self consistent extension
of the inflationary paradigm over the 11 orders of magnitude in
density and curvature, from the big bounce to the onset of slow
roll. In addition, for a narrow window of initial conditions, there
are departures from the standard paradigm, with novel effects
---such as a modification of the consistency relation between the
ratio of the tensor to scalar power spectrum and the tensor spectral
index, as well as a new source for non-Gaussianities--- which could
extend the reach of cosmological observations to the deep Planck
regime of the early universe.

\end{abstract}


\maketitle

%\newpage
%\tableofcontents
%\newpage

\section{Introduction}
\label{s1}

Thanks to a powerful interplay between theory and observations, our
understanding of the early universe has advanced very significantly
in recent years. Consequently, cosmology offers a natural arena to
develop and test approaches to quantum gravity.

In this paper we will focus on inflation, the leading scenario to
successfully account for the tiny inhomogeneities observed in the
cosmic microwave background (CMB)\,\, (see, e.g.,
\cite{ll-book,sd-book,vm-book,sw-book,gr-book}). An impressive
feature of this scenario is that it involves a rather small set of
assumptions. As noted in \cite{as3,asrev}, these can be summarized
as follows:
\begin{itemize}
\item Sometime in its early history, the universe underwent a
    phase of rapid expansion during which the Hubble parameter
    was nearly constant;
    \item During this phase, the universe was well
        described by a Friedman, Lema\^itre, Robertson
        Walker (FLRW) solution to Einstein's equations with
        a scalar field as matter source, together with
        small inhomogeneities which are well approximated
        by first order perturbations;
\item Fourier modes of the quantum fields representing
    perturbations were initially in the Bunch-Davies (BD) vacuum
    at least for a certain set of wave numbers%
\footnote{More precisely, if $k_o$ denotes the co-moving wave
number of the mode which has just re-entered the Hubble radius
now then,
at the time $t(k_o)$ at which this mode exited the Hubble radius
during inflation, Fourier modes with co-moving wave numbers in
the range $\sim\, (k_o,\, 2000k_o)$ were in the Bunch-Davies
vacuum.}; and,
\item Soon after any mode exits the Hubble radius, its quantum
    fluctuation can be regarded as a classical perturbation and
    evolved via linearized Einstein's equations.
\end{itemize}

Analysis of these inflationary perturbations implies that there must
be tiny inhomogeneities at the surface of last scattering whose
detailed features have been observed in the CMB. Furthermore, these
tiny inhomogeneities serve as seeds that grow into a large scale
structure which is in excellent qualitative agreement with
observations. Therefore, even though the underlying assumptions are
by no means `obvious' or `compelling', the success of the subsequent
predictions is impressive.

Yet, as is well known (see e.g. \cite{brandenberger}), the scenario
is conceptually incomplete in several respects. In particular, as
Borde, Guth and Vilenkin \cite{bgv} showed, inflationary space-times
inherit the big-bang singularity. Physically, this occurs because
one continues to use general relativity even in the Planck regime in
which it is simply not applicable. It is widely expected that new
physics in this regime will resolve the singularity, significantly
changing the very early history of the universe. Will inflation
arise naturally in the resulting deeper theory? Or, more modestly,
can one at least obtain a consistent quantum gravity extension of
this scenario?

To fully address this question one has to face two distinct sets of
issues. The first set has its origin in \emph{particle physics.}
What is the physical origin of the scalar field that plays the role
of the inflaton? It should arise naturally in the fundamental
unified theory. Can we derive the inflationary potential from some
first principles? Is there a single inflaton, as in the simplest
models that have been successful so far, or many? How do they
interact with one another? Even more importantly, what are the
interactions that produce the elementary constituents of the
standard model of particle physics during reheating? These issues
remain open, but fall in the realm of theories aimed at unification.
In this and the companion papers, we do not address them.

Rather, we will continue to use the simplest, single inflaton model
and focus on the second set of issues \emph{related to gravity.} In
the systematic evolution starting from the Planck regime, does a
slow roll phase compatible with the WMAP data arise generically in
the background geometry, or, is an enormous fine tuning needed? In
classical general relativity, if we evolve the modes of
observational interest back in time, they become trans-Planckian at
some stage. Is there a quantum field theory (QFT) on \emph{quantum}
space-times that is needed to adequately handle physics at that
stage? Can one arrive at the BD vacuum at the onset of the slow roll
from more fundamental considerations? Or, is an even more elaborate
fine tuning of the quantum state of \emph{perturbations} necessary
in the Planck regime? The natural state of quantum perturbations at
the onset of inflation may be indistinguishable from the BD vacuum
for currently available data but, because of the pre-inflationary
dynamics especially in the Planck regime, it could well carry
certain excitations over the BD vacuum. These can then source
non-Gaussianities for the subsequent evolution during inflation
\cite{chen,holman-tolley,agullo-parker,ganc,agullo-navarro-salas-parker}
which, in turn, could give rise to novel effects in the CMB and
galaxy distribution \cite{halo-bias1,halo-bias2,halo-bias3} which
could be observed in future missions. Can the state at the onset of
the slow roll be close enough to the BD vacuum to agree with current
observations and yet be sufficiently different to give rise to such
effects?

This is the third in a series of papers whose goal is to address
these questions that originate in quantum gravity. The first paper
\cite{aan1} presented a succinct summary of underlying framework and
the principal observational consequences for a broad audience. The
second \cite{aan2} was primarily addressed to the quantum gravity
community and focused on conceptual and mathematical aspects of the
framework. In this paper we use that framework to analyze
phenomenological issues of interest to cosmologists within the
standard inflationary scenario.

The main ideas can be summarized as follows.

By now there is a large body of results in loop quantum
cosmology (LQC) that provides a detailed description of the
Planck scale physics in a number of cosmological models. These
include the k=0 and k=1 (FLRW) models
\cite{mb1,abl,aps1,aps2,aps3,acs,ps,apsv,warsaw1}, possibly
with a non-zero cosmological constant \cite{bp,kp1,ap}, the
anisotropic Bianchi I, II and IX models
\cite{awe2,madrid-bianchi,awe3,we} and the simplest of the
inhomogeneous models ---the Gowdy space-times--- widely studied
in exact general relativity
\cite{hybrid1,hybrid2,hybrid3,hybrid4,hybrid5}. In all cases,
the big bang singularity is resolved and replaced by quantum
bounces (for a review, see, e.g., \cite{mbrev,asrev}). Thus,
the first conceptual limitation of the standard general
relativistic models is overcome. It is therefore natural to use
LQC as the point of departure for extending the cosmological
perturbation theory. How do new effects in the Planck domain
---such as the robust superinflation phase that follows the
density bounce--- alter the standard inflationary scenario? Do
generic initial conditions at the bounce lead to the desired
slow roll inflation later, or, is a fine tuning necessary?

To incorporate cosmological perturbations, however, one cannot just
mimic the standard procedure used in general relativity: Since loop
quantum gravity (LQG) is still incomplete, we do not yet have the
analog of full Einstein's equations to perturb. A number of LQC
inspired strategies have been devised to overcome this obstacle.%
\footnote{For a short summary of the underlying ideas, merits and
limitations, see section II.A in \cite{aan2}.}
In this series of papers we follow a mainstream strategy in LQG:
\emph{First truncate the classical theory in a manner appropriate to
the physical problem under consideration, then carry out
quantization paying due attention to the underlying quantum geometry
of LQG, and finally work out the consequences of the resulting
framework}. This procedure has been successful not only in the
cosmological models referred to above but also in the investigation
of quantum black holes in LQG \cite{abck,abk,blv} and derivation of
the graviton propagator using spin foams \cite{graviton,cr-rev}.

For the present application, we will truncate the full phase space,
keeping only the FLRW backgrounds with \emph{first order}
inhomogeneous perturbations. To facilitate easy comparison with the
standard quantum theory of cosmological perturbations, we will
employ a hybrid scheme used in
\cite{hybrid1,hybrid2,hybrid3,hybrid4,hybrid5,lqc-preinflation2}, in
which one uses the standard LQC for homogeneous modes and
`Fock-like' states for the first order perturbations.
There is however a key difference from the standard cosmological
perturbation theory: Because the background space-time now has
\emph{quantum} geometry, we are led to use QFT on \emph{quantum}
cosmological space-times. This theory was introduced in \cite{akl}
emphasizing conceptual issues and developed in \cite{aan2} to
incorporate the infinitely many modes of perturbations and the
associated regularization and renormalization techniques which, as
discussed below, are key to checking the self-consistency of our
initial truncation. Since we only have a probability amplitude for
various FLRW geometries ---rather than a sharply defined classical
FLRW metric--- at a fundamental level evolution has to be described
using a \emph{relational time variable} rather than cosmic or
conformal time. Thus, quantum gravity introduces genuinely new
conceptual elements. But one can systematically `descend' to a
description in terms of cosmic and conformal time of an effective
metric that incorporates the quantum gravity corrections.

Since quantum perturbations now propagate on quantum geometries
which are all regular, free of singularities, \emph{the framework
automatically encompasses the Planck regime.} What is the status of
the `trans-Planckian issue' which, in heuristic discussions, is
often associated with modes of trans-Planckian frequencies? Recall,
first, that the quantum Riemannian geometry underlying LQG is subtle
\cite{alrev,ttbook}. For example, while there is a minimum non-zero
eigenvalue of the area operators, there is no such minimum for the
volume operators, even though their eigenvalues are also
discrete \cite{rs,al5}.%
\footnote{Properties of the eigenvalues of length operators
\cite{tt-length,eb-length,msy-length} have not been analyzed in
comparable detail. But since their definitions involve volume
operators, it is expected that there would be no `length gap'.}
As a consequence, there is no fundamental obstacle preventing
trans-Planckian modes \emph{of perturbations} in our truncated
theory. Indeed, in the homogeneous LQC models that have been
analyzed in detail, the momentum $p_{(\phi)}$ of the scalar field
$\phi$ is generally \emph{huge} in Planck units. This poses no
problem and, in particular, on the physical Hilbert space the total
energy density is still guaranteed to be bounded by $0.41 \rho_{\rm
Pl}$ where $\rho_{\rm Pl}$ is the Planck density. Similarly, for
perturbations, there is no a priori difficulty with trans-Planckian
momenta (or frequencies) in the truncated theory considered here.
The real danger is rather that, in presence of such modes, the
\emph{energy density} in perturbations may fail to be negligible
compared to that in the quantum background geometry. This is a very
non-trivial issue especially in the Planck regime following the
bounce and requires a careful treatment of regularization and
renormalization of the stress energy tensor of quantum
perturbations. If the energy density does become comparable to that
in the  background, then we would not be able to neglect the
back-reaction and our truncation would fail to be self-consistent. \emph{This is the trans-Planckian problem we face in our theory of
quantum perturbations on inflationary quantum geometries.}

The paper is organized as follows. Section \ref{s2} summarizes the
underlying conceptual and mathematical framework (constructed in
\cite{aan2} using prior results from
\cite{aps1,aps2,aps3,acs,aps4,as3,pf}) for the cosmology community.
This discussion encompasses the scalar as well as the tensor modes.
Section \ref{s3} discusses the initial conditions for the wave
function $\Psi_o$ representing the background quantum geometry, and
for the wave function $\psi$ representing the perturbations
propagating on the quantum geometry $\Psi_o$. The choices can be
motivated by general symmetry principles and physical
considerations. Sections IV and V present results of numerical
simulations of the pre-inflationary dynamics for the quantum
corrected background and perturbations. The free parameter that
dictates the evolution of the background is the value $\phi_{\B}$ of
the inflaton at the bounce. In section \ref{s5} we vary this
parameter (within a range that is numerically feasible) and explore
consequences on the observed power spectrum at the end of inflation.
We find an unforeseen and interesting structure. \emph{The LQC power
spectrum is essentially independent of the value of $\phi_{\rm B}$.}
But recall that there is only a finite interval in the space of
co-moving wave numbers $k$ that is relevant for the CMB
observations. \emph{This interval is completely determined by the
value of $\phi_{\B}$, moving steadily to the right along the
$k$-axis as $\phi_{\rm B}$ increases.} If $\phi_{\B}$ is sufficiently low, the quantum state at the onset
of the slow roll is sufficiently different from the BD vacuum to
lead to observable effects for future measurements along the lines
of \cite{halo-bias1,halo-bias2,halo-bias3}.

In section \ref{s6} we analyze the issue of self consistency of the
truncation strategy: Does the stress energy in the perturbations
remain small compared to that in the background throughout the
evolution from the big bounce to the onset of the slow roll? The
main result is that \emph{the self-consistency criterion is met} for
a large class of initial conditions. In these cases, our main
approximation ---the initial truncation--- is viable. Had it failed,
it would have been inconsistent to keep only the first order
perturbations. Then one would have had to wait for significant
advances in full LQG to extend the inflationary paradigm to the
quantum gravity regime. Thus, technically as well as conceptually it
is quite non-trivial that this does not happen and one obtains a
self consistent extension of the standard inflationary paradigm all
the way to the deep Planck regime of the big bounce for a wide range
of initial conditions.

Of course, self-consistency by itself does not imply that our
truncated solution is necessarily close to an exact one (because the
sum of all higher order terms could be large). But this limitation
is common to all perturbation theories, classical or quantum. In
particular, in classical cosmology the total back reaction is
routinely neglected if stress-energy in the first order
perturbations is small compared to that in the background, even
though strictly this does not imply that there is an exact solution
close to the linearized one. If the first order perturbations do not
grow so much in time as to become comparable to the background, the
test field approximation is \emph{self consistent} and is regarded
as trustworthy. We adopt the same viewpoint here.

In section \ref{s7} we summarize the main results and discuss open
issues. The Appendix summarizes the technical differences between
Ref. \cite{aan2} and the underlying framework presented in section
\ref{s2} which is tailored to the inflation.

We will use the following conventions. The signature of the
space-time metric will be  -,+,+,+. We set $c$=1 but keep $G$ and
$\hbar$ explicitly in various equations to facilitate the
distinction between classical and quantum effects. As is usual, we
will set $\kappa = 8\pi G$ and, for the Barbero Immirzi parameter
$\gamma$ of LQG, use the value $\gamma \approx 0.24$ that comes from
the black hole entropy calculations. Finally, we will use Planck
units used in the quantum gravity literature rather than the reduced
Planck units often used in cosmology. (Thus, our Planck mass $\mpl=
\sqrt{\hbar/G}$ is related to the reduced Planck mass $M_{\rm Pl}$
via $\mpl = \sqrt{8\pi} M_{\rm Pl}$.) \emph{Numerical values} of all
quantities are given in dimensionless Planck units $\lp = \mpl =
t_{\rm Pl} =1$.

\section{The underlying framework}
\label{s2}

In discussions of the early universe, one generally begins with a
FLRW solution to Einstein's equation with suitable matter fields as
sources, then considers the space of solutions to the first order
perturbation equations on this background space-time, and finally
quantizes them using standard techniques of QFT in curved
space-times. To incorporate the Planck regime, the FLRW background
must also be treated quantum mechanically. Therefore, we cannot
begin with a solution to Einstein's equations. Instead, we will
first introduce a Hamiltonian framework that encompasses \emph{both}
the FLRW backgrounds and first order perturbations thereon and then
pass to the quantum theory of the \emph{combined system} as a whole.
As explained in section \ref{s1}, this will lead us to a quantum
theory of fields representing linearized perturbations, propagating
on a \emph{quantum} FLRW geometry.

We will assume that the spatial topology is $\mathbb{R}^{3}$ and
focus on the k=0 FLRW case. As in the standard inflationary models,
the matter field will be taken to be a scalar field and for detailed
calculations we will use the simplest potential $V(\phi) =
(1/2)m^{2} \phi^{2}$. Most of the results discussed in this section
are generalizations of those discussed in detail in \cite{aan2}, now
allowing for a inflaton potential $V(\phi)$ and $\mathbb{R}^3$
spatial topology (in place of a 3-torus).

\subsection{The truncated phase space} \label{s2.1}

Let us first truncate the full phase space of general relativity to
the physical problem of interest. This truncated phase space
$\ps_{\tr}$ is of the form $\ps_{\tr} = \ps_{o}\times \t\ps_{1}$
where $\ps_{o}$ is the 4-dimensional phase space for the FLRW
backgrounds and $\t\ps_{1}$ of gauge invariant perturbations.

Because the background fields are homogeneous and the spatial
topology is $\mathbb {R}^{3}$, the spatial integrals that appear in
the expressions of the symplectic structure and Hamiltonians on
$\ps_o$ diverge. Therefore, in the construction of the Hamiltonian
framework of the homogeneous sector, one first fixes co-moving
Cartesian coordinates $x^a$, introduces an elementary cell
$\mathcal{C}$ whose edges are aligned along these coordinates and
have equal coordinate lengths $\ell$, and restricts all integrations
to $\mathcal{C}$. This is an infrared cut-off for the homogenous
sector which is removed at the end by letting $\ell$ tend to
infinity \cite{asrev}.

In LQC, in place of the scale factor $a$ and its conjugate momentum
$\pi_{(a)}$, it is customary to use the following pair:
 \be\label{eq:BGvariables} \nu =\f{4a^{3}\ell^{3}}{\kappa\gamma\hbar};\quad
{\rm and} \quad \b = - \f{\gamma\kappa}{6 a^{2}\ell^{3}}\,
\pi_{(a)}\,
  \ee
so that the fundamental Poisson bracket is given by $\{\b, \nu\} =
\f{2}{\hbar}\, $. Thus, apart from constants, $\nu$ gives the
physical volume of the fiducial cell $\mathcal{C}$ and, on solutions
to Einstein's equations, $\b$ is related to the Hubble parameter $H
=\dot{a}/a$ via $\b = \gamma H$. The homogeneous scalar field and
its momentum will be denoted by $\phi$ and $\pphi$ respectively.
Thus the background FLRW phase space $\ps_{o} $ is coordinatized by
$\left( \nu, \b; \phi, \pphi \right)$ and carries a single scalar
(or Hamiltonian) constraint:
\be \label{hc} \mathbb{S}_{o}[N_{\hom}] = N_{\hom}\, \left[ -
\f{3\hbar \b^{2}\nu}{4\gamma} + \f{2\pphi^{2}}{\kappa\gamma\hbar\nu}
+ \f{\kappa \gamma\hbar\nu}{4}\,V(\phi) \right] = 0\, , \ee
where $N_{\hom}$ represents a homogeneous lapse. Einstein dynamics
is generated by this scalar constraint. The choice $N_{\hom}=1$
yields evolution in cosmic time; $N_{\hom} =a$ in conformal time;
and $N_{\hom} =a^{3}$ in  harmonic time. ($N_{\hom} =
a^{3}\ell^{3}/\pphi$ corresponds to using the inflaton $\phi$ as
time in `patches' of dynamical trajectories along which it is
monotonic.)

The phase space $\t\ps_{1}$ of gauge invariant perturbations is
spanned by 3 canonically conjugate pairs, one representing the
Mukhanov-Sasaki scalar mode and two representing the tensor modes.
It is simplest to work in the (co-moving) momentum space and
represent them by
\be(\Q_{\vk}, \T^{(1)}_{\vk}, \T^{(2)}_{\vk};\,\, \p^{\Q}_{\vk},
\p^{\T_1}_{\vk}, \p^{\T_2}_{\vk})\, . \ee
Since the perturbations are not homogeneous, in the discussion of
$\t{\ps}_1$ one can work directly with the entire $\mathbb{R}^3$;
restriction to the cell $\mathcal{C}$ is not necessary. We choose
this route to avoid an artificial quantization of the momenta $\vk$
that the restriction to the cell would have introduced. Technically,
we assume that the perturbations are square-integrable so that one
can freely pass between $\x$ and $\vk$ spaces. Then the Poisson
bracket relations are given by $\{\Q_{\vk},\,\,
\p^{(\Q)}_{{\vk}^{\prime}}\} = (2\pi)^{3} \delta(\vk +
\vk^{\prime})$, and similarly for tensor modes. For simplicity of
notation, we will often use collective labels $\T_{\vk},
\p^{(\T)}_{\vk}$ for the two tensor modes.

On the truncated phase space $\ps_{\tr}$, the dynamical trajectories
follow integral curves of the `evolution vector field'
$X^{\alpha}_{{\dyn}} = \Omega_{o}^{\alpha \beta}\partial_{\beta}
\mathbb{S}_{o} + \Omega^{\alpha\beta}\partial_ {\beta}
\mathbb{S}_{2}^{\prime}$ where $\Omega_{o}$ is the symplectic
structure on $\ps_{o}$,\, $\Omega_{1}$ on $\t\ps_{1}$ and
$\mathbb{S}_o$ is defined in Eq (\ref{hc}). The function
$\mathbb{S}_{2}^{\prime}$ governing the time evolution of
perturbations is obtained from the second order truncation
$\mathbb{S}_2$ of the full scalar constraint of general relativity
by keeping only the terms which are quadratic in first order
perturbations and deleting the term that is linear in the second
order perturbations.%
\footnote{Here the Greek indices refer to the (infinite dimensional)
tangent space to $\ps_{\tr}$. In terms of Poisson brackets, the
evolution of functions $F$ of background quantities $\nu, \b, \phi,
\pphi$ is given by $\dot{F} = \{F,\, \mathbb{S}_o\}_o$, where the
`dot' denotes derivative with respect to the time variable
determined by the lapse and the Poisson bracket is taken only with
respect to the background variables. The evolution of functions $f$
of gauge invariant perturbations $(\Q_{\vk},\T_{\vk};\,\,
\p^{(\Q)}_{\vk}, p^{(\T)}_{\vk})$ is given by $\dot{f} = \{f,\,
\mathbb{S}_{2}^{\prime}\}_1$ where the Poisson bracket is taken
\emph{only} with respect to the perturbations, treating the
background quantities in the expression of $\mathbb{S}_{2}^{\prime}$
(and possibly $f$) as \emph{given time-dependent variables
determined by first solving the evolution equation for functions
$F$.} }
It is given by $\mathbb{S}_{2}^{\prime} = \mathbb{S}_{2}^{\prime\,
(\Q)} + \mathbb{S}_{2}^{\prime\, \T_{1}} + \mathbb{S}_{2}^{\prime\,
\T_{2}}$, with
\be \label{pert-ham1} \mathbb{S}^\prime_2{}^{(\T)}\,[N_{\hom}] =
\f{N_{\hom}}{2}\,\int\!\f{d^3k}{(2\pi)^3} \,\,
\big(\f{4\kappa}{a^3}\, |\p^{(\T)}_{\vk}|^2 +
\f{ak^2}{4\kappa}\,|\T_{\vk}|^2 \big)\,  \ee
and
\be \label{pert-ham2} \mathbb{S}^\prime_2{}^{(\Q)}\,[N_{\hom}] =
\,\f{N_{\hom}}{2}\, \int\! \f{d^3 k}{(2\pi)^3}\,\, \big(\f{1}{a^3}\,
|\p^{(\Q)}_{\vk}|^2 + {a}\,(\g + k^2)\, |\Q_{\vk}|^2 \big)\, . \ee
In (\ref{pert-ham2}), $\g$ is a function only of the background
variables (and therefore has no $\vk$-dependence):
\be\label{g} \g = \big(\r\, V(\phi) - 2 \sqrt{\r}\, V_{,\phi} +
V_{,\phi\phi}\big)\, a^2 \ee
with $\r = (3\kappa \pphi^2)/((1/2)\pphi^2 + \ell^6 a^6 V(\phi))$
(see Appendix~(\ref{a1}). Since $\pphi = a^3\ell^3\dot\phi$, the
quantity $\r$ is essentially the kinetic \emph{fraction} of the
total energy density.)

The resulting time evolution of tensor modes is the same as that of
a massless scalar field on a given FLRW background. By contrast, the
scalar mode experiences an \emph{external, time dependent potential}
$\g$ in addition to the FLRW geometry. If $V(\phi)$ were to vanish,
we would have $\g=0$ and then the dynamics of the (trivially
rescaled) scalar mode $\Q_{\vk}/2\sqrt{\kappa}$ would be the same as
that of the tensor mode $\T_{\vk}$ (as in \cite{aan2}).

Recall first that the time evolution in the FLRW sector is generated
by the restriction $\mathbb{S}_{o}$ to the FLRW subspace $\ps_{o}$
of the Hamiltonian constraint on the \emph{full} phase space $\ps$
of general relativity. This evolution is reproduced by the first
term $\Omega_{o}^{\alpha\beta} \partial_{\beta} \mathbb{S}_{o}$ of
the dynamical vector field $X^{\alpha}_{\dyn}$. Consequently these
dynamical trajectories are solutions to the full Einstein equations.
However in the second part of the dynamical vector field,
$\mathbb{S}^{\prime}_{2}$ is \emph{not} constrained to vanish.
Indeed, even in the second order truncation of the full constraint
on $\ps$, it is $\mathbb{S}_{2}$ that vanishes, not
$\mathbb{S}^{\prime}_{2}$. The fact that dynamics is governed by a
constraint only on $\ps_{o}$ and not on the full $\ps_{\tr}$ has an
important consequence in the quantum theory.

Finally, because $\mathbb{S}^{\prime}_{2}$ depends not only on
perturbations but also on the background variable $\nu$ (i.e. the
scale factor), the dynamical vector field is \emph{not} of the form
$(\Omega_{o}^{\alpha\beta} + \Omega_{1}^{\alpha\beta})\, \partial_
{\beta} H$ for any function $H$ on $\ps_{\tr}$: In striking contrast
to full general relativity, or its FLRW sector, dynamics is not
generated by any Hamiltonian on $\ps_{\tr}$. In the phase space
language, we have to first obtain the dynamical trajectory on
$\ps_{o}$ and then \emph{lift it} to the truncated phase space
$\ps_{\tr}$ using integral curves of $X^{\alpha}_{\dyn}$. This
procedure just reflects the steps normally followed using space of
solutions: one \emph{first} solves for the background scale factor
$a(\eta)$ and the scalar field $\phi(\eta)$ using FLRW equations,
\emph{fixes} a specific solution as the background space-time and
\emph{then} solves for perturbation equations \emph{on this
background}.
%(We have used the conformal time $\eta$ just for definiteness here).
Therefore, in the quantum theory we are led to first solve for the
background wave function $\Psi_o(\nu,\phi)$ and then `lift this
quantum trajectory' to a wave function $\Psi_o\otimes\psi(\Q_{\vec{k}}, \T_{\vec{k}},
\phi)$ describing the evolution of the quantum wave function $\psi$
on the quantum background geometry $\Psi_o$.

\subsection{Quantum FLRW geometry in LQC} \label{s2.2}

Since the phase space of the truncated system is a product,
$\ps_{\tr} = \ps_o\times \t{\ps}_1$, the Hilbert space of quantum
states has the form $\H = \H_o\otimes \H_1$. In this sub-section we
will focus on the space $\H_o$ of quantum states of the background
geometry and the metric operator thereon. The sector $\H_1$ of
perturbations will be discussed in the next sub-section.

\subsubsection{Quantum dynamics of the FLRW background}
\label{s2.2.1}

Although we have used background variables tailored to LQC, the
classical Hamiltonian theory is the same as in general relativity.
It is in the passage to the quantum theory that we use the LQG
techniques. For simplicity, we will first summarize the basic ideas
and return to two important technical points at the end.

Recall that $\ps_o$ is spanned by two canonically conjugate pairs
$(\nu, \phi;\, \b, \pphi)$, and carries a scalar constraint
$\mathbb{S}_o (\nu,\b) =0$ (see (\ref{hc})). Therefore quantum
states are represented by wave functions $\Psi_o(\nu, \phi)$ and the
Dirac quantization procedure would lead us to impose
$\h{\mathbb{S}}_o \Psi_o= 0$, which takes the form
$(\hbar^2\partial_\phi^2 + \Theta)\, \Psi_o = 0$ for a specific
operator $\Theta$. A careful analysis \cite{dm,almmt,abc} of
constrained systems implies that the physical Hilbert space $\H_o$
is spanned by states that satisfy a mathematically `sharper'
Hamiltonian constraint, which can be intuitively thought of as the
`positive frequency square-root' of $\h{\mathbb{S}}_o \Psi_o= 0$
\cite{aps2}:
\be \label{qhc1} -i\hbar\partial_\phi \Psi_o(\nu, \phi) =
\h{H}_o\Psi_o(\nu, \phi) \ee
where $\h{H}_o$ is a self-adjoint operator whose explicit form will
not be needed in this summary. Thus, the constraint is
`de-parameterized': its form suggests that we can interpret the
scalar field $\phi$ as an `internal' or a `relational' time variable
(a la Leibniz) with respect to which the `true' dynamical variable
$\h\nu$ (and also the perturbations $\h{\Q}_{\vk}, \h{\T}_{\vk}$)
evolve. Thus, in the \emph{physical sector} of the theory, $\phi$ is
just a parameter; there is no longer an
operator associated with it. Now, given  $\Psi_o(\nu, \phi_{\B})$ at
the bounce time $\phi = \phi_{\B}$, Eq (\ref{qhc1}) enables one to
`evolve' this initial state to obtain the quantum state
$\Psi_o(\nu,\phi)$ at all times $\phi$. This quantum evolution is
\emph{non-singular}; the wave function undergoes a bounce when the
density reaches a maximum value $\rcr \approx 0.41 \rho_{\rm Pl}$.
How does LQC evade the standard singularity theorems? The
singularity theorem due to Borde, Guth and Vilenkin \cite{bgv},
tailored to inflation, is evaded because the LQC universe has a
contacting phase in the past, violating their assumption of eternal
expansion. And LQC bypasses the original singularity theorems
\cite{he} in general relativity \emph{even in cases when matter
satisfies all energy conditions} because quantum geometry effects
modify the (geometrical) left hand side of Einstein's equations.

In the classical theory, a FLRW solution corresponds to a trajectory
$\nu(t), \phi(t)$, or, eliminating the parameter $t$, just
$\nu(\phi)$. A solution $\Psi_o(\nu,\phi)$ to (\ref{qhc1})
represents the quantum analog of this trajectory. Since there is no sharp trajectory, we no longer have a single,
sharply defined space-time metric in the background. Consequently,
we do not have a canonical parameter representing the cosmic or
conformal time. \emph{At a fundamental level, all dynamics in the
Planck regime refers to the relational time} $\phi$.

We will be interested in a specific class of states $\Psi_o$. To
specify that class, it is convenient to first set $V(\phi)=0$ to get
a qualitative insight into LQC dynamics. This case has been analyzed
in great detail in the LQC literature. One can now start at late
times, when the classical approximation is clearly excellent, and
consider wave functions which are sharply peaked at a point on the
phase space. Under evolution, these states are known to remain
sharply peaked and the peaks are known to follow trajectories
satisfying certain \emph{effective equations} that encode the
leading-order quantum corrections. These effective solutions are in
excellent agreement with the FLRW trajectories of general relativity
when the matter density and the curvature are less than a thousandth
of the Planck scale \cite{vt,asrev,ach3}. But they strongly deviate
from general relativity in the Planck regime: they define
singularity-free, bouncing trajectories. This behavior has been
established analytically in the  k=0, $\Lambda$=0 case \cite{abc}
and was checked numerically in the k=1, as well as $\Lambda\not=$0
cases \cite{apsv,bp,ap}. The effective equations themselves are
rather general and continue to be meaningful when $V(\phi)\not=0$,
or if we replace the scalar field by other matter sources
\cite{mairi1,mairi2}.

In presence of the quadratic potential $V(\phi) = (1/2) m^2\phi^2$,
sharply peaked wave functions have been constructed whenever
numerical simulations are feasible, which corresponds to the case
when the kinetic energy is dominant at
the bounce \cite{aps4}.%
\footnote{In this case, it takes about $10^{6}$ Planck seconds to
reach the onset of slow roll starting from the bounce (see, e.g.,
Table 1). As the potential energy at the bounce increases, this
period increases rapidly and it has been difficult to maintain the
required accuracy for these longer periods.}
Again, the peaks of these wave functions follow solutions to
effective equations. We do not see any a priori reason why this
general behavior will not continue away from kinetic energy
dominated bounces. More importantly, as we will see in sections
\ref{s4} and \ref{s5}, it turns out that states $\Psi_o$ which
undergo a kinetic energy dominated bounce are the most interesting
ones for potentially new physics. Therefore, while our underlying
framework is valid for all states, in our detailed analysis and
numerical simulations we will \emph{restrict ourselves to those
solutions $\Psi_o(\nu,\phi)$ to (\ref{qhc1}) which are sharply
peaked on an effective trajectory from the bounce (at least) until
the trajectory enters the regime in which the matter density and
curvature are so low that general relativity is an excellent
approximation.}

Since the FLRW sector of the phase space is a system with a
Hamiltonian constraint, we used the Dirac quantization procedure and
reinterpreted the constraint as an evolution equation with respect
to the internal time $\phi$. This is why we were naturally led to
the Schr\"odinger picture in which states evolve and operators
don't. However, in the cosmology literature one generally uses the
Heisenberg picture. The transition can be carried out in the usual
fashion. In the Heisenberg picture, states are frozen in time and it
is the volume operator $\h\nu$ (or scale factor operator $\h{a} =
(\kappa\gamma\hbar\, |\h{\nu}|/4\ell^3)^{1/3}$) that evolves with
respect to $\phi$ and the space-time metric operator is given by
\cite{aan2}:
\be \label{gop} \hat{g}_{ab}dx^a dx^b \, \equiv\,  d\hat{s}^2 \,= \,
\hat{H}_o^{-1}\, {\ell^6\,\hat{a}^6(\phi)}\,\hat{H}_o^{-1}\, d\phi^2
\,+\,\hat{a}^2(\phi)\,\,d\x^2 \ee
where we have used the fact that, in the classical theory, the lapse
corresponding to the scalar field time is $N_{\hom} =
a^3\ell^3/\pphi$, and, on physical quantum states, $\h{p}_{(\phi)}
\equiv -i\hbar\partial_\phi = \h{H}_o$. The operator $\hat{g}_{ab}$
lives on a 4-manifold $M$ coordinatized by $\phi, \x$, where $\phi =
{\rm const}$ surfaces are regarded as `space' sections. The
effective trajectories correspond to using the expectation value
$\bar{a} = \langle \Psi_o\, \h{a} \Psi_o\rangle$ in place of $\h{a}$
in (\ref{gop}).

At a fundamental level, then, the parameter $\phi$ serves as the
time variable and the quantum geometry is determined by the physical
(Heisenberg) operators $\h{a}(\phi)$. However, as we will see in the
next sub-section, an unforeseen simplification occurs in the
dynamics of perturbations even in the Planck regime, enabling us to
cast this exact quantum dynamics in terms of the conformal time of a
quantum corrected metric.\\

\emph{Remark:} Mathematically as well as physically, the
deparametrization procedure outlined above is straightforward in the
case when the scalar field potential $V(\phi)$ vanishes (for
details, see \cite{aps2,aps3,asrev}). In the present case one has to
incorporate some subtleties on both fronts \cite{aps4}.

On the mathematical side, it turns out that the formal quantum
constraint $\h{\mathbb{S}}_o$ again has the operator form $\hbar^2
\partial_\phi^2 + \Theta$ as in the $V(\phi)= 0$ case but now the
operator $\Theta$ is no longer essentially self-adjoint
\cite{asrev}. Therefore, to arrive at (\ref{qhc1}), one has to
choose a self-adjoint extension of $\Theta$, and $H_o$ is then the
square-root of the corresponding operator $|\h{\Theta}|$ which is
self-adjoint by construction. This overall situation is the same as
in the case when $V(\phi) =0$ but there is a positive cosmological
constant $\Lambda$. That case is well-understood \cite{ap} and, for
states of physical interest, the results have been shown to be quite
insensitive of the choice of the self-adjoint extension. In the
present case, the existing results indicate that the situation
should be similar.

On the physical side, in the $V(\phi) =0$ case $\phi$ serves as a
\emph{global} time parameter along classical dynamical trajectories.
This is no longer the case when $V(\phi)\not=0$. Consequently, one
has to work with `patches' of dynamical trajectories such that
$\phi$ is monotonic in each patch. In simple examples, the
associated subtleties in the quantum theory have been discussed in
the literature (see, e.g., \cite{cr-time,mb-time}). But for our
purposes, the situation is simpler because we can work just in one
`patch'. This is because we will start at the bounce with $\phi$
increasing and by the time it `turns around' after climbing up the
potential, the energy density and curvature are so low compared to
the Planck scale that general relativity is a good approximation.
Thus, $\phi$ is in fact single valued in the regime in which full
quantum treatment of the background geometry is needed.

\subsubsection{Effective equations}
\label{s2.2.2}

The effective equations of LQC, which track the evolution of the
peaks of wave functions $\Psi_o$, have a number of consequences that
seem surprising if one's intuition is based largely on inflationary
dynamics within general relativity. In this sub-section we will
summarize these features. While our focus is on the quadratic
potential, most of these results hold for general
(regular) potentials.\\

\bu \emph{The Hubble parameter:} Recall that on solutions to
Einstein's equations, the momentum $\b$ conjugate to $\nu$ is
related to the standard Hubble parameter $H =\dot{a}/a$ via $b=
\gamma H$. However, on solutions to the LQC effective equations we
have
\be \label{H} H \,=\, \f{1}{2\gamma\lambda}\, \sin 2\lambda \b
\,\,\approx\,\, ({0.93}) \,\, \sin 2\lambda \b \, , \ee
where $\lambda^2 = 4\sqrt{3}\pi \gamma\lp^2 \approx 5.2$ is the
`area-gap' that sets the discreteness scale of LQC. Thus, in
striking contrast to general relativity, \emph{the Hubble parameter
$H$ is bounded in LQC} by $0.93$ in Planck units.\, $\b$ ranges over
$(0, \pi/\lambda)$; where $\b=\pi/\lambda$ corresponds to the
infinite past and $\b=0$ to the infinite future. General relativity
is recovered in the limit $\lambda \rightarrow 0$, i.e., when
quantum geometry effects can be ignored. In this limit, $b$ ranges
over $(0, \infty)$ and $b=\infty$ corresponds to the big-bang.
Finally, note that while $H$ is monotonic in general relativity (in
the absence of spatial curvature and provided $\rho
>0$), (\ref{H}) implies that this is not the case in LQC.

\bu \emph{The Friedmann equation:} Quantum geometry corrections
modify the left side of Einstein's equations. In particular, the
Friedmann equation becomes
\be \label{lqc-fe2} \f{\sin^2 \lambda \b}{\gamma^2\lambda^2}\,\,
=\,\, \f{8\pi G}{3}\, \rho \,\,\equiv\,\, \f{8\pi G}{3}\,
\big(\f{{\dot\phi}^2}{2} + V(\phi) \big)\, .\ee
To compare with the standard Friedmann equation $H^2 = (8\pi
G/3)\,\rho$, it is often convenient to write (\ref{lqc-fe2}) using
(\ref{H}):
\be \label{lqc-fe} \f{1}{9}\,( \f{\dot{\nu}}{\nu} )^2\, \equiv H^2 =
\f{8\pi G}{3} \,\,\rho\, \big(1 - \f{\rho}{\rcr}\big) \,  \ee
where $\rcr = 3/8\pi G\gamma^2\lambda^2 \approx 0.41 \rp$. Thus the
familiar \emph{linear relation between the Hubble parameter and the
matter density no longer holds.} Nonetheless, as is obvious from Eqs
(\ref{H}) - (\ref{lqc-fe}), away from the Planck regime
---i.e., when $\lambda b \ll 1$, or, $\rho \ll \rcr$--- we recover
classical general relativity.

\bu \emph{Bounces:} In general relativity, the Friedmann equation
implies that, if the matter density is positive, $\dot{a}$ cannot
vanish (in the absence of curvature or a cosmological constant).
Consequently, in any given solution, the universe is \emph{either}
contracting, \emph{or} an expanding. By contrast, (\ref{lqc-fe}) now
implies that $\dot{\nu}$ vanishes at $\rho=\rcr$; this is the
quantum bounce. To its past, the solution represents a contracting
universe with $\dot{\nu} <0$ and to its future, an expanding one
with $\dot{\nu} >0$.

In LQC, $\b$ is monotonically non-increasing (if we ignore the
exceptional de Sitter solutions that exist for certain potentials),
evolving from $\b = \pi/\lambda$ in the infinite past to $0$ in the
infinite future. Eqs (\ref{lqc-fe2}) and (\ref{lqc-fe}) imply that
$\b=\pi/2\lambda$ at the bounce. Thus, each solution undergoes
precisely one bounce.

\bu \emph{Boundedness of physical quantities:} Intuitively one can
think of the big bounce of LQC as the replacement of the big bang in
general relativity. However, all physical quantities remain bounded
at the big bounce.

As we have already noted, the matter density achieves its maximum
value $\rcr$ at the bounce. When the potential $V(\phi)$ is bounded
below, the Ricci scalar ---the only non-trivial curvature scalar in
these models--- is bounded above by $\approx 62$ and $|\dot{H}|$ is bounded
above by $\approx 10.29 $. If the potential $V(\phi)$ is bounded below, say
$V \ge V_o$, then it follows from (\ref{lqc-fe2}) that
${\dot\phi}^2$ is bounded by $2\rcr - 2V_o$. If $V$ grows
unboundedly for large $|\phi|$, then $|\phi|$ is also bounded. For
example, for $V= m^2\phi^2/2$, we have $m  |\phi|_{\rm max} \approx 0.90$.
Finally, one can also show that if $\nu \not=0$ initially, it cannot
vanish in finite proper time along any solution.

Perhaps the most striking contrast with general relativity occurs in
the behavior of the Hubble parameter $H$: It \emph{vanishes} at the
bounce while in general relativity it \emph{diverges} at the
singularity and is large throughout the Planck regime. As one would
expect, $H$ is bounded above in LQC, $|H| \lesssim 0.93$, and
achieves its upper bound in every solution at the end of
super-inflation ($\b = \pi/4\lambda$) that follows immediately after
the bounce.

To summarize, solutions to the effective equations in LQC are
everywhere regular \emph{irrespective of whether one focuses on
matter density, curvature or the scale factor.}

\subsection{QFT on quantum FLRW space-times}
\label{s2.3}

In striking contrast to the FLRW solutions to Einstein's equations,
the quantum geometry $\Psi_o$ is regular; there is no big bang
singularity. The quantum perturbations $\h{\Q}_{\vk}, \h{\T}_{\vk}$
propagate on this \emph{regular quantum geometry.} Therefore this
framework is well suited to incorporate Planck scale physics by
facing various quantum gravity issues directly. Our task is to
specify the dynamics of quantum fields $\h{\Q}_{\vk}, \h{\T}_{\vk}$
on any given \emph{quantum} background $\Psi_o(\nu,\phi)$. At first
this task seems formidable. However, the detailed framework of QFT
on quantum FLRW space-times introduced in \cite{akl} and further
developed in \cite{aan2} shows that there is an unforeseen
simplification.

\subsubsection{Dressed, effective metric} \label{s2.3.1}

Let us start with the tensor perturbations. In the classical theory,
the dynamics of $\T_{\vk}$ is the same as that of a zero rest mass
scalar field on the background FLRW metric. In the quantum theory,
we have the following surprising result.

\begin{itemize}
\item \emph{Within the test field approximation} inherent to our
    truncation strategy, the dynamics of $\h{\T}_{\vk}$ on any
    one of our background quantum geometries $\Psi_o$ is
    completely equivalent to that of the quantum field
    $\h{\T}_{\vk}$ propagating on a smooth but quantum corrected
    metric $\t{g}_{ab}$ given by
\be \label{qcg} \tilde{g}_{ab} dx^a dx^b \equiv d\tilde{s}^2 = -
(\tilde{p}_{(\phi)})^{-2}\, \ell^6\, \tilde{a}^6(\phi)\, d\phi^2
+ \tilde{a}^2(\phi)\, d{\x}^2 \, .\ee
where
\be \label{qpara} (\tilde{p}_{(\phi)})^{-1}  = \langle
\hat{H}_o^{-1}\rangle \quad\quad {\rm and} \quad\quad
\tilde{a}^4 = \f{\langle \hat{H}_o^{-\f{1}{2}}\,
\hat{a}^4(\phi)\, \hat{H}_o^{-\f{1}{2}}\rangle}{\langle
\hat{H}_o^{-1}\rangle}\, . \ee
Here the expectation value is taken in the quantum geometry
state $\Psi_o$ and $\h{a}(\phi)$ is the (Heisenberg) operator on
${\H}_o$ describing the scale factor. This is an exact
mathematical equivalence.
\end{itemize}

Let us now turn to the scalar perturbations. In the classical
theory, the dynamics of $\Q_{\vk}$ is governed not only by the
background FLRW metric $g_{ab}$ but also by a time dependent
external potential $\g$ (see \ref{pert-ham2}). In the quantum theory
we have:

\begin{itemize}

\item Again, \emph{within the test field approximation}, the
    dynamics of $\h{\Q}_{\vk}$ on any one of our background
    quantum geometries $\Psi_o$ is completely equivalent to that
    of the quantum field $\h{\Q}_{\vk}$ propagating on a smooth
    but quantum corrected geometry $\t{g}_{ab}$ given by
    (\ref{qcg}) and a quantum corrected time-dependent potential
    $\t\g$, given by
\be \label{qpot} \t{\g}(\phi) = \f{\langle \h{H}_o^{-\f{1}{2}}\,
\h{a}^2(\phi)\, \h{\g}(\phi) \h{a}^2(\phi)\, \h{H}_o^{-\f{1}{2}}
\rangle}{\langle \hat{H}_o^{-\f{1}{2}}\, \hat{a}^4(\phi)\,
\hat{H}_o^{-\f{1}{2}}\rangle}\, . \ee
\end{itemize}

For tensor modes, this result was derived systematically in
\cite{aan2} and the generalization to the scalar modes follows the
same steps. The essence of the result is that while the quantum
perturbations $\h{\T}(\x, \phi)$ and $\h\Q (\x,\phi)$ propagate on
the quantum geometry of $\Psi_o$, they do not experience all the
details of quantum fluctuations encoded in this state. The net
effect of the quantum geometry on their dynamics is neatly encoded
in just three time-dependent quantities constructed from $\Psi_o$:
$\t{p}_{(\phi)},\, \t{a},\, \t{\g}$. This is similar to what happens
when light propagates in a medium. While it interacts with the atoms
of the medium, the net effect can be encoded in just a few
parameters such as the refractive index. Quantum geometry can be
thought of as the medium through which the quantum fields $\h{\Q},
\h{\T}$ propagate and the net effect of the medium is to provide
quantum corrections (\ref{qpara}) and (\ref{qpot}) to $\eta,\, a$
and $\g$.

Next, note that in this exact result the effective quantities that
enter are not just expectation values of the corresponding
operators; they are much more sophisticated and, in particular, they
are sensitive also the quantum fluctuations in $\Psi_o$. Indeed
their definitions are quite intricate and could not have been
arrived at without a detailed calculation. In particular, in the
classical theory $\pphi(\phi)$ and $a(\phi)$ not only determine the
metric $g_{ab}$ but also the external, time dependent potential $\g$
experienced by the scalar perturbation $\Q_{\vk}$. In the quantum
corrected description, by contrast, the potential $\t{\g}$ is
\emph{not} determined by $\t{p}_{(\phi)}$ and $\t{a}$; it has to be
calculated independently. Conceptually this is an important
implication of the underlying quantum geometry.

We will refer to $\t{g}_{ab}$ as the \emph{dressed, effective
geometry} to distinguish it from the metric $\bar{g}_{ab}$ defined
by the effective trajectories, discussed in section \ref{s2.2}, that
track just the peak of the wave function $\Psi_o$. Similarly, we
will refer to $\t{\g}$ as the \emph{dressed, effective, external
potential}. The result on exact equivalence of dynamics holds only
for the \emph{dressed} effective metric $\t{g}_{ab}$. However, in
practice, from the bounce until the effective trajectory enters the
general relativistic regime, the wave function $\Psi_o$ is so
sharply peaked on the effective trajectory that the difference
between $\bar{g}_{ab}$ and $\t{g}_{ab}$ would be too small to track
accurately in numerical simulations (since even in sophisticated
simulations the numerical errors are much larger). Therefore in
numerical simulations in section \ref{s5} we will not distinguish
between $\t{g}_{ab}$ and $\bar{g}_{ab}$.

Finally, although its coefficients (in the $\phi, \x$ chart) depend
on $\hbar$,\, $\t{g}_{ab}$ is a smooth tensor field. Therefore, it
is now straightforward to pass to the conformal time $\t\eta$ it
defines. We have:
\be \label{conformal} \tilde{g}_{ab} dx^a dx^b\, \equiv\,
d\tilde{s}^2 \, = \, \tilde{a}\, (-d \tilde{\eta}^2 + d\x^2) \, ,\ee
where
\be d\tilde{\eta} = [\ell^3 \tilde{a}^2]\, \t{p}_{(\phi)}^{-1}\,
d\phi \, .\ee
This description is especially convenient in the general
relativistic regime to make direct contact with the cosmology
literature.

To summarize, the dynamics of quantum fields $\h{\T}_{\vk}$ on a
quantum geometry $\Psi_o$ is mathematically the same as that of
their dynamics on curved background geometry (\ref{conformal}). For
the scalar mode $\h{\Q}_{\vk}$, the classical `external potential'
$\g$ has also to be replaced by a quantum corrected external
potential given by (\ref{qpot}). This equivalence greatly simplifies
our task of defining the appropriate Hilbert space and operators for
$\h{\Q}_{\vk}, \, \h{\T}_{\vk}$ because we can import into quantum
field theory on \emph{quantum} FLRW space-times of \cite{akl} the
rich set of techniques that have been developed in the QFT on
\emph{classical} FLRW space-times. We will summarize the main
results in section \ref{s2.3.2}. However, we emphasize that the
equivalence holds \emph{only} if the perturbations can be regarded
as test fields; i.e., their back reaction can be neglected. For now
we will assume that this is the case. At the end, in section
\ref{s6}, we will carry out a self-consistency test by verifying
that the assumption is in fact met.

\subsubsection{Hilbert space $\H_1$ of perturbations and operators thereon}
\label{s2.3.2}

In this subsection we briefly summarize the characterization of the
physical Hilbert $\H_1$ of perturbations and the appropriate
regularization of relevant composite operators, such as energy
density. We will work with the adiabatic regularization approach
\cite{parker66, parker-fulling74}, which is particularly convenient
to perform explicit computations in the cosmological context. For
simplicity of presentation we will first discuss the tensor modes
and then summarize the modifications needed to incorporate the
scalar mode.

We can follow the standard procedure used in the cosmology
literature since the perturbations can be regarded as propagating on
a FLRW metric $\t{g}_{ab}$.%
\footnote{For a more mathematical treatment involving
representations of Weyl algebras, see \cite{aan2}.}
The symmetries of this dressed, effective background $\t{g}_{ab}$
allow us to expand the field operator $\h \T(\x,\t\eta)$ in Fourier
modes
\be \label{fieldexp} \h \T(\x,\t\eta) = \int \!\f{d^3k}{(2\pi)^3}
\big( \h A_{\vk} \, e_k(\t\eta)+\h A^{\dagger}_{-\vk} \,
e^{\star}_k(\t\eta) \big) \, e^{i \vk\cdot\x} \, , \ee
where, as usual, $k = |\vk|$.
\footnote{Since ${\T}(\x,\t\eta)$ is purely inhomogeneous in the
classical theory, it is natural to require the same property for
${\h{\T}}(\x,\t\eta)$. Then, the integral excludes the point $k=0$.
But we will require that $e_k(\t\eta)$ is continuous in $k$ and
$k\,e_k(\t\eta)$ admits a limit as $k \to 0$.}
The operators $\h {\T}(\x,\t\eta)$ satisfy the Heisenberg equation
of motion if and only if the mode functions $e_k(\t\eta)$ are
solutions of the wave equation
\be \label{we1} e''_k(\t\eta)+2 \frac{\t a'}{\t a}\,
e'_k(\t\eta)+k^2 \, e_k(\t\eta)=0 \, ,\ee
were the `prime' denotes the derivative with respect to $\t\eta$.
These solutions $e_k(\t\eta)$ are to provide a generalization of the
positive frequency basis $e^{-i k t}/\sqrt{2 k}$ used in Minkowski
space-time. The canonical commutation relations for the field
operator $\h \T(\x,\t\eta)$ and its conjugate momentum imply
\be [\h A_{\vk} ,\h A^{\dagger}_{{\vk}^\prime}] = i \hbar (2\pi)^3
\, \delta(\vk-{\vk}^\prime)  \, \langle
e_k(\t\eta),\,e_{k^\prime}(\t\eta)\rangle^{-1} \quad {\rm and} \quad
[\h A_{\vk} ,\h A_{{\vk}^\prime}]=0\, , \ee
where
\be\langle e_k(\t\eta),\,e_{k^\prime}(\t\eta)\rangle := \frac{ \t
a^2}{4\kappa}(e_k(\t\eta)\,e^{\star\prime}_{k'}(\t\eta) -
e_k^\prime(\t\eta)e^{\star}_{k'}(\t\eta)) \, . \ee
Therefore, if we require that the basis functions $e_k(\t\eta)$ in
(\ref{fieldexp}) to satisfy the normalization condition $\langle
e_k(\t\eta),\,e_{k}(\t\eta)\rangle= i$, then $\h A_{\vk}$ and $\h
A^{\dagger}_{\vk}$ satisfy the familiar commutation relations of
creation and annihilation operators. Note that, because
$e_k(\t\eta)$ and $e_{k'}(\t\eta)$ are solutions of (\ref{we1}), the
scalar product $\langle e_k(\t\eta),\,e_{k'}(\t\eta)\rangle$ is
constant in time. Therefore it suffices to impose it at some initial
instant.

The first steps in the construction of the Hilbert space $\H_1$ of
tensor perturbations can be summarized as follows: i) Choose a
family of normalized solutions $e_k(\t\eta)$ to (\ref{we1}); ii)
Define the associated vacuum state $|0\rangle$ as the state
annihilated by all $\h A_{\vk}$; and, iii) Construct the Fock space
generated by a repeated action of creation operators $\h
A^{\dagger}_{\vk}$ on the vacuum. With further important
qualifications discussed below, this Fock space will be the required
$\H_1$.

The vacuum state constructed in this way is \emph{invariant under
the translational and rotational isometries} of $\t{g}_{ab}$. This
property will be will be important in our later discussion and
follows from the symmetries of the 2-point function
\be \label{2point} \langle 0\mid \h\T(\x_1, \t\eta_1)\, \h\T(\x_2,
\t\eta_2) \mid 0\rangle = \hbar \int\! \f{d^3k}{(2\pi)^3}\,\, \big[
e^{i\vk\cdot(\x_1-\x_2)}\, e_k (\t\eta_1) e^\star_k(\t\eta_2)\big]\,
. \ee
which suffices to completely characterize this Fock representation.
However, it is clear from the above construction that the vacuum
state is far from being unique: Different choices for the bases
$e_k$ lead to different definitions of operators $\h A_{\vk}$ and
$\h A^{\dagger}_{\vk}$, and therefore different vacua. In the
absence of additional physical inputs ---such as invariance under
the much larger isometry groups available in Minkowski or de Sitter
space-times--- one cannot single out a preferred vacuum. More
importantly, there is no a priori guarantee that the vacuum chosen
by a basis $e_k$ would belong to the Fock space determined by a
different basis $\underbar{e}_k$. There are two sets of potential
problems: ultra-violet and infra-red.

In cosmological backgrounds the ultra-violet issues can be faced by
imposing appropriate regularity condition on the basis $e_k$ that
select the Fock space. In the FLRW backgrounds, one can impose the
{\em adiabatic condition}
\cite{parker66,parker69,parker-fulling74},
i.e. require that the $e_k$ must approach the Minkowski space-time
positive frequency modes, $e^{-i k t}/\sqrt{2 k}$, \emph{at an
appropriate rate} in the limit in which the physical momentum $k/\t
{a}$ is much larger than the energy scale $E_{\rm curv}$ provided by
the space-time curvature. (For a succinct discussion of the
technical statements, see the companion paper \cite{aan2}.) The rate
of approach determines the \emph{adiabatic order} of the basis $e_k$
and controls the ultraviolet behavior of states in the Fock space it
selects. Since the stress-energy tensor is a dimension 4 operator,
for it to be well-defined one needs the basis $e_k$ to be of 4th
adiabatic order. We impose this restriction.

Had the spatial hypersurfaces been compact ---with, say,
$\mathbb{T}^3$ topology rather than $\mathbb{R}^3$--- the Fock
representations arising from any two 4th order adiabatic bases $e_k$
and $\underbar{e}_k$ would have been unitarily equivalent. In
particular, all the 4th order adiabatic vacua would lie in the same
Hilbert space, which we could take as our $\H_1$. However even in
this case there would be no `preferred' vacuum state: each regular
basis would define one and, without additional physical input, there
is a no preferred basis. However, the total number of particles
$\langle 0|\, \h{\underbar{N}}\,|0\rangle$ defined by (the creation
and annihilation operators associated with) a regular basis
$\underbar{e}_k$ in the vacuum state $|0\rangle$ defined by another
regular basis $e_k$ would be finite.

In this paper, the spatial topology is $\mathbb{R}^3$ and there is a
further, infrared difficulty. Now, the adiabatic regularity only
ensures that the spatial \emph{number density} of the `under-barred'
particles in the vacuum $|0\rangle$ is finite. Furthermore, since
each vacuum is translationally invariant, this number density is
constant in space. Therefore, the \emph{total} number is infinite
for the trivial reason that the spatial volume is infinite! This
infrared infinity is generally regarded as physically spurious.
However, mathematically it implies that, generically, the vacuum
$|\underbar{0}\rangle$ defined by a regular basis $\underbar{e}_k$
would have infinite norm in the Fock space defined by another
regular basis $e_k$, whence the two Fock representations are
unitarily inequivalent. However, they can be regarded as
\emph{physically equivalent} in the sense that the expectation
values of any `under-barred' observable \emph{that refers to a
compact region} is well-defined in $|0\rangle$ (and hence on a dense
subspace of the corresponding Fock space). This region could be
$\mathbb{R}_{\rm LS}$, the portion of the (infinite) surface of last
scattering that is accessible to observations (physically,
$\mathbb{R}_{\rm LS}$ is intertwined with the background quantum
geometry $\Psi_o$). For example, for the number operator
$\h{\underbar{N}} (\mathbb{R}_{\rm LS})$, corresponding to the
number of `under-barred particles' in the spatial region
$\mathfrak{R}_{\rm LS}$,  $\langle 0|\,
\h{\underbar{N}}(\mathfrak{R}_{\rm LS})\, |0\rangle$ would be finite
for any regular bases $e_k$ and $\underbar{e}_k$.

Thus, in the $\mathbb{R}^3$ topology under consideration, for
definiteness, we will fix a basis $e_k$ which is regular up to the
4th adiabatic order and take $\H_1$ to be the Fock space it defines
(specific examples of convenient bases will be discussed in section
\ref{s4}). All operators of physical interest to us will be
well-defined on $\H_1$. Furthermore, observables such as energy
density, or particle number and energy in a compact region, so
constructed, will have well-defined expectation values on Fock
spaces constructed from any other basis $\underbar{e}_k$ which is
also regular up to 4th adiabatic order.

The expansion (\ref{fieldexp}) of $\h{\T}(\x,\t{\eta})$ immediately
implies that it is a well-defined operator valued distribution on
$\H_1$, i.e. that $\int d^3x \, \h{\T}(\x,\t{\eta}) f(\x)$ is a
well-defined (self-adjoint) operator on $\H_1$ for every smooth
function $f(\x)$ of compact support. However, since observables such
as energy density involve a product of these operator valued
distributions, they have to be regularized. It is here that the 4th
order adiabatic regularity plays a crucial role. It provides a
natural, mode by mode subtraction scheme that removes the
ultra-violet divergences in a local and \emph{state independent}
manner, while respecting the covariance of the underlying theory.

Let us begin with the energy density operator for tensor modes.
Recall first that in the classical theory the stress-energy tensor
is given by
\be T_{ab}= \f{1}{4\kappa}\,\, \big[\t\nabla_a\T \, \t\nabla_b\T
-\frac{1}{2} \, \t{g}_{ab} \, \t{g}^{cd} \, \t\nabla_c\T \,
\t{\nabla}_d\T \big]\, .\ee
Therefore, (\ref{fieldexp}) implies that the expectation value of
the energy density operator $\h{\rho}^{(\T)}(\x,\t{\eta})$ in the vacuum
state associated with a regular basis $e_k$ is \emph{formally} given
by
\be \langle 0| \h\rho^{(\T)}|0\rangle_{\rm formal}:=\langle 0|
\h{T}_{ab}\t{t}^a \t{t}^b |0\rangle_{\rm formal} =\frac{\hbar}{8
\kappa \tilde{a}^2} \int \frac{d^3k }{(2\pi)^3} \big[
|e^\prime_k|^2+k^2 |e_k|^2 \big]\, , \ee
where $\t{t}^a$ is the unit normal to the homogeneous slices w.r.t.
$\t{g}_{ab}$. The formal expression is ultraviolet divergent. The
adiabatic regularization scheme provides a specific mode by mode
substraction, yielding
\be \label{renenergy1} \langle 0|\h \rho^{(\T)}(\x,\t\eta)| 0\rangle
=\frac{\hbar}{8 \kappa \tilde{a}^2} \int \frac{d^3k }{(2\pi)^3}
\left[ |e'_k|^2+k^2 |e_k|^2 -4\kappa \, C^{(\T)}(k,\t\eta)\right]\,
 ,\ee
where the subtraction term $C^{(\T)}(k,\t\eta)$ is given by
\be \label{C-tensor} C^{(\T)}(k,\t\eta)=\frac{k}{\tilde{a}^2}+
\frac{{\t{a}}'^2}{2 {\t{a}}^4 k}+\frac{4{\t{a}}'^2 {\t{a}}''+\t{a}
{\t{a}}''^2-2 \t{a} {\t{a}}'\, {\t{a}}^{'''}}{8 {\t{a}}^5 k^3} \, .
\ee
This subtraction tames the ultraviolet divergences. But there is a
further subtlety, again because the spatial topology is
$\mathbb{R}^3$ (rather than $\mathbb{T}^3$) and $\T(\x,\t{\eta})$ is
effectively a zero rest mass field: the right side of
(\ref{renenergy1}) has an infrared divergence, directly inherited
from the $1/k^3$ term in $C^{(\T)}(k,\t\eta)$. Therefore, we need to
introduce an infrared cut-off. A natural strategy to handle this
issue is to absorb modes with wavelengths larger than the radius
$\mathfrak{R}(\t{\eta})$ of observable universe into the definition
of the homogeneous background itself. Given an instant of time
$\t{\eta}$, this provides us with a minimum physical frequency (or
maximum physical wavelength) for the modes that are to be treated as
arising from perturbations. Thus, we are led to the following
strategy: Perturbations whose physical wavelength is shorter than
$\mathfrak{R}(\t{\eta})$ will contribute to $\langle 0^{\rm
obv}|\,\h{\rho}\,|0^{\rm obv}\rangle_{\rm ren}$ while those with
larger wave lengths will be treated as contributing to the
background. This procedure is supported by the fact that, if we were
to use a 3-torus spatial topology, the current observational limits
on the radius of the torus is close to $\mathfrak{R}(\t{t})$
\cite{torus}. Therefore, had we used a torus with this radius, the
infrared cut-off we use would have been implemented automatically.

Note that, in terms of co-moving wave numbers $k$ used in our
analysis, this infrared cut-off is in fact \emph{time-independent}:
It is given by $k_{o} = k_\star/8.58$ where $k_\star$ is the
reference mode used by the WMAP (see section \ref{s3.1} below). As
discussed in section \ref{s6.2} of \cite{aan2}, an infrared cut-off
with fixed \emph{co-moving} $k_{\rm min}$ leads to a renormalized
energy density satisfying various criteria that are generally used
to select viable renormalization procedures in QFT in curved
space-times. Thus, the strategy is consistent with the conceptual
framework of the renormalization theory.

The final expression of the expectation value of the renormalized
energy density operator is given by
\be \label{renenergy2} \langle 0|\h \rho^{(\T)}(\x,\t\eta)|
0\rangle_{\rm ren} =\frac{\hbar}{8 \kappa \tilde{a}^2} \int_{k_{\rm
min}}^\infty \frac{d^3k }{(2\pi)^3} \left[ |e'_k|^2+k^2 |e_k|^2
-4\kappa \, C^{(\T)}(k,\t\eta)\right]\, .\ee
where $C^{(\T)}(k,\t\eta)$ is again given by (\ref{C-tensor}). Note
that the right hand side is independent of $\x$ because the vacuum
is translationally invariant. The same procedure yields the matrix
elements of $\h\rho_{\rm ren}$ on the obvious dense subspace of
$\H_1$ obtained by operating on the vacuum by finite linear
combinations of products of creation operators. These of course
depend on both $\x$ and $\t\eta$. The matrix elements of the entire
stress-energy operator can be constructed in the same fashion.  We
have focused on energy density because, as noted below, it is the
integrand of the Hamiltonian operator; its expectation values
determine whether our truncation approximation is valid \cite{aan2}. \\

\emph{Remark:} As shown in \cite{aan2}, the Hamiltonian operator
that governs the evolution of $\h{\T}_{\vk}$ in conformal time
$\t\eta$ has the \emph{formal} expression:
\be \label{pert-ham} \h H_{1,{\rm formal}}^{(\T)} = \,\, \int\!
\f{d^3 k}{(2\pi)^3} \, \big(\frac{2 \kappa}{\t{a}^{2}}\, |\h
\p^{(\T)}_{\vk}|^2 + \frac{\t{a}^2 \, k^2}{8 \kappa} |\h
\T_{\vk}|^2\big)  \, . \ee
(In light of (\ref{pert-ham1}), this expression can be anticipated
\emph{once we know} the equivalence between quantum fields $\h\T$
propagating on the quantum geometry $\Psi_o$ and $\t{g}_{ab}$.) Now,
the standard relation between the energy density and the Hamiltonian
generating evolution in the conformal time implies $\h H_{1,{\rm
formal}}^{(\T)}=  \t{a}^4\int d^3x \,  \h \rho^{(\T)}_{\rm formal}$.
Therefore, even if we replace $\h\rho^{(\T)}_{\rm formal}$ with
$\h\rho^{(\T)}_{\rm ren}$ with its infrared cut-off $k_{\rm min}$,
the vacuum expectation value of the Hamiltonian operator still has
the trivial divergence simply because the integrand is constant and
the total volume is infinite. However, as discussed above, we can
restrict the space integral just to the spatial region
$\mathbb{R}_{\rm obs}(\t\eta)$ that is observable at time $\t\eta$
and obtain a Hamiltonian operator $\h{H}_{\rm obs}(\t\eta)$ tailored
to $\mathbb{R}_{\rm obs}(\t\eta)$. It provides the desired evolution
of operators smeared with (space-time) test functions with support
anywhere in the observable universe. (For a more mathematically
complete discussion, see \cite{aan2}.)\\

Finally, let us consider the scalar perturbations $\h\Q$.
Conceptually the situation is parallel. The mode functions $q_k$
(analogous to the $e_k$ for tensor perturbations) now satisfy the
equation
\be \label{we2} q''_k(\t\eta)+2 \frac{\t a''}{\t a}\,
q'_k(\t\eta)+(k^2 +\t \g) \, q_k(\t\eta)=0 \, .\ee
The energy density operator (that is needed to test the validity of
the truncation approximation \cite{aan2}) is given by:
\be \label{renenergy3} \langle 0|\h \rho^{(\Q)}
(\x,\t\eta)| 0\rangle_{\rm ren} =\frac{\hbar}{2
\tilde{a}^2} \int^\infty_{k_{\rm min}} \frac{d^3k }{(2\pi)^3} \left[
|q'_k|^2+ (\t \g+k^2)\, |q_k|^2 - \, C^{(\Q)}(k,\t\eta)\right]\,
,\ee
where the subtraction term $C^{(\Q)}(k,\t\eta)$ is now given by
\be \label{C-scalar} C^{(\Q)}(k,\tilde\eta)=\frac{k}{\tilde{a}^2}+
\frac{{\tilde{a}}'^2+\tilde a^2 \, \tilde {\g}}{2 \, {\tilde{a}}^4
k}+\frac{- \tilde a^3 \,  \tilde{\g}^2+2 \, \tilde a^2 \tilde a' \,
\tilde{\g}' + 4 \,  {\tilde{a}}'^2 {\tilde{a}}''+\tilde{a} \,
({\tilde{a}}''^2-2 \tilde{a}' \, (\tilde{\g} \, {\tilde{a}}'+
{\tilde{a}}^{'''}))}{8
{\tilde{a}}^5 k^3} \, . \ee%

In summary, we have \emph{lifted} the adiabatic techniques from
quantum field theory in curved space-times to quantum fields
$\h{\T}$ and $\h{\Q}$ propagating on FLRW quantum geometries
$\Psi_o$. These techniques led us to quantum states of perturbations
that have a good behavior in the ultraviolet and to a procedure to
systematically regularize products of operator valued distributions
that are of direct physical interest. Because of $\mathbb{R}^3$
topology, there are some subtleties associated with infrared
divergences. However, they can be handled by restricting attention
to spatially compact regions of direct physical interest.

\section{Initial conditions}
\label{s3}

Now that we have well-defined Hilbert spaces $\H_o$ and $\H_1$ and
physical operators thereon, given an initial state on
$\H_o\otimes\H_1$, we can evolve it all the way through the
inflationary era and calculate power spectra and spectral indices.
Our primary goal is to explore if there exist viable candidates of
initial states that lead to: i) a slow roll inflationary phase that
is compatible with the 7 year WMAP data, ii) predictions for power
spectra and spectral indices that are compatible with current
observations, and, iii) deviations from the BD vacuum at the onset
of inflation. If the answer to the first two questions is in the
affirmative, we will have a viable extension of the inflationary scenario
to the Planck regime. If the answer to the third question is also in
the affirmative, we will have means to test LQC signatures of
pre-inflationary dynamics for future observations. Since the bounce
replaces the big bang in LQC we will specify the initial state
$\Psi_o\otimes\psi$ at the bounce.

The emphasis in this paper is on investigating the \emph{existence}
of such initial conditions rather than on their \emph{uniqueness}.
Nonetheless, at the end, we will provide strong motivation for our
choices using symmetries, regularity and the novel `repulsive force'
in the Planck regime of LQC, created by the underlying quantum
geometry. We hope that future investigations will develop these
considerations into a more systematic procedure to arrive at our
initial conditions from first principles.

\subsection{Initial conditions for the background quantum state $\Psi_o$}
\label{s3.1}

As explained in section \ref{s2.2}, in this paper we focus on
quantum states $\Psi_o$ of the background geometry which are sharply
peaked on effective trajectories of LQC. Therefore, to specify
initial conditions for $\Psi_o$ we have to examine the effective
solutions in some detail. Our task is two-fold: First isolate the
initial data at the bounce for effective solutions, and then single
out the portion $\mathfrak{P}$ of the permissible initial data set
that yields trajectories compatible with the 7 year WMAP data.

Let us begin by noting that in the context of inflationary models,
the seven year WMAP data \cite{wmap} is generally parameterized
assuming that the state of perturbations at the onset of inflation
is the BD vacuum. In this parametrization, a combination of theory
and observations fixes the inflaton mass to be \cite{as3}
\be \label{m} m = 1.21 \times 10^{-6}\,\, . \ee
It also implies that the reference co-moving mode $k_{\star}$ used
by WMAP%
\footnote{$k_\star$ is given by ${k_{\star}}/{a_o}  =  2\times
10^{-3}\, {\rm Mpc}^{-1}$,  or, $k_{\star} = 8.58\, k_o$, where
$a_o$ refers to the scale factor today and, as before, $k_o$ refers
to the wave number that has just re-entered the Hubble radius today.
It is only the combination $k_{\star}/a_o$ that has direct physical
meaning; $2\pi a_o/k_{\star}$ is the physical wave length of this
reference mode today.}
exits the Hubble radius during slow roll at a time $t(k_{\star})$ at
which the Hubble parameter $H$, the slow roll parameter $\epsilon =
- \dot{H}/H^{2}$, the inflaton $\phi$, and its time derivatives
$\dot\phi$ have the following values \cite{as3}:
\ba \label{wmaponset} H(t(k_{\star})) = 7.83 \times 10^{-6}; \quad
&{\rm and}& \quad \epsilon(t(k_{\star})) = 8\times 10^{-3}
\nonumber\\
\phi(t(k_{\star})) = \pm 3.15 \quad &{\rm and}& \quad
\dot{\phi}(t(k_{\star}))  = 1.98\times 10^{-7}. \ea
where the `dot' refers to the cosmic time. Because of observational
error bars, these values are uncertain in a $\approx 2\%$ window. We
will use these values in our analysis and ask at the end whether the
quantum state of perturbations at the onset of the slow roll is
indistinguishable from the BD vacuum. In most of the parameter space
that dictates the initial conditions at the bounce, the answer will
be in the affirmative. For a narrow window in which it is in the
negative, one would have to re-calculate values of these parameters
(see Section~\ref{s5.1}).

Our task then is to find the class of initial conditions for the
effective equations for which dynamical trajectories enter the tiny
region of the phase space defined by (\ref{wmaponset}) and the
associated error bars, \emph{some time} in their future evolution.
As in general relativity, the space of initial data in the effective
theory is 4 dimensional: specification of  $\nu,\b; \phi, \pphi$ at
any time determines a unique solution to effective equations.
(Recall that $\nu \sim a^3$ and  $H \sim \sin 2\lambda \b$.) At the
bounce, we have $\lambda \b = \pi/2$ and furthermore, as in general
relativity, we have two symmetries on the space of solutions: i) a
rescaling symmetry , $\nu(t) \to \alpha \nu(t),\, \phi(t) \to
\phi(t)$, and, ii) and the time reflection symmetry $\nu(t) \to
\nu(-t), \, \phi(t) \to \phi(-t)$. The first allows us to restrict
ourselves to solutions with $\nu|_{\B} =\nu^o$ for some fixed
constant $\nu^o$, while the second lets us focus just on solutions
with $\dot{\phi}|_{\B}>0$;\, solutions with arbitrary $\nu|_{\B}$
and also those with $\dot{\phi}|_{\B} <0$ can be obtained using
these symmetries. Finally, the LQC Friedmann equation
(\ref{lqc-fe2}) implies that the value $\phi_{\rm B}$ of $\phi$ at
the bounce point determines $|\dot{\phi}|_{\rm B}$ there. Thus, the
free data at the bounce is just the value $\phi_{\rm B}$ of the
background inflaton. Next, since the total energy density at the
bounce is fixed, $\rho = \rcr$ on all solutions, it follows that
this free data $\phi_B$ is restricted to lie in a \emph{finite}
interval:
\be \label {phibounce} |\phi_{\B}| \in (0, \phi_{\rm max}),\quad
{\rm where} \quad \phi_{\rm max} = \f{0.90}{m} \approx 7.47 \times
10^{5}. \ee
This interval provides us with the \emph{relevant parameter space}
in this paper.

For definiteness, we will assume $\phi_{\B} >0$ since the sign does
not make a qualitative difference to the analysis. In this case, it
was shown in \cite{as3} that a sufficient condition for the
trajectory to enter the tiny phase space region compatible with the
`WMAP slow roll' is that $\phi_B \ge 0.93$. This excludes only a
tiny part of the full range of allowed values. In \emph{this} sense,
in the effective theory, `almost all' initial conditions at the
bounce point are compatible with the 7 year WMAP data. We would like
to emphasize that this behavior is very non-trivial. At first sight
it may seem that the statement just says that, as in general
relativity, inflationary trajectories are attractors in effective
LQC. However, the implication is \emph{much} stronger because
compatibility with the seven year WMAP data is a sharp
\emph{quantitative} requirement: We need the trajectory to achieve
$\phi= 3.15$ and $\epsilon = 8\times 10^{-3}$ at the time when $H
=7.83 \times 10^{-6}$ within the WMAP error bars. Second, as we saw
in section \ref{s2.2.2}, pre-inflationary dynamics of effective LQC
have several features that are very distinct from what occurs in
general relativity. Therefore, a priori one cannot assume that even
the attractor property of general relativity must carry over.
Finally the pre-inflationary dynamics covers some 11 orders of
magnitude in matter density and curvature. Therefore it is rather
striking that `almost all' effective trajectories starting at the
bounce would enter the tiny region of phase space compatible with
the onset of the desired slow roll.

From now on we will restrict ourselves to this part, $\phi_{\B} \ge
0.93$,  of the parameter space. The restriction involved is tiny.
However, this \emph{does not} imply that the initial conditions for
the quantum state $\Psi_{o}$ at the bounce are generic. The
permissible quantum states are \emph{very special} because we have
asked that $\Psi_o$ be sharply peaked at a point on the constraint
surface of the phase space of the effective theory at the bounce
time. What is generic is only the point at which they are peaked.

We will conclude by clarifying a subtlety about what it means to
have the initial state $\Psi_o(\nu, \phi_{\B})$ at the bounce to be
peaked at a point $(\nu^o,\b = \pi/2\lambda;\, \phi_{\B}^o,
\pphi^o)$ on the constraint surface. Note first that for any given
$\phi_{\B}^o$ and $\nu^o$, the Hamiltonian constraint
(\ref{lqc-fe2}) in the effective theory determines $\pphi^o$. The
subtlety is that, as noted in section \ref{s2.2}, the Hamiltonian
constraint is de-parameterized on the physical sector: Since $\phi$
serves as `internal time', it is a parameter rather than an operator
on the physical Hilbert space $\H_o$. However, because of the
quantum Hamiltonian constraint, $\h{p}_{(\phi)} \Psi_o = -\hat{H}_o
\Psi_o$,\,\, $\h{p}_{(\phi)}$ is a well-defined operator on $\H_o$.
Therefore, the initial state $\Psi_o(\nu,\, \phi_{\B})$ is to be
chosen so that the expectation value of $\h{\nu}$ is $\nu^o$,\, that
of $\h{H}_o$ is the $\pphi^o$ determined by the chosen
$\phi_{\B}^o$,\, and fluctuations in \emph{both} quantities are very
small. For further details on construction of such states, see
\cite{aps1}.

\subsection{Initial conditions for the quantum state $\psi$ of perturbations}
\label{s3.2}

For linear test fields in de Sitter space-time, one can single out a
unique state by demanding that it be regular \emph{and} invariant
under the full de Sitter group. This is the BD vacuum. Since in the
slow roll inflationary phase the Hubble parameter is approximately
constant, space-time geometry is well approximated by the de Sitter
metric. This motivates the usual choice of the BD vacuum as the
quantum state of perturbations at the onset of inflation. However,
in the pre-inflationary phase, especially near the bounce,
space-time geometry is very far from the de Sitter geometry. As we
noted in section \ref{s2.2.2}, the Hubble parameter vanishes at the
bounce. It then increases very quickly to attain a Planck scale
value (in less than a Planck second for kinematic energy dominated
bounces!) \cite{as3}. Therefore, it is not possible to even
formulate the notion of an approximate BD vacuum at or near the
bounce.

However, we can still carry over the central idea that led to the
choice of the BD vacuum in the standard inflationary scenario and
ask that the initial state $\psi(\T_{\vk},\, \Q_{\vk})$ of
perturbations be \emph{regular} and \emph{maximally symmetric}.
Since the state $\Psi_o$ and hence the dressed effective metric
$\t{g}_{ab}$ is invariant under the FLRW isometries, the symmetry
requirement is precisely that $\psi (\T_{\vk},\, \Q_{\vk})$ be
invariant under spatial translations and rotations. We are thus led
to ask that \emph{$\psi(\T_{\vk},\, \Q_{\vk})$ be a 4th order
adiabatic vacuum at the bounce time $\t\eta = \t\eta_{\B}$.} In the
Schr\"odinger picture, the state will evolve but retain its 4th
order regularity and invariance under rotations and translations
(see Eq (\ref{2point})). In the Heisenberg picture, of course, it is
manifest that the conditions are not tied to any specific time.

Although we have imposed maximal possible symmetry requirements that
are compatible with pre-inflationary physics, as our discussion of
section \ref{s2.3} shows, the space of permissible states is still
infinite dimensional. A second requirement on the choice of the
initial $\psi$ is motivated by our truncation approximation.
Clearly, for the approximation to be meaningful, the initial
perturbed state $\psi$ should be such that its contribution $\langle
\psi|\, \h{\rho}(\x,\t\eta_{\B})\, |\psi\rangle_{\rm ren}$ to the
energy density at the bounce time $\t\eta_{\B}$ \emph{should be
negligible compared to $\rcr$,} the energy in the background quantum
geometry $\Psi_o$. In contrast to the regularity and symmetry, this
last requirement \emph{is} tied to a specific time, $\t\eta =
\t\eta_{\B}$.

A key question is whether states satisfying both these conditions
exist. Detailed calculations using 4th order adiabatic regularity
show that they do. A particularly convenient choice for numerical
simulations can be arrived at as follows. For definiteness, let us
consider tensor modes. Then, in the large $k$ limit, \emph{any} 4th
adiabatic order basis $e_k(\t\eta)$ has to approach the explicitly
known approximate (WKB) solutions $f^{(4)}_k(\t\eta)$ to (\ref{we1})
faster than $(\t{a}/k)^{9/2}$. The $f^{(4)}_k(\t\eta)$ are given by
\cite{aan2}
\be \label{approxsol}
f_{k}^{(4)}(\t\eta)=\frac{1}{\t{a}(\t\eta)\,\sqrt{2 \,
W^{(4)}_k(\t\eta)}} \, e^{-i \int^{\t\eta} W^{(4)}_k(\tau) \,
d{\tau} } \, \ee
where $W^{(4)}_k(\t\eta)= \sum_{i=1}^4 W_i$, with
\ba W_0&=& k\, ; \quad W_1 =0\, ; \quad W_2= -\frac{1}{2\, k}
\frac{{\t{a}}''}{\t{a}} \nonumber\\
W_3 &=&0\, ; \quad W_4=\frac{2 {\t{a}}'' {\t{a}}'^2 - 2 {\t{a}}''^2
\t{a} - 2 \t{a} {\t{a}}' \t{a}^{'''} + \t{a}^2 \t{a}^{''''}} {8 k^3
\t{a}^3} \ea
The leading order term of (\ref{approxsol}) corresponds to the
positive frequency solution in Minkowski space and the rest of the
terms are higher order contributions that vanish at different rates
when $(\t{a}/k) \to 0$. Therefore, a natural strategy is to
construct an `obvious' 4th order regular basis $e^{\rm
obv}_k(\t\eta)$, tailored to the bounce time $\t\eta_{\B}$, as
follows. Although the definition of 4th order adiabaticity is only
an asymptotic one, we can ask that $e^{\rm obv}_k(\t\eta)$ and
$f^{(4)}_k(\t\eta)$ share the same \emph{initial data} at the bounce
time $\t\eta =\t\eta_{\B}$:
\be \label{obvious} e_k^{\rm obv}(\t\eta_{\B})=
f_k^{(4)}(\t\eta_{\B})\, ; \quad \quad {\rm and} \quad  \quad
\partial_{\t\eta} e^{\rm obv}_k(\t\eta_{\B})=\partial_{\t\eta}
f^{(4)}_k (\t\eta_{\B})\, . \ee
(This relation will not hold at future times because $e_k^{\rm obv}$
satisfies (\ref{we1}) exactly while $f^{(4)}_k$ satisfies it only
approximately.) The vacuum state $|0^{\rm obv} \rangle$ determined
by $e_k^{\rm obv}$ will be referred to as the `obvious 4th adiabatic
order vacuum'.

There is one subtlety: For this construction to work,
$W^{(4)}_k(\t{\eta}_{\B})$ in (\ref{approxsol}) must be non-negative
since it appears under a square-root in the expression of
$f_k^{(4)}(\t{\eta}_{\B})$. This is the case when the infra-red
cut-off is not too low. In this case, detailed numerical
calculations have shown that $\langle 0^{\rm obv}|\, \h{\rho}(\x,
\t\eta_{\B})\,|0^{\rm obv}\rangle_{\rm ren}$ is less than a few
percent of the background energy density $\rcr$. If the infrared
cut-off is lower, on the other hand, one has to modify the procedure
to obtain the initial data for $e_k^{\rm obv}$: It is to be given by
a suitable smooth extension of $f^{(4)}_k$ and its time derivative
from the high $k$ values where $W^{(4)}_k(\t{\eta}_{\B})$ is
non-negative to the lower $k$ values where it becomes negative.
Consequently, $|0^{\rm obv}\rangle$ is not as `canonical'. However,
the required extension can again be carried out keeping the
renormalized energy density in perturbations only a few percent of
that in the background.

The non-trivial question is whether the energy density in the
perturbation continues to remain negligible for $\t\eta
>\t\eta_{\B}$. The issue is especially non-trivial in the Planck
regime. However, using numerical evolution we will show in section
\ref{s5} that the answer is in the affirmative for the state
$|0^{\rm obv}\rangle$ at least when $\phi_{\B} > 1.23$. Furthermore,
we will argue analytically that the same is true for states in an
open (infinite dimensional) neighborhood of $|0^{\rm obv}\rangle$.
Thus, states satisfying all our initial conditions at the bounce do
exist. \\

\emph{Remark:} Physically, the IR cut-off is dictated by the
physical radius $\mathfrak{R}_{\rm LS}$ of the observable universe
at the surface of last scattering (see section \ref{s2.3.2}). More
precisely it is the \emph{co-moving} radius $\mathfrak{R}_{\rm IR}$
such that $\mathfrak{R}_{\rm LS} = \big(\t{a}(\t\eta_{\rm
LS})\big)\, \mathfrak{R}_{\rm IR}$. Since $\mathfrak{R}_{\rm LS}$ is
observationally fixed, the value of $\mathfrak{R}_{\rm IR}$ is
sensitive to the number of e-foldings between the bounce and the
surface of last scattering. Now, since the background states
$\Psi_o$ under consideration all meet conditions that are compatible
with the 7 year WMAP data at the onset of slow roll, the number of
e-foldings between this onset and the surface of last scattering is
the same for them. What differs from one $\Psi_o$ to another is the
number of e-foldings between the bounce and the onset of slow roll;
they grow rapidly as $\phi_{\B}$ increases. Therefore the subtlety
in the choice of mode functions mentioned above arises only for low
values, $\phi_{\B} \le 1.18$, for which the number of e-foldings
between the bounce and the onset of inflation is low.

\subsection{Physical consideration}
\label{s3.3}

In this paper, our primary goal is to explore the extent to which
the inflationary scenario can be extended to the Planck regime. We
saw that, in effective LQC, a very large portion $\mathfrak{P}$ of
the allowed initial data at the bounce leads to dynamical
trajectories that encounter a slow roll phase compatible with the 7
year WMAP data. This led us to choose an initial state $\Psi_o$ of
the background geometry to be sharply peaked at a point of
$\mathfrak{P}$. The initial conditions on the states $\psi$  of
perturbations were motivated by three considerations: i) Regularity;
ii) Symmetry; and iii) Compatibility with our truncation
approximation. The first two conditions led us to restrict $\psi$ to
be a 4th adiabatic order `vacuum' state and the third led us require
that the energy density in $\psi$ be negligible compared to that in
background quantum geometry $\Psi_o$ at the bounce. This set of
conditions is not overly restrictive in the sense that they allow an
infinite set of initial states. At the same time, because $\Psi_o$
is required to be \emph{sharply peaked} at a point in
$\mathfrak{P}$, and because $\psi$ has to satisfy the three
conditions at the bounce, the allowed initial states
$\Psi_o\otimes\psi$ constitute a very tiny subset of all states in
$\H_o\otimes \H_1$. In this sense they are \emph{very special}. We
will now provide heuristic considerations to help clarify the
physical motivation that underlies our choices. This discussion
could also serve to further narrow down the initial conditions and
opens some avenues for further work.

A common strategy, articulated by Penrose \cite{penrose-weyl} in
particular, is to use known physics and our knowledge of the current
universe to draw conclusions about the special nature of initial
conditions. In LQC this general idea has been used to narrow down
the choice of the initial quantum state $\Psi_o$ of the background
geometry \cite{asrev}. As noted in section \ref{s3.1}, one begins
with the observation that, away from the deep Planck regime, the
universe is extremely well-described by a FLRW solution of
Einstein's equation. One therefore takes states in $\H_o$ which are
sharply peaked on a FLRW solution and evolves them \emph{back in
time}. One finds that the states continue to be sharply peaked with
small uncertainties in \emph{both} `conjugate' variables $\h{H}_o$
and $\h{\nu}$, all the way to the bounce. It is this property that
motivates our initial conditions on $\Psi_o$ at the bounce.

The choice of $\psi$ can be further narrowed down by similar
considerations. In any of our vacua $|0\rangle$ for perturbations,
the expectation values of $\h{\T}_{\vk}, \, \h{\Q}_{\vk}$ and their
conjugate momenta vanish, encoding the idea that departures from
homogeneity and isotropy are small. But we could also ask that the
states be `as peaked on the zero values as is allowed by the
uncertainty principle'. There is a precise sense in which the state
$|0^{\rm obv}\rangle$ we discussed in some detail in section
\ref{s3.2} satisfies this condition. Heuristically, this condition
would say that the initial state $\Psi_o\otimes\psi$ is as
homogeneous and isotropic as is permissible in the quantum physics
of the truncated theory. Therefore this condition will be referred
to as \emph{quantum homogeneity and isotropy} at the bounce. A
precise formulation and implications of this requirement will be
discussed in detail in \cite{aan4}.

As we explained in the beginning of this section, our goal is only
to show the existence of initial conditions that lead to a
self-consistent completion of the standard inflationary paradigm and
we will see in the next two sections that quantum homogeneity and
isotropy at the bounce meets this goal. In future and more
comprehensive investigations one could inquire if one can
\emph{arrive} at this condition from deeper considerations.
Specifically, since the background quantum geometry $\Psi_o$ has a
pre big bang branch, one can ask if the initial conditions at the
bounce naturally arise from the prior evolution. We will conclude by
providing some heuristics that suggest a promising direction to
address this issue.

First, if $\phi_{\B} \ge 1.15$ (the value used in most numerical
plots in this paper), the universe that is observable at the surface
of last scattering expanded from a ball of radius $r_{\B} < 10 \lp$
at the bounce. Therefore, to account for the CMB observations, one
needs quantum homogeneity and isotropy at the bounce only at this
small scale. At first, this smallness of scale may appear to make
the requirement quite weak. Indeed, such an argument is sometimes
made in the standard inflationary scenario where one assumes the
validity of general relativity all the way to the Planck scale and
argues that a similar small region expands out to fill the universe
that is observable at the surface of last scattering. However, it
has been pointed out by many authors, on the initial ball the
required homogeneity and isotropy has to be \emph{extraordinarily}
high and a natural mechanism to achieve this does not appear in the
standard scenario. We will now give qualitative arguments to the
effect that \emph{the underlying quantum geometry of LQG may provide
the missing element.}

Prior to the bounce, the universe would be collapsing thereby
producing strong inhomogeneities that lead to complicated growth of
curvature. However, the LQC models that have been analyzed in detail
suggest the following overall picture. Quantum geometry effects lead
to a new repulsive force. This force is completely negligible until
a curvature scalar approaches the Planck regime. But then it grows
very quickly, overwhelms the classical gravitational attraction that
would have made the curvature scalar singular, and dilutes its
value. Thus, in addition to the `global' density bounce we have
focused on, there would be many `local' bounces associated with
local growth of curvature. The quantum geometry corrections start
to be significant when the curvature is $ \sim 10^{-2}\, \lp^{-2}$,
(i.e. about a thousandth of the maximum value) which corresponds to
a radius of curvature of about $10\, \lp$. Therefore, one may expect
the `dilution' mechanism to be effective in ironing out the
`wrinkles' in the curvature also on that scale. Thus, the missing
element in achieving the required quantum homogeneity and isotropy
on the $10\,\lp$ scale could well be provided by the repulsive
forces that originate in quantum geometry. (See~\cite{mairi1,mairi2}
for a semi-classical mechanism leading to desired initial conditions
in a different context.) This mechanism would not preclude
inhomogeneities and anisotropies on larger scales at the bounce. But
these would correspond to modes whose wavelength at the surface of
last scattering is much greater than the radius of the observable
universe and would not be observable. If these considerations are
borne out, the pre-bounce history of the background would not have a
direct relevance to the post-bounce evolution of the portion of the
universe that is observable.

Of course a much more detailed investigation of this `dilution
effect' is needed to determine if these qualitative ideas are viable
and can be developed into a detailed quantitative argument.\\

\emph{Remark:} If we just focus on the tensor perturbations, then
the requirement of quantum homogeneity and isotropy at the bounce
implies that the state $\Psi_o (\nu)\otimes \psi(\T_{\vk})$
satisfies \emph{a quantum version} of Penrose's Weyl curvature
hypothesis \cite{penrose-weyl} in the following sense. In its
original formulation, space-time geometry was assumed to be
classical, and the condition was that the initial \emph{singularity}
was very special in that the Weyl curvature vanishes there. In LQC
of course there is no singularity and hence it is natural to impose
a condition at the bounce surface at which the matter density
achieves its maximum value. Furthermore, since the electric and the
magnetic parts of the Weyl tensor, $\h{E}_{ab}, \h{B}_{ab}$,  do not
commute in the truncated theory, and their commutator is a c-number,
there is \emph{no} state on which they can both vanish. Thus, the
Heisenberg uncertainty principle tells us that it is not meaningful
to ask that the Weyl tensor operator be zero even at an instant of
time. This is completely analogous to what happens in the Maxwell
theory in Minkowski space, where there is no state at all in the
Fock space that is annihilated by the Maxwell field operator
$\h{F}_{ab}$, or equivalently, the electric \emph{and} magnetic
field operators $\h{E}^a, \h{B}^a$. The expectation values of both
these operators can vanish but this can be achieved on a \emph{very}
large class of states. But we can impose, in addition, the next
natural requirement: i) the product of uncertainties in the electric
and magnetic fields be saturated; and, ii) the uncertainties be
divided equally between the two fields. Some care is required in
formulating these conditions in a precise manner. When this is done,
one finds that there is exactly one state that satisfies the
condition: The standard Maxwell vacuum! In the truncated theory now
under considerations, it is natural to elevate Penrose's Weyl
curvature hypothesis to the quantum theory by asking for states
$\psi$ in which the expectation values of $\h{E}_{ab}, \h{B}_{ab}$
vanish, the product of uncertainties is minimized and the
uncertainties are equally divided between the electric and magnetic
parts. This \emph{quantum version of the Weyl curvature hypothesis}
is satisfied in a weaker sense (i.e., only in the adiabatic limit)
by states $\Psi_o\otimes \psi$ which meet our quantum homogeneity
and isotropy requirement. (For details, see \cite{aan4}.)

\section{Pre-inflationary dynamics of the quantum corrected background}
\label{s4}

This section is divided into two parts. In the first we discuss the
interplay between the dynamics of the background curvature and that
of modes of perturbations in physical terms. In the second we
describe results of numerical simulations of the evolution of the
effective background geometry.

To discuss dynamics in detail, one has to choose a time parameter
defined by the effective background metric. In the cosmology
literature one generally uses conformal time $\eta$ which moves the
big bang to $\eta= -\infty$. In cosmic time $t$, it occurs at $t=0$.
Now, we would like to start our evolution at the LQC big bounce
which replaces the big bang of general relativity. To compare the
results with those of general relativity, it is technically more
convenient to use the cosmic time $\t{t}$ defined by the
$\t{g}_{ab}$ and set the bounce time to be $\t{t}_{\B} =0$.
Therefore, while our analytical considerations use conformal time
$\t\eta$, our numerical evolutions are all carried out in cosmic
time $\t{t}$.

\subsection{Why pre-inflationary dynamics matter}
\label{s4.1}

\begin{figure}[htbp]
\begin{center}
\includegraphics[width=16cm]{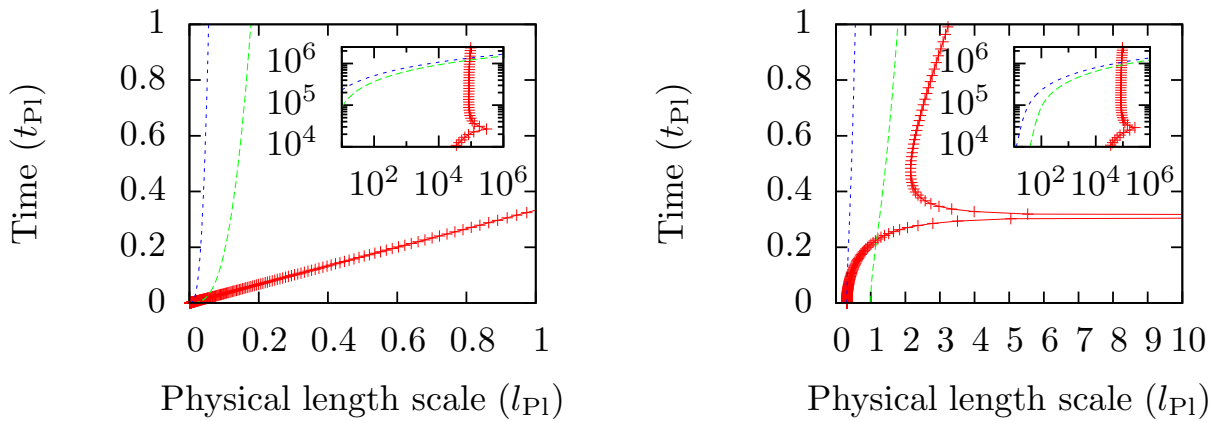}
\caption{\label{fig:bgplot}
Time evolution of the curvature radius (red solid line with dashes)
and of a wave length of interest to observations (green dashed line).
\emph{Left Panel: General relativity.} The modes of interest have wave lengths
less than the curvature radius all the way from the big bang ($\t{t}=0$) until
after the onset of slow roll, shown in the inset ($\t{t} \sim 10^{5} t_{\rm Pl}$).
\emph{Right panel: LQC.} The bounce occurs at time $t=0$ and we have set
$\t{a}|_{\t{t}=0} =1$. The blue dotted line (extreme left) shows the evolution
of the mode whose wave length $\lambda_{\rm phy}\mid_{\t{t}=0}$ at the big
bounce equals the curvature
radius. This mode and modes with smaller wavelengths remain within the
curvature radius until the onset of inflation. On the other hand,
modes with physical wave length \emph{greater} than the curvature radius at
the bounce (for example green dashed line) enters the curvature radius soon after the bounce
and remain within the curvature radius until after the onset of inflation,
shown in the inset ($\t{t} \sim 10^{5} t_{\rm Pl}$). These modes can be
excited due to
curvature while their wave length is greater than the curvature radius and
will not be in the BD vacuum at the onset of inflation. (There are two
points (at $\t{t}\approx 0.3 t_{\rm Pl}$ and $\t{t}\approx 5\times 10^4 t_{\rm Pl}$)
in the LQC evolution at which the $w=1/3$ in the effective equation of state,
whence the scalar curvature vanishes and the radius of curvature
goes to infinity.)}
\end{center}
\end{figure}

There is a common lore that inflation would simply dilute away all
the effects of pre-inflationary dynamics. This belief stems from the
following physical considerations. For definiteness let us consider
tensor perturbations. To bring out the physics that governs
evolution, it is convenient to simplify the dynamical equation by
noting the following mathematical fact. Under a general conformal
transformation $\mathring{g}_{ab} = \Omega^{-2} \t{g}_{ab}$ we
obtain $(\widetilde\Box - (1/6) \t{R}){\t\phi} =
\Omega^{-3}(\mathring{\Box} - (1/6) \mathring{R})\phi$ where $\t\phi
= \Omega^{-1} \phi$. Now, if we let $\Omega =\t{a}$, then
$\mathring{g}_{ab}$ is the flat metric defined by the coordinates
$\t\eta, \x$ and we have $\widetilde{\Box} e_k = \t{a}^{-3}\,
[\mathring{\Box} +(1/6)\, a^2\t{R}]\, (a\, e_k) =0$. Therefore, the
fact that $e_k$ satisfies the wave equation $\widetilde{\Box} e_k
=0$ w.r.t. $\t{g}_{ab}$ (see (\ref{we1})) implies that the rescaled
function $\chi_k(\t\eta) = \t{a}(\t\eta) e_k(\t\eta)$, satisfies a
wave equation \emph{w.r.t. the flat metric $\mathring{g}_{ab}$} in
presence of a time dependent potential $(1/6)\,\t{a}^2\, \t{R}$:
\be \label{we3}
\partial^2_{\t\eta} \chi_{k}(\t\eta)+\t{a}^2(\t\eta) \,
\left(\frac{k^2}{\t{a}(\t\eta)^2}- \frac{\t{R}(\t\eta)}{6}\right)\,
\chi_{k}(\t\eta)=0 \, , \ee
where $\t{R}(\t\eta)$ is the scalar curvature. Eq.~(\ref{we3})
brings out the fact that for modes with physical wave numbers
$k/\t{a}$ much larger than the curvature energy-scale $k_R =
(\t{R}/6)^{1/2}$, the effect of curvature on their evolution can be
ignored. Put differently, if the physical wave length $\lambda_{\rm
phy} = 2\pi \t{a}/k$ of the mode is much smaller than the radius of
curvature $(6/\t{R})^{1/2}$, the mode propagates as if it is in flat
space-time (defined by $\t{\eta}_{ab}$); its dynamics is trivial. In
the standard inflationary scenario, modes that can be observed in
the CMB have physical wave length smaller than the
curvature radius at the onset of inflation.%
\footnote{For (quasi-) de Sitter space-times this is equivalent to
requiring $\lambda \ll \mathfrak{R}_H$ where $\mathfrak{R}_H = 1/H$
is the Hubble radius. This is the condition generally discussed in
the inflationary literature. However, away from slow roll,
$\mathfrak{R}_H$ is not so simply related to the radius defined by
the curvature scale.}
If one were to use general relativity to evolve back in time, then
this inequality can continue to hold all the way to the big bang
(see the left panel in Fig.~\ref{fig:bgplot}). Therefore, the modes
would remain in the same quantum state throughout the
pre-inflationary era, whence one would conclude that the
pre-inflationary dynamics does not have any effect on modes that are
observable in the CMB.

What is the situation in LQC? Because of the exact equivalence
between the evolution of perturbations $\h{\T}$ on the quantum
geometry $\Psi_o$ and on the dressed-effective metric $\t g_{ab}$
given by Eq.~(\ref{qcg}), these general considerations continue to
hold all the way back to the bounce. However, the pre-inflationary
dynamics of the scale factor $\t{a}(\t\eta)$ and the curvature
$\t{R}(\t\eta)$ is now \emph{qualitatively different}. The key
question then is: Do the physical wavelengths $\lambda_{\rm phy}$ of
the relevant modes ever become comparable to $(6/{\t{R}})^{1/2}$
during the pre-inflationary evolution? If they do, their dynamics
would be non-trivial.

A detailed analysis shows that this is the case for modes whose
physical wave length $\lambda_{\rm phy}$ exceeds the curvature
radius at the bounce time (see the right panel of
Fig.~\ref{fig:bgplot}). As Parker \cite{parker66,parker69} showed
already in the sixties, modes that experience curvature are excited.
Indeed, during slow roll, this is the phenomenon that leads to the
inhomogeneities that seed the large scale structure. Qualitatively
the situation is similar in the \emph{pre-inflationary epoch} as
well: modes with $\lambda_{\rm phy}\mid_{t=0}\, > (6/{\t
R})^{1/2}|_{t=0}$ are excited in the Planck regime that immediately
follows the bounce. As a consequence, at the onset of inflation, the
quantum state of perturbations is populated by excitations of these
modes  over the BD vacuum. \emph{Thus, pre-inflationary dynamics can
change the initial conditions for perturbations at the onset of the
slow roll.} Can this lead to physically distinct consequences? The
answer is in the affirmative. It has been shown that the predictions
for the CMB and for the distribution of galaxies are sensitive to
the quantum state of perturbations at the beginning of inflation
\cite{chen,holman-tolley,agullo-parker,ganc,agullo-navarro-salas-parker}.
Furthermore, concrete observational tests to probe these
consequences have recently been proposed
\cite{halo-bias1,halo-bias2,halo-bias3}.

To summarize, pre-inflationary dynamics can lead to predictions that
are distinct from that of the standard inflation. In principle,
there may be excitations over the BD vacuum at the onset of
inflation which are so large as to be in conflict with the observed
power spectrum and spectral index of the scalar perturbations. If
this occurs, the corresponding initial conditions at the bounce
would be ruled out observationally. If the predictions are
compatible with current observations, the departure from the BD
vacuum can still lead to effects that could be observed in the
future. This possibility is of special interest because it relates
the Planck scale dynamics to observations. A priori we do not know
if any or all of these possibilities would be realized for the
permissible values of the free parameter of the theory, $\phi_B$.
These questions can be answered only by a detailed analysis of the
pre-inflationary dynamics.

\subsection{Quantum corrected background geometry}
\label{s4.2}

Recall from section \ref{s2.3.1} that, because $\Psi_o$ is so
sharply peaked, within numerical errors, one can replace the dressed
effective metric $\t{g}_{ab}$ with the $\bar{g}_{ab}$ that traces
the evolution of the peak of $\Psi_o$ and satisfies effective
equations. This is a set of three equations. The first is the
Hamiltonian constraint (\ref{lqc-fe}) that must be satisfied at any
instant of time and the other two provide evolution:
\ba \label{dyn} &&\ddot{\nu} = \f{24\pi \nu}{\rcr}\,\big[(\rho
-V(\phi))^2 + V(\phi) (\rcr-V(\phi))\big]\\
&&\ddot\phi + \f{\dot{\nu}}{\nu}\, \dot\phi + V_{,\phi} =0\, , \nonumber\ea
where as usual the `dot' denotes derivative with respect to comic
time $\t{t}$. As discussed in section \ref{s3.1}, at the bounce time
the only free parameter is the value $\phi_{\B}$ of the background
inflaton which is constrained to lie in the interval $\phi_{\B} \in
(0.93,\, 7.41\times 10^{5})$. For each choice of $\phi_{\B}$ in this
interval, we obtain a solution to the effective equations which
enters the slow roll compatible with the 7 year WMAP data some time
in its future evolution.

\begin{figure}[htbp]
\begin{center}
\includegraphics[width=15cm]{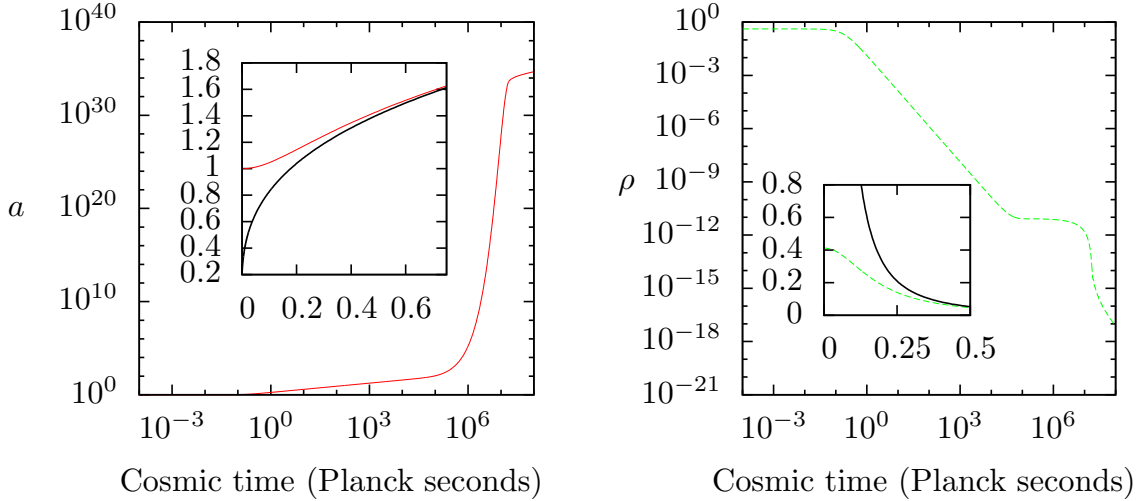}
\caption{\label{fig:avst} Evolution of the scale factor $\t{a}$
(left) and the energy density $\rho$ (right) as a function of
cosmic time for $\phi_{\B} = 1.15$. Slow roll
occurs around $\t{t} \sim 10^6\, t_{\rm Pl}$ and during this phase
the scale factor increases exponentially and the density
remains nearly constant. The insets in both plots highlight the
behavior near $\t{t} =0$. The lower curve in the left inset and the
upper in the right inset show the behavior in general relativity in
which the scale factor goes to zero and the energy density diverges.
By contrast in LQC (upper curve in the left inset and the lower curve
in the right inset) both quantities remain finite at the bounce.}
\end{center}
\end{figure}
The dynamical equations (\ref{dyn}) constitute a set of coupled
ordinary differential equations. They were solved using
\emph{Mathematica}. As an accuracy check, the satisfaction of the
Hamiltonian constraint (\ref{lqc-fe}) was monitored throughout the
evolution. For reasons mentioned in section \ref{s3.1} and will
become clearer in this subsection, in this discussion we will pay
special attention to the case when the bounce is kinetic energy
dominated.

The bounce occurs at $\t{t} =0$ and we use the convention
$\t{a}|_{\t{t} = 0} = 1$ (rather than $\t{a}_{\rm today} =1$).
Immediately after the bounce there is a phase of super-inflation
---i.e. faster than exponential expansion--- because $\dot{H}>0$
at the bounce and $H>0$ immediately after the bounce in all
solutions. At the end of this super-inflation phase, the Hubble
parameter achieves its maximum value $H_{\rm max} \approx 0.93$. For
bounces in which the kinetic energy dominates, the super-inflation
phase is dominated by the Planck scale dynamics because the Hubble
parameter grows from zero to its maximum value in a fraction of a
Planck second. This extremely short lived phase is followed by a
much longer phase during which the LQC corrections to dynamics
become progressively weaker as the inflaton loses kinetic energy
because the friction term (due to the Hubble parameter) is large; it
lasts till $ \t{t} \sim 10^{4}\, t_{\rm Pl}$. At the end of this
phase, the kinetic energy in the inflaton is equal to the potential
energy and the total energy density is about $10^{-10}\, \rho_{\rm
Pl}$ (and hence the curvature has also fallen by ten orders of
magnitude). The inflaton is still going up the potential at the end
of this phase. It continues to lose kinetic energy and at $\t{t}
\sim 10^{5}\, t_{\rm Pl}$, it stops climbing, turns around and
starts descending the potential. Soon there after, when $\t{t} \sim
10^5\, -\, 10^{6}\, t_{\rm Pl}$, it enters the slow roll inflation
compatible with the 7 year WMAP data. The slow roll phase lasts for
about the same time interval as it takes to reach that phase
starting from the bounce through pre-inflationary dynamics.
Fig.~\ref{fig:avst} illustrates the dynamical evolution of the scale
factor and energy density using $\phi_{\B} = 1.15$. (See also
\cite{as3}.)

\begin{figure}
\begin{center}
 \includegraphics[width=13cm]{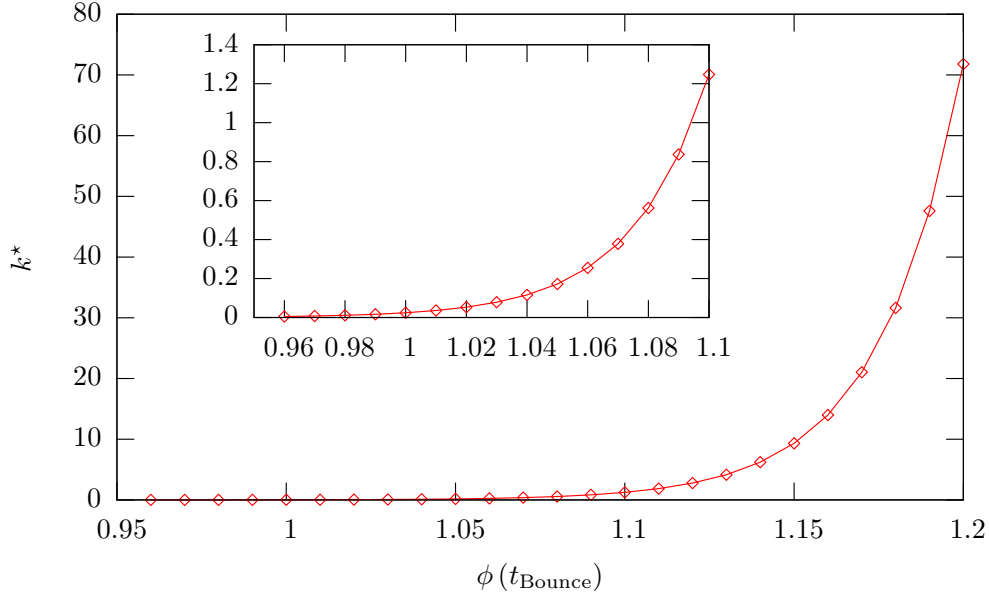}
\caption{\label{fig:kstar}
The value of $k_\star$ as a function of $\phi_{\rm B}$.
As the inset shows, $k_\star$ always increases with $\phi_{\B}$
but the increase is especially pronounced for $\phi_{\B} > 1.1$.
}
\end{center}
\end{figure}

To make contact with the WMAP phenomenology, we need to locate
the reference mode $k_\star$. We fix a value of $\phi_{\B}$ and
obtain the corresponding solution $\t{a}(\t{t}), \phi(\t{t})$.
We then monitor values of the Hubble parameter $H(\t{t})$ and
locate the time $\t{t}(k_\star)$ at which it assumes the
reference value $H(\t {t}(k_\star)) = 7.83 \times 10^{-6}$
determined by the WMAP data (see (\ref{wmaponset})). Then
$k_\star$ is the co-moving wave number of the mode whose
\emph{physical} wave number equals the Hubble parameter at this
time: $k_\star/\t a(\t{t}(k_\star)) = 7.83 \times 10^{-6}$.
This procedure determines both $k_\star$ and the time $\t
{t}(k_\star)$ at which the reference mode $k_\star$ exits the
Hubble radius for any given value of $\phi_B$. As
Fig.~\ref{fig:kstar} shows, $k_\star$ increases very rapidly
with $\phi_{\B}$ . We also calculate
$\epsilon(\t{t}(k_{\star}))$ and verify that it is in a small
window around $8\times 10^{-3}$ in  order to ensure that the
slow roll phase is compatible with the 7 year WMAP data within
its error bars. As a check on numerics, we also verify that
$\phi(\t{t}(k_{\star}))$ and $\dot\phi(\t{t}(k_{\star}))$ are
in agreement with (\ref{wmaponset}). Finally, we calculate the
physical wave length $\lambda_\star$ of the mode $k_\star$ at
the bounce point and the number of e-foldings between the
bounce, $\t{t}=0$ and $\t{t} = \t{t}(k_{\star})$. Results for
an interesting range of values of $\phi_{\B}$ are collected in
Table 1.

\begin{table}
\begin{center}
\begin{tabular}{| c | c | c | c | c | c | }
\hline
$\phi(\tilde{t}_B)$ &  $k_{\star}$  &
$\lambda_{\star}(\tilde{t}_B)$ &  $\tilde{t}_{k_{\star}}$ &
$\ln[a(\tilde{t}_{k_{\star}})/a(\tilde{t}_B)]$
\\ \hline \hline
0.934 & 0.0016 & 4008 & $1.8 \times 10^{5}$& 5.2 \\ \hline
1 & 0.024 & 261 & $5.2 \times 10^{5}$ & 8.0\\ \hline
1.025 & 0.063 & 98.8 &$ 6.4 \times 10^{5}$& 9\\ \hline
1.05 & 0.17  & 37.1 &$ 7.6 \times 10^{5}$& 10\\ \hline
1.075 & 0.45  & 13.8 &$ 8.8 \times 10^{5}$& 11\\ \hline
1.1 & 1.2  & 5.1 & $1.0 \times 10^{6}$& 12\\ \hline
1.15 & 9.17  & 0.63 & $1.25 \times 10^{6}$& 13.9\\ \hline
1.2 & 70.7  & 0.09 & $1.48 \times 10^{6}$& 16 \\   \hline
\,\,1.3\,\, & \,\,\, $4.58 \times 10^3 $ \,\,\,  & \,\,\,
$1.36\times 10^{-3} $ \,\,\, &\,\,\,\,\,\,$1.97 \times 10^{6}$
\,\,\,\,\,\, & 20.2 \\   \hline
1.5 & $2.7\times 10^7 $  & $2.3\times 10^{-7} $& $2.9 \times
10^{6}$& 28.9\\   \hline
\end{tabular}
\end{center}
\caption{\label{tab:1} This table shows the value of the reference
co-moving momentum $k_\star$ used in the WMAP data, the
corresponding physical wavelength $\lambda_\star(\t{t}_\B)$ at the
bounce, the time $\t{t}(k_{\star})$ at which the mode $k_\star$
exits the Hubble radius during inflation, and
$\ln[\t{a}(\t{t}(k_{\star})/\t{a}(\t{t}_\B)]$, the number of e-folds
of expansion between the bounce and $\t{t}({k_\star})$. We focus on
the range for $\phi_{\B}$ that is relevant to explore whether
pre-inflationary dynamics can lead to deviations from the BD vacuum
at the onset of the slow roll.}
\end{table}

Finally, we can use Table 1 to analyze the effect of the
pre-inflationary dynamics on the initial state of perturbations at
the onset of inflation. The argument is semi-heuristic but its main
conclusions provide reliable guidelines because it is based on
rather simple physical considerations. At the bounce, the value of
the background scalar curvature is universal, $\t R_{\B} = 62$. This
provides a \emph{new energy scale} $k_{\LQC}$ and the associated
wave-length $\lambda_{\LQC}$ :
\be k_{\LQC} = \Big(\f{62}{6}\Big)^{\f{1}{2}} \approx 3.21, \quad
{\rm and} \quad \lambda_{\LQC} = \f{2\pi}{k_{\LQC}} \approx 0.977\,
, \ee
(see Eq. (\ref{we3}) for the origin of the factor $6$ and
e.g.~\cite{mairi3} for a discussion of the observational
consequences of states with preferred scales). Since $\t{a}|_{\rm B}
=1$, \, $k_{\LQC}$ and $\lambda_{\LQC}$ are \emph{physical} energy
and length scales at the bounce time. Thus, modes whose physical
wave-length is larger than $\lambda_{\LQC}$ at the bounce will
experience curvature sometime during their evolution and, as
Fig.~\ref{fig:bgplot} shows, this will occur in a short time
interval soon after the bounce. Will any of these modes be in the
observable range for CMB? Recall that, at the surface of last
scattering, the physical wave length $\lambda_\star$ of the
reference mode $k_\star$ is about $(1/8.58) \times \mathfrak{R}_{\rm
LS}$ where $\mathfrak{R}_{\rm LS}$ is the radius of the observable
universe at that time. Therefore, we will have modes which are both
observable in the CMB, \emph{and} which will experience curvature
during their pre-inflationary evolution provided the value of
$\phi_{\B}$ is such that
\be \lambda_\star\,\, \ge\,\, \f{\lambda_{\LQC}}{8.58}\,\, \approx
\,\, 0.11\, . \ee
Table 1 shows that this is possible only if $\phi_{\B} < 1.2$. If
$\phi_{\B} \gg 1.2$, none of the modes that are in the observational
range will encounter significant curvature during their
pre-inflationary evolution. Therefore, the initial 4th order
adiabatic vacuum ---say, $|0^{\rm obv}\rangle$ discussed in section
\ref{s3.2}, for definiteness--- we start out with, will not be
excited in these modes. Therefore the state $|0^{\rm obv}\rangle$ we
constructed at the bounce will be indistinguishable from the obvious
4th adiabatic order vacuum constructed at the onset of inflation.
But since the physical wave lengths of the observable modes are all
within the Hubble radius at the onset of slow roll, it turns out
that the BD vacuum is observationally indistinguishable from the
obvious 4th adiabatic order vacuum defined at the onset of
inflation. Thus, the net result is that $|0^{\rm obv}\rangle$ we
started out at the bounce will have negligible excitations over the
BD vacuum $|\BD\rangle$ at the onset of inflation in the observable
modes. If, on the other hand, $\phi_B < 1.2$, the state at the onset
of inflation would carry excitation over $|\BD\rangle$ in modes that
are observable in the CMB and this can serve as a  source of new
effects \cite{agullo-parker, halo-bias1,halo-bias2,halo-bias3}. This
is why we have focused on kinetic energy dominated bounces.

Of course this conclusion can only be taken as a general guideline.
Detailed calculations are necessary to verify that they did not miss
a subtle but important point and to sharpen the conclusions through
precise quantitative results. In particular, there is no a priori
guarantee that the power spectrum will be compatible with the WMAP
observations for \emph{any} value of $\phi_{\B}$. Nor is it clear
that the underlying truncation scheme is self-consistent, i.e., that
the back reaction of the quantum perturbations can indeed be
neglected all the way from the bounce till the onset of slow roll.
Section \ref{s5} addresses the first issue, and section \ref{s6},
the second.

\section{Pre-inflationary dynamics of perturbations}
\label{s5}

In this section we will summarize the results of numerical evolution
of quantum fields representing scalar and tensor perturbations all
the way from the bounce to the end of slow roll inflation. We will
then calculate the scalar and tensor power spectra and compare them
with those of the standard inflationary scenario. This discussion
will provide a \emph{conceptual completion} of the scenario to
include the Planck regime. In addition, we will find that the
pre-inflationary dynamics can provide corrections to the standard
inflationary scenario with potentially observable consequences.

This section is organized as follows. In \ref{s5.1} we first recall
the WMAP phenomenology and then summarize the modifications that
arise if the state is not the BD vacuum. In \ref{s5.2} we sketch the
plan of the calculations. Illustrative choices of initial conditions
are discussed in \ref{s5.3} and the results of numerical evolution
are presented in \ref{s5.4}. Section \ref{s5.5} discusses a sample
of checks that were performed on the numerics and \ref{s5.6}
summarizes the overall picture. In broad terms, detailed numerical
calculations bear out the physical scenario developed in section
\ref{s4.2} and provide detailed quantitative information on the
scalar and tensor power spectra in LQC and their relation to those
in the standard inflation.

\subsection{WMAP: interplay between theory and observations}
\label{s5.1}

As noted in section \ref{s3.1}, the parametrization of the WMAP data
is normally carried out assuming that the state of perturbations at
the onset of inflation is the BD vacuum. We will start with this
parametrization. For values of $\phi_{\B}$ that lead to the BD
vacuum at the onset, as in the standard theory, we will have a
self-consistent analysis. When there are significant departures from
the BD vacuum, one has to revisit the issue and find a
parametrization that is consistent with the new quantum state at the
onset.

Let us begin with the scalar perturbations. To compare with
observations, one generally calculates the power spectrum ${\mathcal
P}_{\mathcal{R}}(k)$ of co-moving curvature perturbations
$\mathcal{R}_{\vec{k}}$ (see Appendix~\ref{a1}). This is because they are conserved to an
excellent approximation once the mode exits the Hubble radius during
the slow roll: $\mathcal{R}'_{\vec{k}}\to 0$ for $k/a <H $. This
insensitivity to the details of the post-inflation dynamics of the
background geometry greatly facilitates the task of relating the
spectrum of perturbations at the end of inflation with the CMB
temperature fluctuations. It is therefore tempting to carry out the
entire analysis starting from the bounce using
$\mathcal{R}_{\vec{k}}$. Unfortunately, these first order gauge
invariant perturbations fail to be well-defined at the `turn around'
during the inflaton evolution where $p_{(\phi)}=0$. On the other
hand, the Mukhanov-Sasaki variables $\Q_{\vk}$ introduced in section
\ref{s2} are well defined \emph{throughout} the evolution from the
bounce to the end of inflation. Therefore, we numerically evolve
$\Q_{\vk}$ starting from the bounce, calculate the power spectrum
${\mathcal P}_{\Q}(k)$, and convert it to the power spectrum
${\mathcal P}_{\mathcal{R}}(k)$ at the end of inflation via
\be {\mathcal P}_{\Q}(k)= \left|\f{z}{a}\right|^2\, {\mathcal
P}_{\mathcal{R}}(k),\quad {\rm where} \quad z(\t\eta)= -
\f{6}{\kappa}\,\, \f{p_{(\phi)}}{\pi_{(a)}} \equiv
\f{\gamma}{a^2\ell^3}\,\, \f{\pphi}{\b}\, . \ee
Given a 4th adiabatic order vacuum $|\Omega\rangle$ of the scalar
perturbation, the power spectrum ${\mathcal P}_{\Q}(k)$ for the
Mukhanov-Sasaki variables is defined in terms of the associated two
point function in the momentum space:
\be \label{PSQ} \langle \Omega |\,\h \Q_{\vec{k}}\h
\Q_{\vec{k}'}\,|\Omega \rangle := (2\pi)^3\,
\delta(\vec{k}+\vec{k}')\,\, \frac{2\pi^2}{k^3} \, {\mathcal
P}_{\Q}^{\Omega}(k)~, \ee
where all quantities are evaluated a few e-folds after the time $\t
\eta_k$, defined by $k/\t a(\t \eta_k)=H(\t \eta_k)$, at which the
mode $k$ `exits the Hubble radius'. As discussed above, at the onset
of the slow roll we will first set $|\Omega\rangle = |{\BD}\rangle$,
the BD vacuum,  and introduce the modifications that are needed for
a general state $|\Omega\rangle$ at the end. As is well-known, a
basis $q_k^{\BD}(\t{\eta})$ that defines $|{\BD}\rangle$ is given by
\be\label{BDmodes} q_k^{\BD}(\t\eta)=\frac{1}{\t{a}}\,
\sqrt{\frac{-\t\eta\, \pi}{4}} \,\, H^{(1)}_{\mu}(-k\t\eta)~, \ee
where  $H^{(1)}_{\mu}(-k\t\eta) $ is a Hankel function (of the first
kind). The index $\mu$ is determined by the slow roll parameters,
$\mu=3/2+2 \epsilon+\delta$, with
\be \epsilon:=-\f{\dot H}{H^2} \approx \f{1}{2 \kappa}
\big(\f{V_{,\phi}}{V}\big)^2 \quad {\rm
and}\quad \delta:=\f{\ddot H}{2 \dot{H} H} \approx\f{1}{\kappa}\Big[
\big(\f{1}{2}\,\f{ V_{,\phi}}{ V}\big)^2-
\big(\f{V_{,\phi\phi}}{V} \big)\Big]\, . \ee
Here the `dot' refers to the derivative with respect to the cosmic
time $\t{t}$ and we have ignored quantities quadratic in slow roll
parameters to approximate them by functions of the potential
$V(\phi)$. The mode functions $q_k(\t\eta)$ determine the power
spectrum via ${\mathcal P}^{\BD}_{\Q}(k)=\hbar ({k^3}/{2\pi^2})\,\,
|q^{\BD}_k|^2$, with the modes evaluated a few e-folds after
$\t\eta_k$. Using the asymptotic properties of Hankel functions, the
background equations and the relation between ${\mathcal
P}^{\BD}_{\Q}(k)$ and ${\mathcal P}^{\BD}_{\mathcal{R}}(k)$, one
arrives at well-known expression%
\footnote{In the cosmological literature it is common to set
$\hbar=1$ but retain and express $\kappa$ in terms of the `reduced'
Planck mass $M_{Pl}^2= \hbar/\kappa$.},
\be \label{BDPS} {\mathcal
P}^{\BD}_{\mathcal{R}}(k)=\frac{\hbar\kappa}{2\epsilon(\t \eta_k)}
\left(\frac{H(\t \eta_k)}{2\pi}\right)^2\, ,\ee
where, on the right side, the $k$-dependence is encoded in the time
$\t\eta_k$ at which $H$ and $\epsilon$ are evaluated. Since $H$ and
$\epsilon$ change very slowly during slow roll, ${\mathcal
P}^{\BD}_{\mathcal{R}}(k)$ depends very weakly on $k$;\, \emph{it is
almost scale-invariant}. The weak $k$ dependence is parameterized by
the scalar spectral index $n_s(k)$,
\be \label{BDns} n_{s}^{\BD}(k)-1:=\frac{d\ln  {\mathcal
P}^{\BD}_{s}(k)}{d\ln k}= -2 (2 \epsilon(\t \eta_k)+\delta(\t\eta_k))~.
\ee
(Note that in Eqs. (\ref{BDmodes}), (\ref{BDPS}) and (\ref{BDns}),
one ignores terms which are quadratic and higher order in slow roll
parameters.) The WMAP observations \cite{wmap} provide
\be \label{observations} {\mathcal P}_{\R}(k_{\star})=(2.430\pm
0.091) \times 10^{-9}\, \quad {\rm and} \quad
n_s(k_{\star})=0.968\pm 0.012~, \ee
where $k_{\star}$ is the WMAP reference mode introduced before. For
the potentials $V(\phi)$ commonly employed in inflation (quadratic,
quartic, exponential, etc.), the parameters $\epsilon$ and $\delta$
are not independent. For the case of a quadratic potential
considered here, $\delta=0$, and observational
data~(\ref{observations}) gives the values reported in
(\ref{wmaponset})
\be H(\t\eta(k_{\star}))\,=\, 7.83 \times 10^{-6}\, \quad {\rm and}
\quad \epsilon(\t\eta({k_*}))\, =\,8\times 10^{-3} \, . \nonumber~
\ee

To conclude this sub-section, let us discuss the modifications that
occur if the quantum state of the scalar perturbation is
$|\Omega\rangle\neq |{\BD}\rangle$. In view of the initial
conditions discussed in section \ref{s3}, $|\Omega\rangle$ can be
taken to be a `vacuum' which is determined by a 4th adiabatic order
basis $q_k(\t\eta)$ which differs from $q^{\BD}_k(\t\eta)$:
\be \label{modesrelation} q_k(\t\eta)= \alpha_k \,
q_k^{BD}(\t\eta)+\beta_k \,  q_k^{BD\, \star}(\t\eta) ~. \ee
Here, the Bogoliubov coefficients $\alpha_k$ and $\beta_k$ are time
independent and are functions only of $k := |\vec{k}|$. Since the
basis $q_k(\t\eta)$ is also normalized, we have $|\alpha_k|^2 -
|\beta_k|^2=1$. Physically, $|\beta_k|^2$ represents the number
density of the BD `excitations' with momentum $\vk$, per unit
co-moving volume in position as well as momentum space, contained in
the state $|\Omega\rangle$. The power spectrum and the spectral
index determined by  $|\Omega\rangle$ are given by
\ba\label{eq:pow0} {\mathcal P}^{\Omega}_{\mathcal{R}}(k) &=&
{\mathcal P}^{\BD}_{\mathcal{R}}(k) \,\,|\alpha_k+\beta_k|^2~, \nonumber\\
n^{\Omega}_s(k)-1 &=& n_s(k)^{BD} - 1 + \frac{d\ln
|\alpha_k+\beta_k|^2}{d \ln k}~. \ea
If the coefficients $\alpha_k$ and $\beta_k$ are not trivial, i.e.
$\beta_k$ differ significantly from zero, our procedure will fail to
be self-consistent because the background quantities extracted from
observation will differ from (\ref{wmaponset}), the values assumed
at the start of our numerical evolution. We then have to seek a new
solution. In the detailed numerical simulations we have carried out
so far ---with $\phi_{\B} \ge 1.15$--- we find that even when the
state $|\Omega\rangle$ at the onset of the slow roll differs from
the BD vacuum, the difference is sufficiently small for the
parametrization of the data using the BD vacuum to be adequate.\\

\emph{Remark:} If $\phi_{\B}$ is significantly less than $1.15$, the
BD parametrization will not be adequate. In these cases, one can use
a simple `cyclic method'. As before, one can start with the inflaton
mass (\ref{m}) obtained from the BD vacuum and calculate the quantum
state $|\Omega\rangle$ at the onset of inflation. If this state
differs significantly from the BD vacuum, i.e. if the $\beta_k$
coefficients are large, one can recalculate the inflaton mass using
background quantities at time $\t{\eta}(k_\star)$ obtained from
(\ref{eq:pow0}). One can then use this value of the inflaton mass
and \emph{recalculate} the state at the onset of inflation. If this
state $|\Omega_1\rangle$ resulting from this first iteration agrees
with $|\Omega\rangle$ we have a self consistent solution. If not,
one has to continue the iteration procedure until there is self
consistency. Of course, there is no \emph{a priori} guarantee that
this iterative procedure will converge. However, details of the
pre-inflationary LQC dynamics summarized in section \ref{s4.1}
suggest that not only will it converge, but the convergence may be
reached just after the first iteration. Recall that: i) we are
interested in only those modes which are observable in the CMB,
\emph{and}, ii) only the modes whose physical wave lengths exceed
the curvature radius during evolution have BD excitations at the
onset of inflation. As we saw in section \ref{s4.1}, this
circumstance occurs only during a short interval close to the bounce
(see the right panel in Fig.~\ref{fig:bgplot}). At that time, the
background is dominated by quantum-geometry effects, which are
(approximately) universal, i.e., largely insensitive to the value of
the mass $m$ in the potential $V(\phi)$. Therefore, after a change
in the value of $m$ as a consequence of the first iteration, one
does not expect a significant change to the number of created
quanta. Hence, the coefficients $\alpha_k$ and $\beta_k$ should
remain (approximately) unchanged. This argument is supported by a
few preliminary numerical computations for $0.93 \le \phi_{\B} <
1.15$ where one obtains convergence after one iteration. These low
$\phi_{\B}$ appear to exhibit a number of interesting features that
are relevant to non-Gaussianities which will be discussed in detail
in a separate publication.

\subsection{Plan of the calculations}
\label{s5.2}

The scalar and tensor perturbations $\h{\Q}_{\vk},\, \h{\T}_{\vk}$
propagate on the quantum corrected, effective solution
$\t{a}(\t{t})$ and $\phi(\t{t})$, satisfying:
\ba \label{sevo} \ddot{\h{\Q}}_{\vk} + 3 {H} \dot{\h{\Q}}_{\vk} +
\frac{(k^2+ \t\g)}{\t a^2}\, {\h{\Q}_{\vk}} &=& 0, \\
\label{tevo}\ddot{\h{\T}}_{\vk} + 3 {H} \dot{\h{\T}}_{\vk} +
\f{k^2}{\t a^2}\, {\h{\T}_{\vk}} &=& 0 \, , \ea
where, as usual, the `dot' denotes derivative with respect to the
cosmic $\t{t}$,\, ${H} = \dot{\t{a}}/\t{a}$ is the Hubble parameter
and $\t{\g}$ is defined in (\ref{qpot}). As discussed in section
\ref{s3}, we will assume that the quantum state of perturbations is
a vacuum $|\Omega\rangle$ of 4th adiabatic order. We provide
concrete examples of these states in section \ref{s5.3}. Each of
these states is determined by a basis which we will denote by $q_k$
for scalar perturbations and by $e_k$ for tensor perturbations. They
satisfy (\ref{sevo}) and (\ref{tevo}) respectively and are
normalized:
\ba q_k(\t{t})\,\dot{q}^{\star}_{k}(\t{t}) - \dot{q}_k(\t{t})\,
q^{\star}_{k}(\t{t})  &=&  \frac{i}{\t a^3}, \nonumber\\
e_k(\t{t})\,\dot{e}^{\star}_{k}(\t{t}) - \dot{e}_k(\t{t})\,
e^{\star}_{k}(\t{t})  &=&  \frac{4i\kappa}{\t a^3}\ea
Then, as discussed in section \ref{s5.1}, the power spectra are
given by
\be  \label{pow} \mathcal{P}_{\T}(k)=\hbar \frac{k^3}{2\pi^2}\, |e_k|^2 \,
\quad {\rm and} \quad
\mathcal{P}_{\mathcal{R}}(k)=\hbar \frac{k^3}{2\pi^2}
\Big(\frac{\dot\phi}{H}\Big)^2 |q_k|^2 \, ,\ee
where the mode functions are evaluated at the end of inflation.
Thus, to obtain the power spectra, we need to evolve the two sets of
bases functions. Numerical calculations were carried out and we will
present plots for both scalar and tensor perturbations. However, for
brevity we will focus on the scalar modes in most of our narrative.

To bring out the similarities and differences between the chosen 4th
adiabatic order vacuum $|\Omega\rangle$ and the BD vacuum
$|\BD\rangle$  we will also compute the Bogoliubov coefficients
$\alpha_k$ and $\beta_k$ relating the two vacua, defined in
(\ref{modesrelation}). Using this relation, its time derivative and
the normalization condition, it is straightforward to obtain
expressions for $\alpha_k$ and $\beta_k$:
\ba \label{bog} \alpha_k= i \t a^3(\t{t}) [\dot q_k(\t{t}) \, q^{\BD
\star}_k(\t{t})- q_k(\t{t}) \, \dot q^{\BD \star}_k(\t{t}) ] \, ,
\nonumber\\
\beta_k=-i \t a^3(\t{t}) [\dot q_k(\t{t}) \, q^{\BD}_k(\t{t})- q_k(\t{t})
\, \dot q^{\BD }_k(\t{t}) ] \, . \ea
Note that although the right sides of (\ref{bog}) contain
time-dependent terms, $\alpha_k, \beta_k$ themselves are time
independent and they satisfy the identity $|\alpha_k|^2 -
|\beta_k|^2 =1$. These properties will be used to monitor the
numerical accuracy of our simulations.

The numerical evolutions were performed using the internal {\em
Mathematica} numerical integrators (adaptive $4^{\rm th}$ order
Runge-Kutta).

\subsection{Illustrative examples of states} \label{s5.3}

As discussed in section \ref{s3.2}, the quantum states
$|\Omega\rangle$ of interest are vacua of 4th adiabatic order which
satisfy an additional initial condition: At the bounce time
$\t{t}=0$, the renormalized energy density in the state
$|\Omega\rangle$ is negligible compared to the universal energy
density $\rcr$ in the background. These conditions allow an infinite
class of states $|\Omega\rangle$. However, in each numerical
simulation we need to work with a specific choice.

Our detailed simulations were carried out using the state
$|\Omega\rangle = |0^{\rm obv}\rangle$ which was defined in section
\ref{s4.2}. As discussed there, the state is rather simple to
construct. The definition is tied to an instant of time $\t{t}$ and
is `local' in the sense that it is sensitive only to the scale
factor $\t{a}$ and its first four time derivative evaluated at
$\t{t}$. Consider a mathematical example in which $\t{a}$ is
constant during small time intervals around times $\t{t}_1$ and
$\t{t}_2$. In this case, the prescription would lead to the standard
vacuum in Minkowski space-time at these two times and the `particle
creation' from $\t{t}_1$ to $\t{t}_2$ would be exactly as in
Parker's original work of \cite{parker66,parker69} which initiated
the study of quantum fields in cosmological space-times. As a second
example, consider the case in which the $\t{t}$ derivative of $H =
\dot{a}/a$ vanishes in an small interval around a time $\t{t}_o$.
Then the state would be indistinguishable from the BD vacuum for
high frequency modes, i.e., modes for which terms of the order
$\mathcal{O}((k/H\t a)^6)$ can be neglected. Finally, as explained
in section \ref{s4.2}, $|0^{\rm obv}\rangle$  has the attractive
feature that, in a precise sense, it minimizes the uncertainties in
the fundamental canonically conjugate fields at an instant of time.
These properties motivated our use of the `obvious vacuum', tailored
to the bounce time, in detailed numerical calculations.

However, to develop intuition for whether the main results are
sensitive to the specific choice of state, we also carried out
several simulations using three other states. We will summarize
these choices both for completeness and because some of them may be
useful in future investigations.

\begin{itemize}

\item Zero energy-state at the bounce: In the absence of a
    larger group of isometries to single out a preferred quantum
    state, we can further constrain the family of 4th-order
    adiabatic vacua by imposing additional physical conditions.
    Considering the key role played by back-reaction in our
    truncation scheme, one natural requirement is to ask that
    the initial state have vanishing expectation value of the
    renormalized energy density. In the context of adiabatic
    regularization it is possible to construct 4th adiabatic
    order states with this property at any given
    time~\cite{zerostates}. These states are tailored to a given
    time, in the sense that their energy density is not zero at
    later times. Although this condition does not select a
    unique 4th-order vacuum, it narrows down considerably the
    possibilities.

\item Zero stress-energy state at the bounce: A stronger
    prescription to select a natural vacuum state is to demand
    that the state have vanishing expectation value of the
    renormalized energy-momentum tensor at a given time.  When
    the infra-red cut-off in the momentum integrals is
    sufficiently large, this selects a unique state at the
    bounce \cite{zerostates}. However, for smaller cut-offs (and
    general times), this state does not exist. As in the case of
    the obvious 4th order adiabatic vacuum at the bounce (see
    subsection \ref{s3.2}), for those cases some of the modes
    defining that state need to be modified. This introduces a
    freedom in the definition and the resulting state no longer
    has zero expectation value of the energy-momentum tensor.
    Nevertheless, this prescription is useful, in particular
    when the cut-off is sufficiently large (i.e. $\phi_B$ is
    sufficiently large).

\item It follows from our discussion in the beginning of section
    \ref{s4.1} that, if we re-scale tensor modes via $\chi_k
    =\t{a}\ e_k$, then $\chi_k$ can be regarded as modes
    propagating in Minkowski space but with \emph{`time
    dependent frequency'} $w(\tilde{\eta})$. (This statement
    extends also to the scalar modes.) Following the procedure
    used in Minkowski space to construct the preferred vacuum
    state, one can imagine using `instantaneous positive
    frequency modes' at a given time, with frequency
    $w(\tilde{\eta})$. However, this naive choice needs to be
    modified both for low and high $k$. For low $k$ the reason
    is that the `frequency' $w(\tilde{\eta})$ becomes imaginary.
    For large $k$ the frequency is positive, however the modes
    do not define a 4th adiabatic order state and hence, if (for
    example) the energy density is to be well defined, at large
    $k$ the prescription needs to be changed. While there is
    large freedom in those modifications, if one restricts
    oneself to observable modes, the freedom becomes largely
    irrelevant. Therefore, this methods is well suited for quick
    calculations of power spectra and provides a useful way to
    analyze the effects of different initial conditions.

\end{itemize}

In all these three cases, we found that the main features of the
power spectra and the energy density in the perturbations were the
same as those calculated with $|0^{\rm obv}\rangle$ and reported
sections \ref{s5.4} and \ref{s6}. In this sense the results are
robust and not tied to the choice $|\Omega\rangle = |0^{\rm
obc}\rangle$.

\subsection{Numerical evolution} \label{s5.4}

\begin{figure}[htbp]
\begin{center}
\includegraphics[width=10cm]{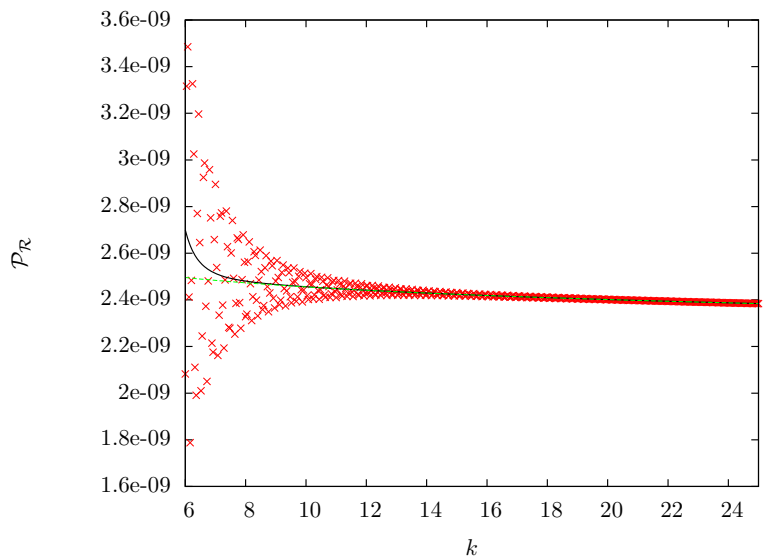}
\caption{\label{PSvsk} Power-spectrum for the scalar perturbations (crosses)
for $\phi_{\B} = 1.15$. The solid line shows the average of the points and the
(green) dashed line is the standard inflationary
power-spectrum which assumes the BD vacuum at the onset of smooth roll.
The deviation from the Bunch-Davies power-spectrum at low $k$ is
{\it not} numerical noise. Rather, modes with these frequencies experience
background curvature during their pre-inflationary evolution and are thus
excited. The state $|\Omega\rangle$ contains excitations over the BD vacuum
$|\BD\rangle$ with quanta of these modes. This leads to a highly oscillatory
power spectrum discussed in the text.}
\end{center}
\end{figure}

Recall that the parameter space for the background is dictated by
the value $\phi_{\B}$ assumed by the background inflaton at the
bounce. Ideally one would like to explore as much of this parameter
space as possible. However, because of the exponential relationship
between $\phi_B$ and $\t{a} \big(\t{t}(k_\star)\big)$ (see
Tab.~\ref{tab:1} and Fig.~\ref{fig:kstar}), even a relatively modest
value of $\phi_B$ results in extremely long integration times, which
rapidly becomes computationally prohibitive. Therefore, we have had
to restrict our attention to $\phi_B \lesssim 2.0$. Fortunately, as
we discussed in section \ref{s4.2}, this interval covers the region
of the parameter space that is physically most interesting. For, our
numerical simulations bear out the expectation based on those
physical considerations: the main results are essentially
insensitive to the value of $\phi_{\B}$ once $\phi_{\B}$ exceeds
$1.2$.

From an observational point of view, the most significant result is
the scalar power-spectrum, plotted in Fig.~\ref{PSvsk}, for
$\phi_B=1.15$ and the `obvious' 4th adiabatic order vacuum, i.e. for
$|\Omega\rangle = |0^{\rm obv}\rangle$. We will now comment on
various features of the plot.

First, the physical argument in section \ref{s4.2} suggested that
modes with $k \gg k_{\LQC} \sim 3.21$ will not be excited because
their physical wave length would be smaller than the curvature
radius throughout the pre-inflationary evolution. Therefore, these
modes would be in the BD  vacuum at the onset of inflation. This
reasoning is borne out because for large $k$ the LQC power spectrum
for these modes with $|\Omega\rangle = |0^{\rm obv} \rangle$ as the
initial state at the bounce is essentially the same as that obtained
in the standard inflationary scenario with $|\BD\rangle$ as the
state at the onset of inflation. Second, for modes with lower $k$
values, the LQC power spectrum shows a highly oscillatory behavior.
This phenomenon has been noted before (see e.g. \cite{brandenberger,
easther, danielsson}). Its origin can be traced back to
Eq.~(\ref{eq:pow0}):
\be\label{eq:pow3}
 P^{\Omega}_{\mathcal R}(k)\,=\,P^{\BD}_{\mathcal R}(k) |\alpha_k+\beta_k|^2
 \,=\, P^{\BD}_{\mathcal R}(k)\left( 1 + 2 |\beta_k|^2
 + 2 {\mathbb Re}\left(\alpha_k \beta_k^\star\right)\right)~,
\ee
where in the second equality we have used the normalization
condition $|\alpha_k|^2 - |\beta_k|^2 = 1$. The oscillatory behavior
of the power-spectra arises from the final interference term, due to
the rapidly changing relative phase. This oscillation is so fast in
$k$ that in any realistic observations --which have a finite $k$
resolution-- they would be `averaged out'. Therefore, in \cite{aan1}
we introduced `bins' in the $k$ space with a width $0.5\lp^{-1}$ and
averaged the oscillations in each bin. It turns out that the
resulting plot is indistinguishable from the one obtained by simply
neglecting the interference term in Eq.~(\ref{eq:pow3}). If we do
so, the LQC power spectrum $P^{\Omega}_{\mathcal R}(k)$ is given
simply by rescaling $P^{\BD}_{\mathcal R}(k)$ with $(1 + 2
|\beta_k|^2)$. Since the number density of excitations in the mode
$k$ is given by $|\beta_k|^2$, this simplified version of
Eq.~(\ref{eq:pow3}) brings to forefront the fact that the
modification of the power spectrum can be traced back directly to
the creation of excitations by curvature. The number of excitations
decay rapidly as $k$ increases because modes with smaller wave
lengths do not experience as much curvature during their
pre-inflationary evolution. Hence the $\beta_k$ coefficients decay
rapidly with $k$, damping the amplitude of the oscillations.

\begin{figure}[htbp]
\begin{center}
\includegraphics[width=10cm]{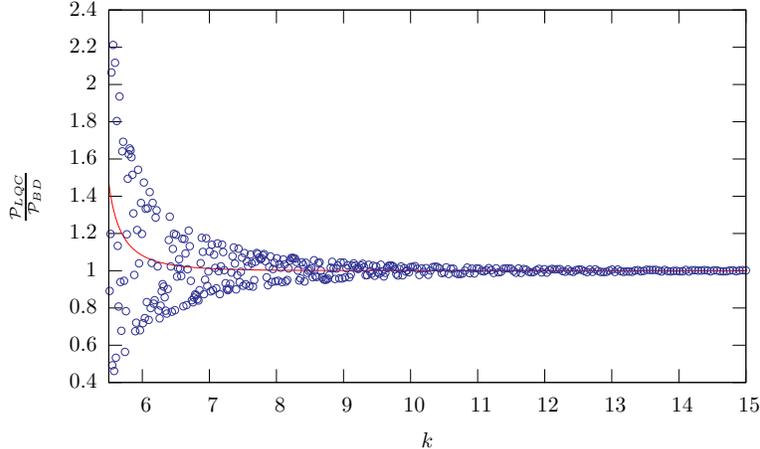}
\caption{\label{fig:ratio}
Ratio of the LQC power spectrum $P^{\Omega}_{\mathcal R}(k)$ for scalar
perturbations to the standard one, $P^{\BD}_{\mathcal R}(k)$ which assumes
the BD state at the onset of slow-roll inflation.
The circles represent the actual ratio which exhibits rapid oscillations
at low $k$ due to a rapidly varying relative phase. The line is the smooth
part of this ratio, which can either be produced by binning the exact data,
or by removing the interference factor, as explained in the text.
This smoothed ratio is simply $1+2 N_{k}$, where $N_k$ is the number of
$\vec{k}$ particles.}
\end{center}
\end{figure}

The relation between the LQC and the BD power spectra is brought out
more clearly in Fig.~\ref{fig:ratio} where we plot the ratio
$P^{\Omega}_{\mathcal R}(k)/P^{BD}_{\mathcal R}(k)$. To define the
BD vacuum, mode functions were not approximated by Hankle functions
for all times. Rather they were taken to be Hankle functions at a
time when the physical frequency of the mode $k_\star$ was
$10^{3}H^{-1}$, to ensure that all the modes of interest were well
inside the Hubble radius, and then evolved numerically. The circles
represent the actual data points which show oscillations while the
solid (red) curve is the plot without oscillations (or, obtained by
binning as discussed above). Since $\phi_B = 1.15$, it follows from
Table 1 that the reference mode used in the WMAP data is $k_\star =
9.17$. It is clear from Fig.~\ref{fig:ratio} that, at this value,
the LQC and the standard inflationary predictions are almost
indistinguishable. Consequently, although the LQC state at the onset
of inflation differs from the standard BD vacuum for low $k$,
nonetheless the LQC prediction is in agreement with the WMAP values
(\ref{observations}) of the amplitude and spectral index which are
evaluated at the time $t(k_\star)$.

\begin{figure}[htbp]
\begin{center}
\includegraphics[width=10cm]{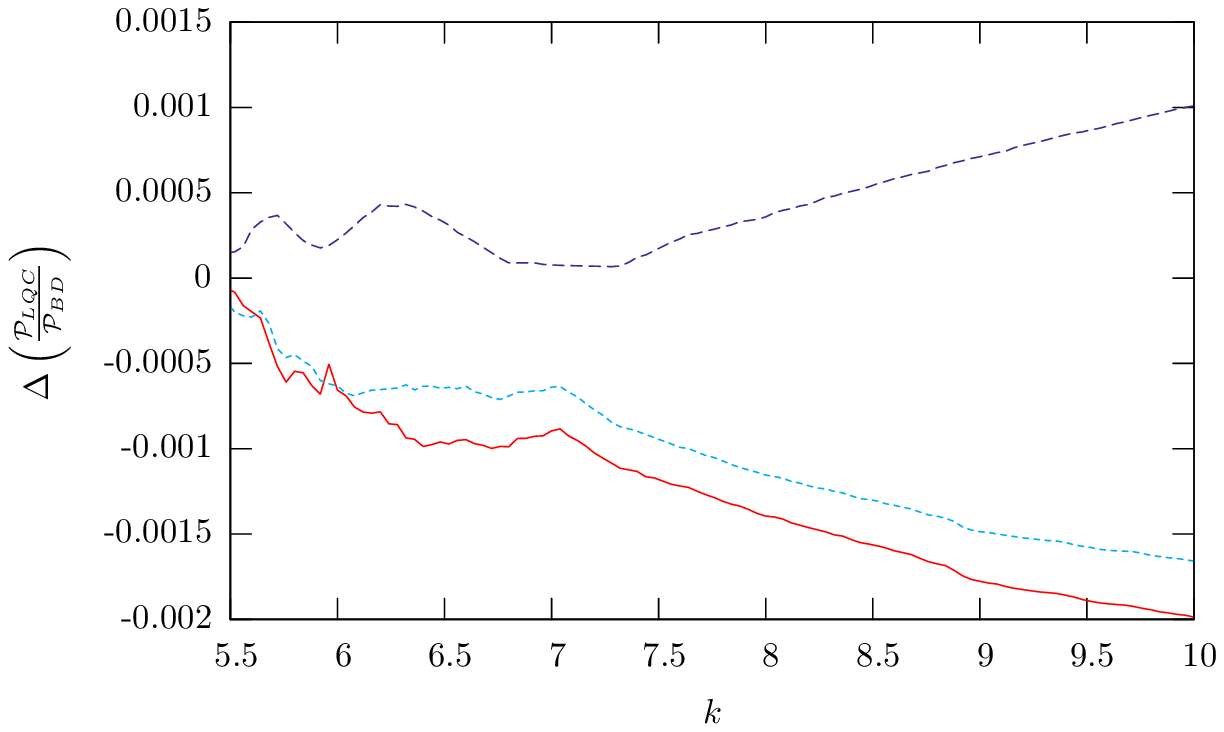}
\caption{\label{fig:PS_diff}
The ratio of the LQC power spectrum $P^{\Omega}_{\mathcal R}(k)$ for
scalar perturbations to the standard one, $P^{\BD}_{\mathcal R}(k)$,
is insensitive to the value of $\phi_\B$. Here we have plotted the
\emph{fractional difference} between ratios calculated using
different values of $\phi_B$, i.e. $\Delta ({ P_{\LQC}}/{ P_{\BD}})
\, \equiv \, \big(\left.({P_{\LQC}}/{ P_{\BD}})\right|_{\phi_\B} - \left.({
P_{\LQC}}/{ P_{\BD}})\right|_{\phi_\B=\mathring\phi}\big) \big( \left. ({
P_{\BD}}/{P_{\LQC}})\right|_{\phi_\B=\mathring\phi}\big)~, $ where
$\mathring\phi = 1.1$
%i.e. $\Delta \left(\frac{ P_{LQC}}{ P_{BD}}\right) \equiv \left(
%\left.\frac{ P_{LQC}}{ P_{BD}} \right|_{\phi_B} - \left. \frac{
%P_{LQC}}{ P_{BD} }\right|_{\phi_B=\phi_c} \right) \left.\frac{
%P_{BD}}{ P_{LQC} }\right|_{\phi_B=\phi_c}~, $ where $\phi_c = 1.1$
and the solid (red) line is for $\phi_\B=1.0$, the dashed (blue) line
is for $\phi_\B=1.05$ and the dotted (cyan) line is for
$\phi_\B=1.15$. Note that the deviation from zero is almost entirely
due to numerical precision, but even so the ratios are the same to
a few parts in $10^3$. }
\end{center}
\end{figure}

So far we focused on the background geometry with $\phi_{\B} =
1.15$. What is the situation with other values? The physical
considerations of section \ref{s4.2} suggest that if we examine the
excitations in any one mode $\vk$ ---ignoring for the moment the
issue of whether it is observable in the CMB--- then the LQC power
spectrum would not be sensitive to the specific value of
$\phi_{\B}$. The same should hold for the ratio
$P^{\Omega}_{\mathcal R}(k)/P^{BD}_{\mathcal R}(k)$ since the
standard inflationary power spectrum makes no reference to
$\phi_{\B}$ at all. Is this expectation borne out?
Fig.~\ref{fig:PS_diff} shows that the answer is in the affirmative.
Specifically, we plot the \emph{fractional difference} between the
scalar power-spectra, calculated using different values of $\phi_B$.
This fractional difference is less than $\approx 0.2\%$ over the
range of $k$ in which there is a significant deviation from the BD
power-spectra. The true difference is, in fact, smaller than this,
since the value at large $k$ is dominated by numerical error (see
Section~\ref{sec:check_of_numerics}).

\begin{figure}[htbp]
\begin{center}
\includegraphics[width=10cm]{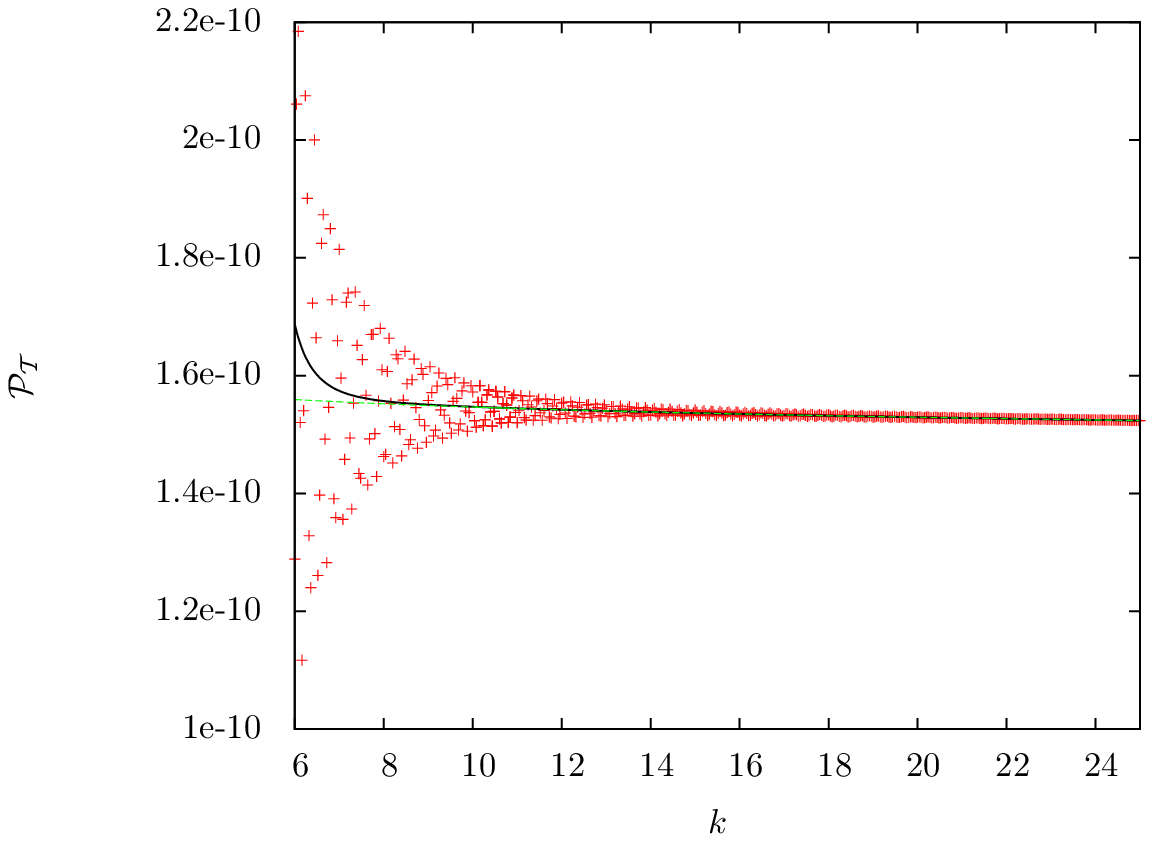}
\caption{\label{fig:tensors} The LQC power spectrum for tensor modes. As
for scalar modes, we have set $\phi_B=1.15$ and used the `obvious'
4th adiabatic order vacuum at the bounce, $|\Omega\rangle =
|0^{\rm obv}\rangle$ (crosses). The solid line is the average and the dashed line is the standard tensor power-spectrum
assuming the BD at the onset of slow-roll inflation.}
\end{center}
\end{figure}

These results lead us to the following interesting overall picture.
If we look at the entire range of wave numbers $k$, the ratio of the
LQC and BD power spectra is \emph{essentially universal,} i.e.,
insensitive to the value of $\phi_\B$ (at least for the range of
$\phi_{\B}$ we analyzed in detail). Modes with $k \gg k_{\LQC} =
3.21$ are in the BD vacuum at the onset of inflation while those
with lower values of $k$ are excited because they experience the
background curvature in the Planck regime immediately after the
bounce. \emph{What changes with $\phi_{\B}$ is the window in the $k$
space that is accessible to observations in the CMB.} As Table 1
shows, each $\phi_B$ determines the co-moving value of $k_\star$,
the reference mode used in by WMAP (consistent with our convention
$\t{a}|_{\t{t} =0} =1$). The observationally relevant window is
given by $(k_o, 2000k_o)$, where $k_o = k_{\star}/8.58$, and moves
rapidly to the right as $\phi_{\B}$ increases.

\begin{figure}
\begin{center}
\includegraphics{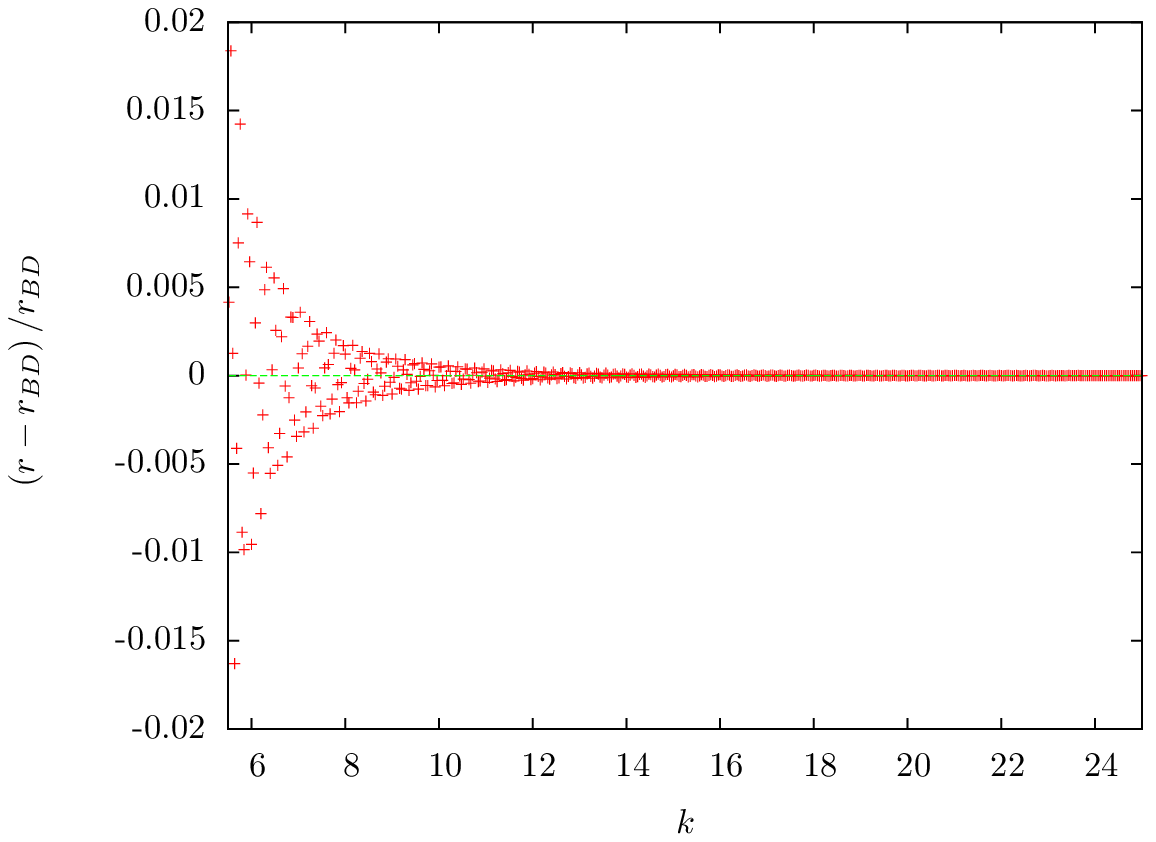}
\caption{\label{fig:tensor_ratios} Plot of the fractional difference
between the LQC tensor to scalar ratio, $r_{\LQC}$ and the usual
prediction from standard inflation, $r_{\BD}$ i.e.
$({r_{\LQC} - r_{\BD}})/{r_{\BD}}~.$
The line is the average of points (see discussion below
Eq.~(\ref{eq:pow3})).}
\end{center}
\end{figure}

\begin{figure}[htbp]
\begin{center}
\includegraphics[width=10cm]{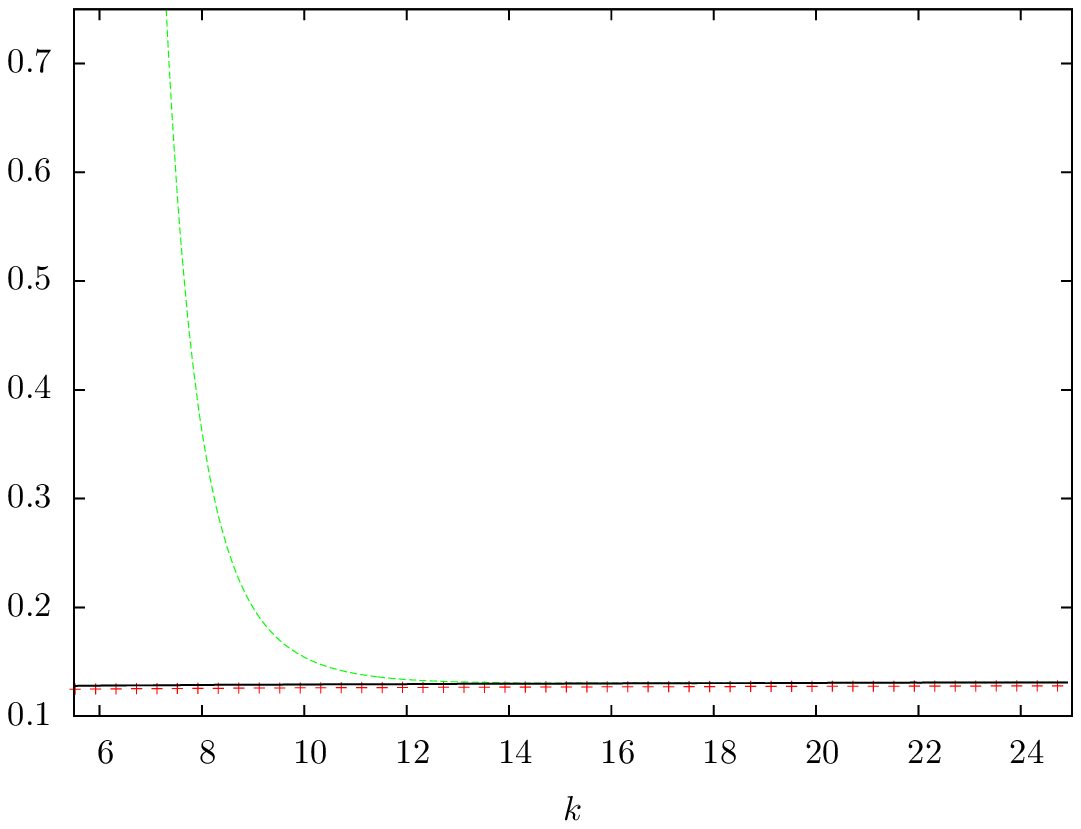}
\caption{\label{fig:consistency} Modification of the standard inflationary
consistency relation $r_{\BD}\approx - 8 n^{\BD}_t$,
due to pre-inflationary dynamics. The green dashed line is $ -8
n_t$, the solid black line is $-8\big( n_t - \big({{d}
\ln ( |\alpha_k|^2 + |\beta_k|^2 ) }/{ {d} \ln k }\big)$ and
the red crosses are the numerically calculated tensor to scalar
ratio.}
\end{center}
\end{figure}

Next, in Fig.~\ref{fig:tensors} we plot the power-spectrum of the
tensor modes. Fig.~\ref{fig:tensors} is very similar to the power
spectrum of the scalar modes shown in  Fig.~\ref{PSvsk}, even though
the evolution equation (\ref{sevo}) for the scalar mode includes a
potential $\t\g$ which is absent in the tensor modes. This is
because from its definition (\ref{g}), it is clear that the
potential is proportional to $m^2$, the square of the inflaton mass,
and $m^2 \sim 10^{-12}$. Since the $\r$ that features in (\ref{g})
ranges between $0$ and $1$ and, for kinetic dominated bounces, $\phi
< 10$ during the entire pre-inflationary evolution, the effects of
the external potential $\t\g$ turns out to be negligible for the
range of $k$'s of interest to observations. As a consequence, the
evolution of the two sets of bases functions is almost
indistinguishable: $|\beta^{\Q}_k|^2 \simeq |\beta^{\T}_k|^2$; in
the Planck regime, the same amount of quanta are created in tensor
and scalar modes. An immediate consequence is that the LQC ratio of
the tensor to scalar power spectra, $r_{\LQC}$,(when averaged) is
the same as in standard inflation
\be r_{\LQC}:=\frac{ 2 P^{\Omega}_{ \T}}{P^{\Omega}_{\mathcal R}}\,
=\, \frac{ 2 P^{\BD}_{\T} \, (1+2 |\beta^{\T}_k|^2)}{P^{\BD}_{\mathcal{R}}\,
(1+2 |\beta^{\Q}_k|^2)}\,\approx\, \frac{ 2 P^{\BD}_{\T}(k)
}{P^{\BD}_{\mathcal R}(k) } \,=\,  r_{\BD} \, . \ee
This result is shown in Fig.~(\ref{fig:tensor_ratios}), where we
plot the ratio $r_{\LQC}/r_{\BD}$.

To conclude, let us comment on an inflationary consistency relation.
An important result of the standard inflationary scenario is the
relation between the tensor-to-scalar ratio and the tensor spectral
index. In slow roll inflation this relation reads $r_{\BD}\approx -
8 n^{\BD}_t$, where the approximation indicates that terms of
quadratic or higher order in the slow roll parameters are neglected.
Since this expression holds independently of the inflaton potential
and relates two independent observable quantities, it serves as a
test of the standard scenario. Forthcoming observation of the effect
of tensor perturbations in the CMB will provide a test of this
relation.  Does LQC modify this relation? We have seen that the
tensor-to-scalar ratio remains unmodified. However, the tensor
spectral index {\em is} modified because of the pre-inflationary
evolution. The LQC tensor spectral index is obtained from the tensor
power spectrum after averaging

\be n_t=\frac{d \ln  P^{\Omega}_{ \T}}{\ln k}= \frac{d\ln P^{\BD}_{
\T}}{\ln k}+ \frac{d\ln (1+2 |\beta^{\T}_k|^2)}{d \ln
k}=n_t^{\BD}-1+\frac{d\ln (1+2 |\beta^{\T}_k|^2)}{d \ln k} \, .\ee
Therefore, the LQC  consistency relation is given by
\be  r_{\LQC}\approx - 8 \left( n_t-\frac{d\ln (1+2 |\beta^{(\T)}_k|^2)}{d
\ln k}\right) \, \ee
where, as before, we have averaged over the rapid oscillations.

Fig.~\ref{fig:consistency} shows that this relation is satisfied to
an excellent approximation in numerical simulations. It shows the
imprint left by the pre-inflationary dynamics which is potentially
observable: a deviations from the standard inflationary prediction
at low $k$'s.

\subsection{Checks of the numerics}\label{sec:check_of_numerics}
\label{s5.5}

\begin{figure}[htbp]
\begin{center}
\includegraphics[width=10cm]{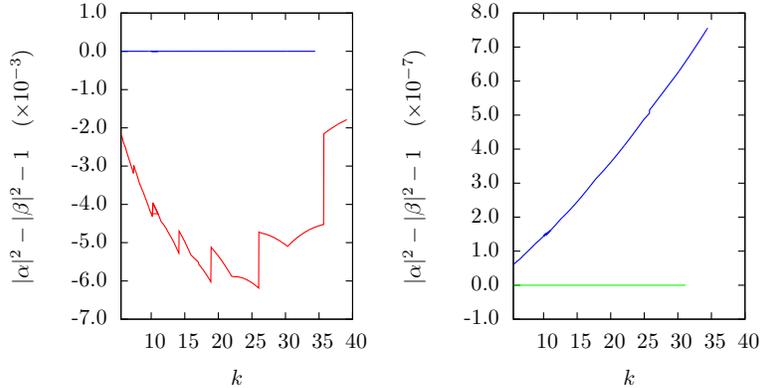}
\caption{\label{fig:ba_precision} These plots show the accuracy to
which the normalization of the modes is maintained throughout the
numerical integration. The Bogoluibov coefficients given in
Eq.~(\ref{modesrelation}) satisfy $|\alpha|^2 - |\beta|^2=1$. These plots
correspond to a (cosmic) time $\approx 5\times 10^4 t_{\rm Pl}$, for
$\phi_{\B}=1.15$. The left plot shows the normalization for
precision/accuracy goals within the numerical integrator
being $10^{-10}$ (lower jagged line) and $10^{-15}$ (upper flat line),
whilst the right plot is for precision/accuracy goals of $10^{-15}$ (upper,
rising line) and $10^{-20}$ (lower flat line). Note the different scales
on the two plots.}
\end{center}
\end{figure}

The accuracy of the simulations has been checked by carrying out
multiple tests. We verified that:
\begin{itemize}
\item The norm of the evolved modes $e_k(\eta)$ and $q_k(\eta)$
    is preserved under the numerical evolution.
\item The Bogoluibov coefficients computed using Eq.~(\ref{bog})
    are time independent even though individual terms on the
    right side of this equation depend on time.
\item The relation $|\alpha_k|^2-|\beta_k|^2=1$ is satisfied at
    all times and for all $k$.
\item The two expressions for the power spectrum,
    Eq.~(\ref{pow}) and Eq.~(\ref{eq:pow0}) agree.

\end{itemize}
\noindent Although at the theoretical level some of these are just
identities, they are excellent tests for the numerics.

\begin{figure}[htbp]
\begin{center}
\includegraphics[width=10cm]{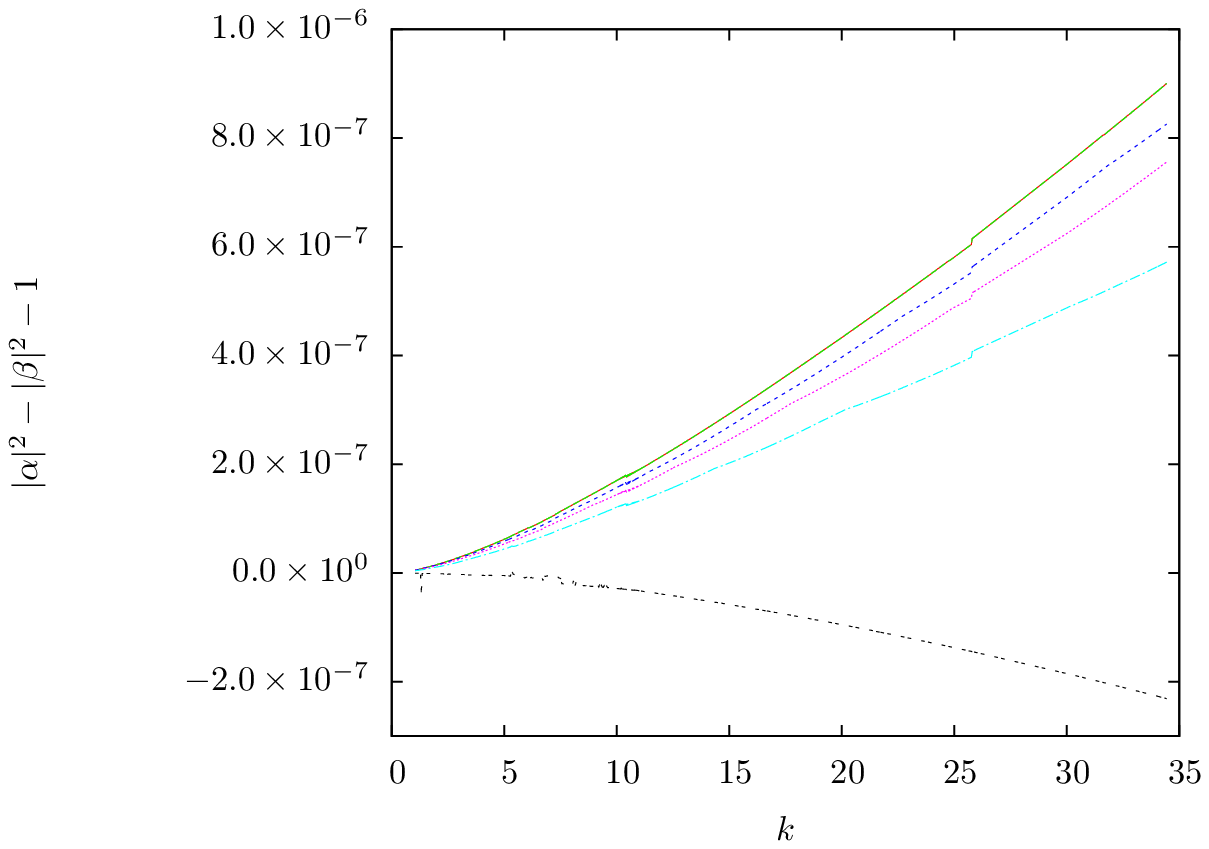}
\caption{\label{fig:ba_time}
Plot of the normalization condition of the Bogoluibov coefficients given
in Eq.~(\ref{modesrelation}).
In Fig.~\ref{fig:ba_precision}, we showed how this normalization converges
with respect to the precision of the numerical integrator and here we
demonstrate that this accuracy is maintained throughout the evolution.
The various lines correspond to different time slices (from top to bottom
(cosmic) time = $ 79,\ 7.9\times 10^2,\ 2.5\times 10^4,\ 5.0\times 10^4,\
10^6,\ 3.9\times 10^6\,\, t_{\rm Pl} $, respectively) for
$\phi_{\rm B}=1.15$, using precision/ accuracy goals of $10^{-15}$.
}
\end{center}
\end{figure}

As an example of convergence with respect to numerical precision, in
Fig.~\ref{fig:ba_precision} we have plotted the normalization
condition, $|\alpha_k|² - |\beta_k|² - 1$ calculated using three
increasing levels of numerical precision (the step size of the
internal integrator being reduced until the relative change at that
step is less than one part in $10^{10}$, $10^{15}$ and $10^{20}$
respectively). As can clearly be seen, we have already converged to
better than one part in $10^6$ using the middle precision. Unless
otherwise stated, all plots have been calculated with the middle
precision level, with convergence tested using the higher precision.
Note that there is a (relative) loss of precision at large $k$,
which becomes important, for example, in Fig.~\ref{fig:PS_diff} and
also in the calculation of the renormalized energy density at late
times.

\begin{figure}[htbp]
\begin{center}
\includegraphics[width=10cm]{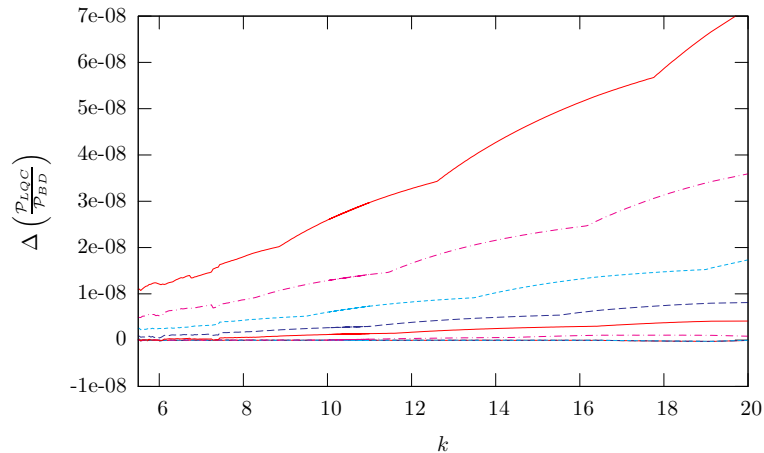}
\caption{\label{fig:PS_time} The ratio of the LQC power-spectrum to
the BD one for scalar perturbations is a constant in time, since
it can be expressed entirely in terms of the Bogoluibov coefficients,
$P_{\LQC}/ P_{\BD}\left(k\right) =  |\alpha_k + \beta_k|^2$. As
a check of our numerics, this ratio was calculated at each out-put time.
Here we have plotted the {\it fractional} change in this ratio (relative
to that calculated at (cosmic) time $\mathring{t}\approx 1580\, t_{\rm Pl}$)
for a range of (cosmic) times (from bottom to top) $t= 177,\ 354,\ 794,\
3550,\ 6310,\ 12600,\ 25100,\ 50100,\ 10^6 $ Planck seconds.
$
\Delta ({ P_{LQC} }/{ P_{BD} }) \equiv \left({P_{\LQC} }/{P_{\BD} }
(\t{t}) - { P_{LQC} }/{ P_{BD} }(\mathring{t} )
\right) \left({ P_{BD}}/{ P_{LQC}}(\mathring{t}) \right)~.
$
The growing deviations at large $k$ is an issue to do
with numerical precision (see also Fig.~\ref{fig:ba_precision} which shows
the normalization of the modes).
}
\end{center}
\end{figure}

Another important check of the numerical accuracy comes from the
fact that the $\alpha_k$ and $\beta_k$ coefficients given in
Eq.~(\ref{modesrelation}) are time independent. An example of the
level to which our numerical integration maintains this is given in
Fig.~\ref{fig:ba_time}. Here the normalization condition is plotted
(as a function of $k$) for various times during the evolution (with
$\phi_B=1.15$). A second (related) test comes from the time
independence of the ratio $P^\Omega_{\mathcal R} / P_{\BD}$, which
is plotted in Fig.~\ref{fig:PS_time}. In both cases one again sees
the loss of precision for large $k$. However, for the range of $k$
we are interested in, both quantities are conserved in time to
better than one part in $10^7$.

\subsection{Summary}
\label{s5.6}

In this section we have analyzed the effects of the pre-inflationary
LQC space-time on the propagation of tensor and scalar
perturbations. The numerical results shown in Fig.~\ref{fig:ratio}
support the physical picture presented in section \ref{s4.2}.
Pre-inflationary geometry provided by LQC has a significant effect
the evolution of modes with low $k$, which have  wavelengths of the
same order or larger than the curvature scale $k_{\LQC}$ at the
bounce. However, the dynamics of modes with $k \gg k_{\LQC}$ is
largely insensitive to the background geometry; they essentially
evolve as if they were in flat space-time.

The relevant questions is then: What is the range of co-moving $k$
corresponding to {\em observable modes}? As discussed in section
\ref{s4.2}, this window depends on the value $\phi_B$ of the
background inflaton at the bounce. The window is given by $k_{\rm
min} = k_o \approx k_{\star}/8.58$ and $k_{\rm{max}}\approx 2000
k_{\rm{min}}$ and, $k_{\star}$ increases with $\phi_B$ (see Table
1). If $\phi_B\gtrsim 1.2$ we have $k_{\rm min}\gtrsim k_{\LQC}$,
and the evolved state is indistinguishable from the BD vacuum at the
onset of inflation for observable modes. Therefore, for
$\phi_B\gtrsim 1.2$ the present analysis provides a justification
for the assumption of the Bunch-Davies vacuum at the onset of
inflation for both tensor and scalar modes (at least for the
observable range of $k$). It is important to emphasize that this
conclusion is non-trivial because for those values of $\phi_B$
observable modes acquire trans-Planckian frequencies near the bounce
and the standard treatment of cosmological perturbation in classical
space-time is not applicable. The introduction of a quantum
space-time was essential to incorporate those modes in a controlled
fashion and to explicitly describe their evolution starting from the
deep Planck regime.

Are there then any deviations from the standard predictions at all?
The answer is in the affirmative for the small window in the
parameter space given by  $\phi_B\lesssim 1.2$. In this case, the
state at the onset of inflation differs significantly from the BD
vacuum, and the results for the spectrum of scalar and tensor
perturbations differ from the standard predictions. For values of
$\phi_B$ not too far from $1.2$, the modes for which the power
spectrum deviates from standard predictions have wave numbers
smaller than $k_\star$. For these modes the WMAP observational error
bars are large. For instance, for $\phi_B=1.15$ we have
$k_{min}\approx 1.07$ and modes for which deviation from standard
predictions appear correspond to $\ell\lesssim 30$ in the WMAP
angular decomposition. As a result, the LQC predictions are
compatible with current data for the power spectrum. However, the
fact that the state differs from the BD vacuum for those modes leads
to a deviation from the standard `consistency relation',
$r_{\BD}\approx - 8 n^{\BD}_t$, for low $k$ that may be seen in
future observations (see Fig.~\ref{fig:consistency}). In addition,
as pointed out in \cite{chen,holman-tolley,agullo-parker}, the
deviation from the BD vacuum at the onset of inflation can serve as
a source of non-Gaussianity generated during inflation. Now, the
presence of scalar BD quanta at the onset of inflation has been
shown to lead to concrete effects in the CMB
\cite{halo-bias1,halo-bias3}, and in the distribution of galaxies
\cite{halo-bias1,halo-bias2} which could be observed in the near
future. If they are seen, these observations will significantly
constrain the initial state for scalar perturbations at the onset of
inflation. One would be able to make specific predictions for
non-Gaussianities originating in LQC, thereby opening a novel avenue
to observationally probe the Planck regime. Because the window,
$\phi_B\lesssim 1.2$, in the parameter space for which new
predictions arise is narrow, should these effects be observed, they
would essentially pin down initial conditions for the background at
the bounce, making more directed analysis and more detailed
prediction feasible.

All these predictions were made under the assumption that our
initial `truncation approximation' is valid; i.e., that the back
reaction of the perturbations on the (quantum) background is
negligible. The goal of the next section is to analyze if this is
the case.

\section{Self-consistency}
%\section{self-consistency of truncation}
\label{s6}

\begin{figure}
\begin{center}
\includegraphics{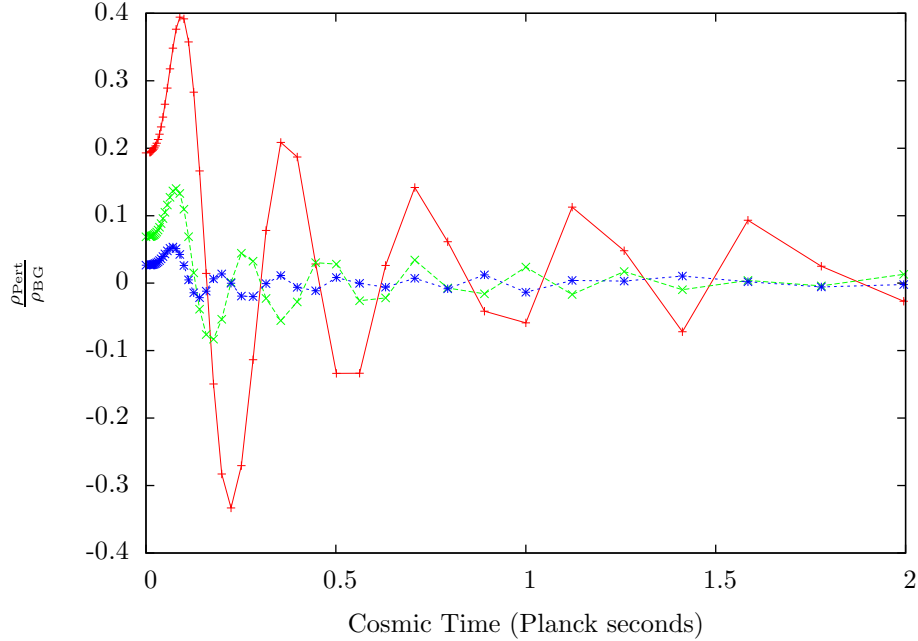}
\caption{ \label{fig:energy_density_ratio1}
Plot of the ratio of the energy density in the perturbations to that
of the background for three values of $\phi_{\B}$. Here the (red) $+$ plot
refers to $\phi_{\B} = 1.21$; the (green) $\times$ to $\phi_{\B}=1.22$;
and the (blue) $\star$ to $\phi_{\B}=1.23$ ). In order for the truncation scheme
to be valid, we need to ensure $\rho_{\rm Pert}/\rho_{\rm BG}
\ll 1$, which is the case for $\phi_{\B}\geq 1.23$. This ratio remains small
not only close to the bounce, but throughout the evolution, as shown in
Fig.~\ref{fig:energy_density_ratio2}. }
\end{center}
\end{figure}

We began with the standard truncation of general relativity (coupled
to a scalar field) that is used in the inflationary scenario and
developed a quantum theory based on this truncation. A key question
is whether this theory admits solutions which are consistent with
the truncation approximation, i.e. in which the back reaction of
perturbations can be neglected all the way from the bounce to the
onset of the slow roll. Of course our conditions at the bounce on
permissible states $\psi$ of perturbations are such that the
approximation is satisfied initially. However, there is no a priori
guarantee that it would continue to be satisfied during the entire
evolution especially since it covers the 11 orders of magnitude in
density and curvature.

In the first part of this section, we will report on the results
obtained by numerical simulations using a specific quantum state of
perturbations, $|\psi\rangle = |0^{\rm obv}\rangle$. In the second
part, we will use analytic considerations to argue that, if there is
a state $\psi^0$ for which the back reaction is negligible, then
there are infinitely many states `close to' $\psi^0$ with this
property.

Our focus will be on scalar perturbations; the situation with tensor
perturbations is completely analogous. So far, our numerical methods
allow us only to put upper bounds on the energy density in
perturbations. But fortunately this is already sufficient to
establish the existence of a large class of consistent solutions.

\subsection{Numerical analysis}
\label{s6.1}

Conceptual aspects of the back-reaction issue were discussed in
section VI.D of \cite{aan2}. It was shown that a necessary and
sufficient condition for the back reaction of perturbations $\psi$
on the background $\Psi_o$ to be negligible is that the expectation
value $\langle \psi|\,\h{\rho}(\x,\t{t})\,| \psi \rangle_{\rm ren}$
should remain negligible compared to the energy density $\rho_{\rm
BG}$ \emph{all the way from the bounce to the onset of slow roll.}%

\begin{figure}
\begin{center}
\includegraphics{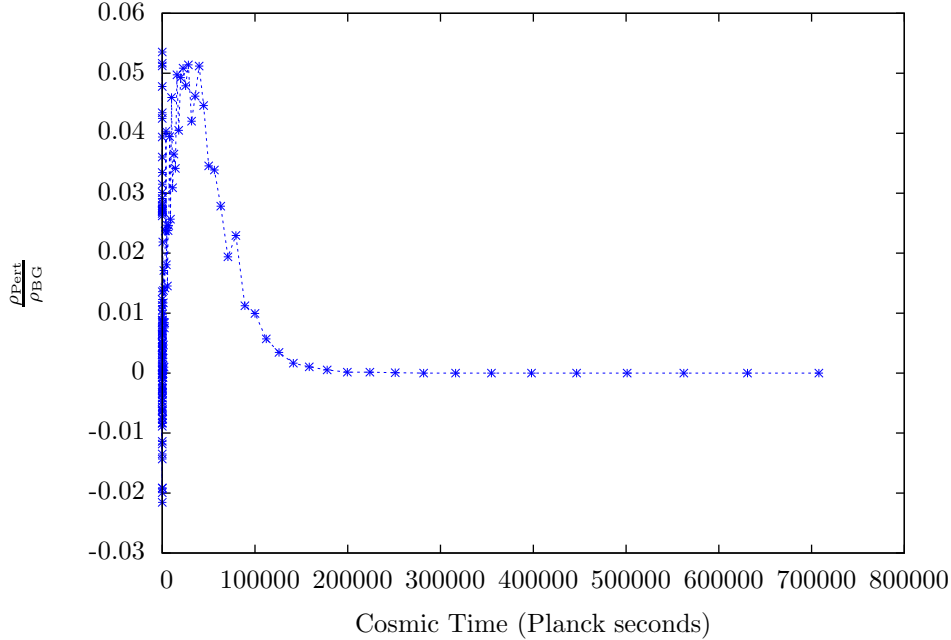}
\caption{ \label{fig:energy_density_ratio2} Time dependence
of the ratio of the energy density in the perturbations to that of the
background (for $\phi_B=1.23$) over the entire evolution from
the bounce (at $t_B=0$) to the onset of slow-roll inflation. The noisy
peak at (cosmic) time $\approx 5\times 10^{4}$ Planck seconds is due to
numerical issues and in this region one should consider these points as
an upper bound on the ratio.
}
\end{center}
\end{figure}

Thus our task is to numerically evaluate the expression
(\ref{renenergy3}) of the renormalized energy density
\be \langle 0^{\rm obv}|\h \rho^{(\Q)}(\x,\t{t})| 0^{\rm
obv}\rangle_{\rm ren} =\frac{\hbar}{2 } \int^\infty_{k_{\rm min}}
\frac{d^3k }{(2\pi)^3} \left[ |\dot q_k|^2+ \frac{(\t \g+k^2)}{\t
a^2}\, |q_k|^2 - \frac{1}{\t a^4} \, C^{(\Q)}(k,\t t)\right]\, ,\ee
where, the subtraction-term is given by (\ref{C-scalar}):
\be C^{(\Q)}(k,\tilde t)=k+ \frac{{\tilde{a}}^2 \dot{\tilde{a}}^2+\tilde a^2 \,
\tilde {\g}}{2 \, {\tilde{a}}^2 k}- \frac{\tilde \g^2+2 \tilde \g
\dot {\t a}^2-3 \dot {\t a}^4+2 \t a \dot {\t a}^2 \ddot {\t a}-\t a^2
\ddot {\t a}^2+2 {\t a} \dot{\t a} (-\dot {\tilde \g}+\t a \dddot {\t a})}{8 k^3}
 \, . \ee%
Recall that the infrared cut-off $k_{\rm min}$ is given by $k_{\rm
min} = k_\star/8.58$. Since the value of $k_\star$ is determined by
$\phi_{\B}$, the co-moving infrared cut-off is simply a fixed number
for any given state $\Psi_o$ of the background geometry. What would
happen if one were to choose a lower $k_{\rm min}$ as the infra-red
cut-off? In this case, the results discussed below will change only
in the value $\phi_{\B}$ above which we have self-consistent
solutions of the truncated theory. Furthermore, the exponential
relation between $\phi_{\B}$ and $k_{\star}$ shown in
Fig.~\ref{fig:kstar}, implies that the required change in $\phi_B$
will only be logarithmic in the change in $k_{\rm min}$.

\begin{figure}
\begin{center}
\includegraphics{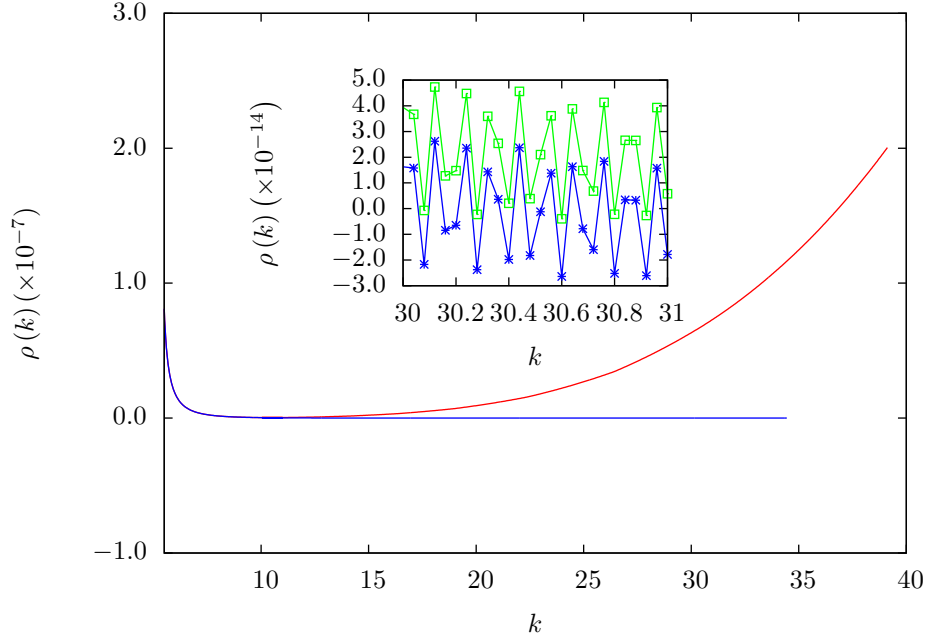}
\caption{ \label{fig:energy_vs_precision} An illustration of the importance of
numerical precision for high $k$. The plot shows the \emph{integrand} $\rho(k)$
of the energy density in perturbations for $\phi_{B} = 1.15$ and at time
$t\approx 5\times 10^4\, t_{\rm Pl}$. The corresponding background energy
density $\rho_{\rm BG}$ is only $\approx 10^{-9}$! The upper (red) solid line was
evaluated with the lowest precision/accuracy goal $10^{-10}$, whilst the lower
(blue) line corresponds to precision accuracy of $10^{-25}$. The inset (where the
y-axis is magnified by another factor of $10^{7}$) shows that there are delicate
cancelations between positive and negative values of $\rho(k)$. In the inset
the upper (green) plot corresponds to precision/ accuracy goal of
$10^{-20}$ while the lower (blue) to $10^{-25}$.}
\end{center}
\end{figure}

Since $k_{\rm min}$ increases with $\phi_{\B}$, one might expect
that an increase in $\phi_B$ will result in a decrease in the
expectation value of the energy density at any given particular
time, simply because of the decrease in the domain of integration.
Our numerics bear this out. The results are plotted in
Fig.~\ref{fig:energy_density_ratio1} for early times near the bounce
and in  Fig.~\ref{fig:energy_density_ratio2} for all times from the
bounce to the onset of slow roll. These results show, in particular,
that for $\phi_B \ge 1.23$ we have,
\be\label{eq:pert_requirement} \frac{\langle \psi |
\hat{\rho}|\psi\rangle_{\rm ren} \left(\tilde{t}\right)}{\rho_{\rm
BG} \left(\tilde{t}\right)} < 0.05 \quad \forall~ \t{t}~. \ee
Thus the test field approximation holds if $\phi_{\B} \ge 1.23$. It
may hold also for lower values of $\phi_{\B}$ because, as indicated
in Figs.~\ref{fig:energy_density_ratio2} and
\ref{fig:energy_vs_precision} and discussed below, what we control
numerically is an upper bound on the energy density in
perturbations.

Numerically these calculations are challenging because the integrand
rapidly builds up a fast (but decaying) oscillation in $k$ (similar
to Fig.~\ref{PSvsk}). In addition, for large $k$, even a very small
loss of precision can result in a significant error, since the
renormalization procedure subtracts two diverging terms. This
problem is exacerbated by the fact that there are times when even
extremely small numerical errors (of the order of one part in
$10^{15}$) can be comparable to the background energy density; see
Fig.~\ref{fig:energy_vs_precision}. But, typically, this is not an
issue for the early phase where quantum gravity dominates and the
background energy density is high, or the very late time evolution
where the energy density in the perturbations is suppressed by the
large scale factor. The problem is serious only for the intermediate
times, $\t{t} \approx 50,000 t_{\rm Pl}$, when the calculations of
the expectation value of the energy density are dominated by
numerical noise. However we have used sufficient precision to ensure
that this noise is always less than half a percent of the background
energy density. This allows us to ensure that
Eq.~(\ref{eq:pert_requirement}) is satisfied. But because of this,
our numerical estimates should be considered as upper bounds.

\subsection{Analytical considerations}
\label{s6.2}

In section \ref{s6.1} we used numerics to provide an explicit
example of a state for which our underlying truncation scheme is
self-consistent. However, one would also like to know whether this
particular example is special or if there are `many' other states in
the Hilbert space $\H_1$ for which our truncation is also
self-consistent. Of course, such questions have to be phrased with
care because $\H_1$ is infinite dimensional. In this section we will
provide an analytical argument to establish the following result: If
a state $|\psi\rangle$ in $\H_1$ is such that $\langle \psi|\,
\h{\rho}\,|\psi\rangle_{\rm ren}$ is very small compared to the
energy density $\rho_{\rm BG}$ in the background, then the same is
true for every state $|\tilde \psi \rangle$ in an open neighborhood of
$\psi$ in $\H_1$. Together with the numerical example $|\psi
\rangle = |0^{\rm obv}\rangle$ given in the previous subsection,
this result establishes that there are `many' states in $\H_1$ for
which the truncation approximation is valid. Since our considerations are all analytic, in this subsection we will work with the conformal time $\t{\eta}$ of the dressed effective metric $\t{g}_{ab}$ (rather than the cosmic time $\t{t}$).

Let $| \psi\rangle $ then be a 4th adiabatic order vacuum in
$\H_1$ such that,
\be \langle \psi| \hat{\rho} | \psi\rangle_{\rm ren}\left(
\tilde{\eta} \right) \ll \rho_{\rm BG}\left( \tilde{\eta} \right)
\qquad \forall ~ \tilde{\eta} \in \left(\tilde{\eta} _{\rm
ini},\tilde{\eta} _{\rm final}\right)~, \ee
where $ \left( \tilde{\eta} _{\rm ini}, \tilde{\eta} _{\rm
final}\right)$ is the dynamical range we are interested in. Now
consider a second 4th adiabatic order vacuum $|\tilde \psi\rangle$ in
$\H_1$. Let $q_k$ and $\t q_k$ be any 4th order bases that
define $|\psi\rangle$ and $|\tilde \psi \rangle$ and let the Bogoluibov
transformation relating them be given by
\be\label{eq:bog} \t q_k = \alpha_k \, q_k \,+\,
\beta_k  \, q_k^\star~. \ee
One can easily check that the difference of the expectation values
of the renormalized energy density with respect to $|\t \psi\rangle$
and $|\psi\rangle$ is given,
\ba\label{eq:deltarho1} \Delta \rho\left[ \t \psi, \psi\right] &\equiv&
 \langle \t \psi |\hat{\rho}|\t \psi\rangle -  \langle \psi|\hat{\rho}|\psi\rangle  \\\nonumber
&=&\frac{\hbar}{(2\pi)^3 \t a^2 } \int d^3k \, \Bigg[ |\beta_k|^2 \left(|q'_k|^2+(k^2+{\t\g}) |q_k|^2\right) + {\rm Re}\left[\alpha_k \beta_k^{\star}  \left({q'_k}^{2}+(k^2+\t\g) q_k^2 \right) \right] \Bigg]  \, , \ea
where we have used the relation $|\alpha_k|^2-|\beta_k|^2=1$. From the previous equation, by taking into account that $-|z|\le {\rm Re}(z)\le |z|$ for any complex number $z$, we can obtain the following bound for $\Delta \rho$

\be \label{eq:bound} \frac{\hbar}{(2\pi)^3 \t a^2} \int d^3k \, B_- \left(| q'_k|^2+(k^2+\t\g) |q_k|^2\right) \le \Delta\rho \le \frac{\hbar}{(2\pi)^3\t a^2} \int d^3k \, B_+ \left(| q'_k|^2+(k^2+\t\g) |q_k|^2\right) \, , \nonumber\ee
where $B_{\pm}= |\beta_k|^2(1 \pm \sqrt{1/|\beta_k|^2 +1})$. Finally, since $|\psi\rangle$ and $|\t \psi\rangle$ are both $4^{\rm
th}$ adiabatic order states, the $\beta_k$ coefficients decay
sufficiently fast with $k$ so that the integrals in these bounds are
convergent. Therefore, the bounds are finite. Furthermore, all time
dependent quantities are continuous in $\t{\eta}$ whence the
difference $\Delta \rho\left[ \t \psi, \psi\right] (\t{\eta})$ has an
upper bound in the closed time interval under consideration. Since
this bounds scales with $B^\pm_k$, we can make sup$\Delta \rho\left[ \t \psi, \psi\right] (\t{\eta})$ arbitrarily small simply by rescaling
the $\beta_k$ by a sufficiently small constant. Thus we have an
(open) neighborhood of the state $|\psi\rangle$, parameterized
(for each mode) by $\left|\beta_k\right|$, in which the change in
the expectation value of the energy density in the perturbations is
also very small compared to the background for all times between the
bounce and the onset of the slow roll.

To summarize, in this section we have shown that if the state
$\Psi_o$ of the background geometry is chosen so that $\phi_{\B} \ge
1.23$, then $|\Psi_o\rangle \otimes |0^{\rm obv}\rangle$ provides a
self-consistent solution to our truncated quantum theory.
Furthermore, there are infinitely many states $|\t \psi\rangle$ `close
to' $|0^{\rm obv}\rangle$ for which $|\Psi_o\rangle\otimes
|\t \psi\rangle$ shares this property. Each of these states provides an
admissible extension of the standard inflationary scenario to the
Planck regime. Finally, so far our numerical techniques have
provided only upper bounds for the renormalized energy density. It
is conceivable that future work will show that the self consistent
solutions exist also for lower values of $\phi_{\rm B}$.

\section{Summary and discussion}
\label{s7}

In the standard inflationary scenario one works with quantum fields
representing first order perturbations on a FLRW solution to
Einstein's equations. One assumes that these quantum fields are in
the BD vacuum at the onset of slow roll, works out their evolution
and calculates the power spectra and other correlation functions at
the end of inflation. The fact that one starts the calculations at
the onset of slow roll is often justified on the basis that
pre-inflationary dynamics will not affect any of the observable
predictions. This belief stems from the argument that modes seen in
the CMB cannot be excited during the pre-inflationary dynamics
because, when evolved back in time starting from the onset of the
slow roll, their physical wave lengths $\lambda_{\rm phy}$ continue
to remain within the Hubble radius $\mathfrak{R}_{\rm H}$ all the
way to the big bang. However, this argument is flawed on two
reasons. First, what matters to the dynamics of these modes is the
curvature radius $\mathfrak{R}_{\rm curv} = \sqrt{6/R}$ determined
by the Ricci scalar $R$, and not $\mathfrak{R}_{\rm H}$, and the two
scales are equal only during slow roll. Thus we should compare
$\lambda_{\rm phy}$ with $\mathfrak{R}_{\rm curv}$ in the
pre-inflationary epochs. The second and more important point is that
the pre-inflationary evolution should not be computed using general
relativity, as is done in the argument given above. One has to use
an appropriate quantum gravity theory since the two evolutions can
well be very different in the Planck epoch. Then modes that are seen
in the CMB could well have $\lambda_{\rm phy} \gtrsim
\mathfrak{R}_{\rm curv}$ in the pre-inflationary phase. If this
happens, these modes \emph{would be} excited and the quantum state
at the onset of the slow roll could be quite different from the BD
vacuum. Indeed, the difference could well be so large that the
resulting power spectrum is incompatible with the amplitude and the
spectral index observed by WMAP. In this case, that particular
quantum gravity scenario would be ruled out. On the other hand, the
differences could be more subtle: the new power spectrum for scalar
modes could be the same but there may be departures from the
standard predictions that involve tensor modes or higher order
correlation functions of scalar modes, changing the conclusions on
non-Gaussianities. In this case, the quantum gravity theory would
have interesting predictions for future observational missions.
Thus, as we emphasized in section \ref{s4}, pre-inflationary
dynamics can provide an avenue to confront quantum gravity theories
with observations.

In this paper we investigated these possibilities in the context of
loop quantum gravity. In the standard inflationary scenario, one
starts by truncating general relativity, keeping only the FLRW
solutions (coupled to a scalar field) and first order perturbations
thereon. Therefore, in our analysis we began in section \ref{s2}
with the truncated phase space $\ps_{\tr} = \ps_o \times \t{\ps}_1$
where $\ps_o$ is the FLRW phase space of homogeneous, isotropic
fields and $\t\ps_1$ is the phase space of gauge invariant
\emph{first order} perturbations. We then passed to the quantum
theory of the \emph{entire} $\ps_{\tr}$. Consequently in the
resulting framework tensor and scalar perturbations $\h{\T}$ and
$\h{\Q}$ propagate on a \emph{quantum FLRW geometry} encapsulated in
a wave function $\Psi_o$. This quantum geometry is
\emph{non-singular;} the big bang of general relativity is replaced
by a quantum bounce. Therefore we can directly face the
trans-Planckian issues. We found that the existence of
trans-Planckian modes is not a problem in itself. The key question
is whether these modes contribute so much to the energy density in
the perturbations that the back reaction of perturbations on
$\Psi_o$ cannot be neglected. This could easily happen especially in
the Planck regime. If it did, our `truncation' strategy would fail
to be self-consistent. Therefore the challenge is to develop
techniques to calculate the energy density in perturbations all the
way from the bounce to the onset of the slow roll and compare it
with background energy density. Finally, as we saw in section
\ref{s2.2.2}, LQC dynamics reveals several features in the Planck
regime that are very surprising from the perspective of general
relativity which has shaped intuition in the standard inflationary
scenario.

The underlying FLRW quantum geometry streamlines the analysis by
providing the necessary control on calculations in the deep Planck
regime. However, we are now faced with the challenge of
investigating the dynamics of quantum fields $\h{\T},\, \h{\Q}$ on
\emph{quantum} geometries. At first this problem seems formidable.
But fortunately there is a key simplification \cite{akl,aan2}:
Mathematically this evolution is completely equivalent to that of
$\h{\T},\, \h{\Q}$ propagating on a dressed, effective metric
$\t{g}_{ab}$, which incorporates the appropriate quantum properties
of the state $\Psi_o$.%
\footnote{While the standard inflationary dynamics of tensor modes
$\h\T$ is the same as that of a massless scalar field on a FLRW
metric $g_{ab}$, that of the Mukhanov-Sasaki field $\h\Q$
representing scalar perturbations involves also an `external
potential' $\g$. For both modes, $g_{ab}$ is replaced by a quantum
corrected metric $\t{g}_{ab}$ in LQC. For the scalar mode, in
addition, $\g$ is replaced by a quantum corrected $\t{\g}$.}
This occurs because the dynamics of test quantum fields does not
experience all the details of the probability amplitude for various
FLRW metrics encapsulated in $\Psi_o$; it is sensitive only to a few
moments of this distribution. The phenomenon is analogous to the
propagation of light in a medium where all the complicated
interactions of the Maxwell field with the atoms in the medium can
be captured just in a few parameters such as the refractive index.
This unforeseen outcome enabled us to `lift' the adiabatic
regularization and renormalization techniques, normally used in QFT
in classical FLRW space-times, to QFT on quantum FLRW geometries,
thereby providing the necessary control on the dynamics of $\h{\T},\,
\h{\Q}$ on the quantum geometry $\Psi_o$.

In section \ref{s3} we specified a class of initial conditions at
the bounce for the quantum state $\Psi_o\otimes\psi$ of the combined
system, $\Psi_o$ describing the background and $\psi$ describing
perturbations. The permissible states $\Psi_o$ are required to be
sharply peaked on certain effective, bouncing trajectories in the
phase space, which are well-understood. It turns out that, except for
a tiny portion, $\phi_{\B} < 0.93$, of the allowed range of values
for the background inflaton at the bounce, all effective
trajectories encounter the slow roll compatible with the 7 year WMAP
data sometime in their future evolution \cite{as3}. This was an
unanticipated result. However, it does \emph{not} mean that the
background states $\Psi_o$ are generic. In fact, they are very
special because they are required to be sharply peaked. For the
states $\psi$ of perturbations, we imposed three conditions: i)
Regularity: They be of 4th adiabatic order so that the expectation
value $\langle \psi|\, \h\rho\, |\psi\rangle_{\rm ren}$ of the
renormalized energy density operator is well defined; ii) Symmetry:
They be invariant under spatial translations and rotations, i.e.,
under symmetries of the background geometry; and, iii) Initial
compatibility with truncation: \emph{At the bounce}, $\langle
\psi|\, \h\rho\, |\psi\rangle_{\rm ren}$ be negligible compared to
the background energy density $\rho_{\rm BG}$. There is a large
class of such states. We discussed in detail a specific example
$|\psi\rangle = |0^{\rm obv}\rangle$, called the `obvious vacuum of
the 4th adiabatic order', in section \ref{s3}, and mentioned a few
other examples section \ref{s5.3}.

In sections \ref{s4} and \ref{s5} we used numerical methods to
evolve these initial conditions and obtained the power spectra for
scalar and tensor perturbations. The key free parameter in this
analysis is the value $\phi_{\B}$ of the background inflaton at the
LQC bounce because it determines the effective trajectory on which
$\Psi_o$ is peaked. For $\psi$, we first carried out a few
simulations using various choices discussed in section \ref{s5.3} to
ensure robustness of the final results, and then focused on the
choice $|\psi\rangle = |0^{\rm obv}\rangle$ because it has several
attractive properties.

As explained in sections \ref{s5} and \ref{s6}, several of the
numerical simulations require very high precision (see, e.g.,
Fig.~\ref{fig:energy_vs_precision}). Furthermore, it turns out that
the integration time increases \emph{very} rapidly with $\phi_{\B}$.
Therefore, we had to restrict our simulations to $\phi_{\B} \lesssim
2$. Fortunately, as summarized below, the interval $\phi_{\B}
\lesssim 2$ covers the physically most interesting range. Our
numerical simulations brought out an unforeseen feature: The LQC
power spectra are essentially insensitive to the precise value of
$\phi_{B}$. However, in the CMB we have access only to a finite
range of wave numbers. The 7 year WMAP data in particular covers a
range $\sim (k_o,\,\, 2000k_o)$ where $k_o$ is the co-moving wave
number of the mode whose physical wave length equals the radius
$\mathfrak{R}_{\rm LS}$ of the observable universe at the surface of
last scattering. In LQC, the value of $k_o$ is dictated by
$\phi_{\B}$ (in our scale factor convention $a_{\B} =1$). Therefore,
in our analysis the window of modes (in the $k$ space) that can be
seen in the CMB is dictated by the value of $\phi_{\B}$, and moves
rapidly to the right along the $k$ axis as $\phi_{\B}$ increases.

Are there values of $\phi_{\B}$ then, for which LQC predicts
deviations from the standard inflationary scenario for modes in the
observable window?  As mentioned in the beginning of this section,
on rather general grounds we expect that a mode would be excited in
the pre-inflationary dynamics if $\lambda_{\rm phy} \gtrsim
\mathfrak{R}_{\rm cuv}$ at any time. A detailed analysis of the
pre-inflationary dynamics of the LQC background geometry shows that
such modes can be seen in the CMB only if $\phi_B < 1.2$ (see
section \ref{s4.2}). In this case, for observable modes, the state
$|\psi \rangle$ has non-trivial excitations over the BD vacuum at
the onset of inflation. If $\phi_{\B} =1.15$ for example, the
reference mode used in WMAP corresponds to $k_\star = 9.17$ in our
conventions and as Fig.~\ref{fig:ratio} shows, the LQC predictions
for the power spectrum and the spectral index at $k = 9.17$ are
indistinguishable from those of standard inflation. Thus, the
prediction is observationally viable. However, for $k < 9.17$
---which correspond to $\ell \lesssim 30$ in the angular
decomposition used by WMAP--- there are significant differences
between LQC and the standard inflation because of the excitations
over the BD vacuum. Because the observational error bars for low
$\ell$ are large, both predictions are observationally viable.
Furthermore, since the LQC excitations over the BD vacuum are the
same for scalar and tensor modes, the ratio $r$ of the tensor to
scalar power spectra is unchanged from that in standard inflation.
However, the standard `consistency relation' $r_{\rm BD} = -8
n_t^{\rm BD}$ \emph{is} modified. Future measurements of tensor
modes should be able to distinguish between the two. This departure
from the BD vacuum also has implications for the CMB and galaxy
distribution
\cite{holman-tolley,agullo-parker,ganc,agullo-navarro-salas-parker}
and observational tests for such effects have already been proposed
\cite{halo-bias1,halo-bias2,halo-bias3}. Thus, there \emph{are}
differences between the LQC and the standard inflationary
predictions if $\phi_{\B} <1.2$.

What if $\phi_{\B} > 1.2$? Then we would have $\lambda_{\rm phy} \ll
\mathfrak{R}_{\rm cuv}$ throughout the pre-inflationary evolution
for all the modes that can be seen in the CMB. Therefore, the
physical reasoning strongly suggests that they would not
`experience' significant curvature and would be in the BD vacuum at
the onset of inflation. In this case, pre-inflationary dynamics
would not modify standard inflationary predictions for observable
modes. Numerical simulations made this semi-quantitative
considerations precise in that our plots provide sharp figures for
excitations over the BD vacuum for various values of $\phi_\B$. In
particular, if $\phi_{\B}
>1.4$, we find that the excitations are so small that the departures
predicted by LQC would be negligible for any of the forthcoming
observation missions. This also implies that the fact that we had to
limit our numerical simulations to $\phi_{\B} \lesssim 2$ is not
physically restrictive.

To summarize, by analyzing the pre-inflationary dynamics in detail
we arrived at two main conclusions. First, there do exist natural
initial conditions at the bounce which lead to a completion of the
standard inflationary scenario to include the quantum gravity
regime. In this completed theory, one has a consistent evolution all
the way from the deep Planck regime that accounts for the
inhomogeneities seen in the CMB. Since the origin of the large scale
structure can be traced back to these inhomogeneities, \emph{now one
can systematically trace back the seeds of this structure to the
quantum fluctuations of the initial state at the LQC bounce itself.}
Second, there is a narrow window in the $\phi_{\B}$ parameter space
for which the state at the onset of inflation would not be the BD
vacuum. While the LQC and the standard inflation predictions are
both compatible with current observations, future observations
should be able to distinguish between the two. \emph{Thus, there is
a potential to extend the reach of observational cosmology all the
way to the Planck scale.} Of course, since the window is narrow, the
`a priori' probability of its being realized in Nature is small.
This is compensated by the fact that, if observations are compatible
with $\phi_{\B}$ being in this window, the initial conditions would
be narrowed down tremendously, making very detailed calculations and
predictions feasible.

Finally, in section \ref{s6} we investigated the issue of whether
our truncation strategy is self-consistent, i.e., whether in the
solutions $\Psi_o\otimes\psi$ we analyzed the energy density in the
perturbations can be neglected compared to that in the background
all the way from the bounce to the onset of inflation. This is a
difficult issue on two accounts and to our knowledge it had not been
analyzed in the Planck regime in any approach. The first difficulty
is conceptual: one needs a systematic framework to compute the
renormalized energy density. As discussed above, we were able to
construct this framework using the well-developed adiabatic
renormalization theory on cosmological space-times because of the
exact mathematical equivalence between QFT on the LQC quantum
geometries $\Psi_o$ and QFT on dressed effective metrics
$\t{g}_{ab}$ determined by $\Psi_o$. The second set of difficulties
comes from numerics: because of rapid oscillations of the integrand
of $\langle \psi|\, \h\rho\, |\psi\rangle_{\rm ren}$ in the $k$
space, and because the background energy density itself diminishes
rapidly as one evolves to the future of the bounce, one requires
very high accuracy and numerical precision. We were not able to
calculate $\langle \psi|\, \h\rho\, |\psi\rangle_{\rm ren}$ over the
entire evolution from the bounce to the onset of inflation to the
desired precision. However, we \emph{were} able to provide an upper
bound on this quantity. For $\phi_B \ge 1.23$, these upper bounds
suffice to guarantee that the back reaction of perturbations can
indeed be ignored for the states $\Psi_o\otimes\psi$ we considered.
This suffices to show that there \emph{is} a satisfactory extension
of the standard inflationary scenario in which it is consistent to
ignore the back reaction also in the pre-inflationary epoch.
However, for lower values of $\phi_{\B}$ ---particularly those for
which the state at the onset of inflation is significantly different
from the BD vacuum--- we have not been able to demonstrate that the
truncation scheme is self consistent. We hope to return to a
detailed analysis of this issue
in the near future.\\

We will conclude with a few remarks.\\

\bu While this paper was focused on inflation, the underlying
framework developed in \cite{aan2} is much more general. Together,
the two investigations offer some general lessons that could be
useful also in other paradigms for the early universe. For example,
we saw that, in the investigation of whether a mode would be
dynamically excited, what matters is not the Hubble radius which
often dominates this discussion but the curvature radius
$\mathfrak{R}_{\rm curv}$ determined by the scalar curvature of the
background. As Fig.~\ref{fig:bgplot} illustrates, the evolution of
$\mathfrak{R}_{\rm curv}$ in the early universe can be quite
involved and therefore non-intuitive, especially in the regime in
which Einstein's equations receive significant corrections.
Therefore, it is important to calculate this quantity throughout the
period of interest. A second issue involves back
reaction~\cite{backreaction}. Since it is negligible during slow
roll, the issue has often been ignored in the investigations of
pre-inflationary dynamics as well as in discussions of alternatives
to inflation. However, the issue is important especially in the
Planck era and our numerical simulations showed that it is quite
subtle. In particular, there is a correlation between the fact that
modes with $\lambda_{\rm phy} \gtrsim \mathfrak{R}_{\rm curv}$ are
excited during evolution and the observation from numerics that
these modes make significant contribution to the energy density in
perturbations. Now, if there is sufficient inflation, then the modes
that are relevant for CMB observations tend to have very small
$\lambda_{\rm phy}$ in the Planck regime. On the other hand, if
there is no inflation at all, it is not easy to achieve sufficient
expansion between the bounce and the CMB epoch for these modes to
have $\lambda_{\rm phy} \ll \mathfrak{R}_{\rm curv}$ in the Planck
era. These general physical considerations suggest that in scenarios
without inflation there is a danger that the back reaction may not
be negligible especially in the Planck regime. Therefore it is
important to check the consistency of first order perturbation
theory in those cases.

\bu In recent years, there have been a number of interesting
investigations of possible LQC corrections to inflationary dynamics
(for a brief summary, see section II of \cite{aan2}.) Some of these
have focused on pre-inflationary dynamics studied in this paper (see
especially \cite{lqc-preinflation1,lqc-preinflation2}). The
distinguishing features of our analysis are: i) It is based on the
general truncation strategy which has been successfully employed in
LQG in other problems;  ii) The issue of the initial state has been
investigated in a stream-lined fashion; iii) Numerical simulations
are better controlled and considerably more extensive. Furthermore,
the physics behind the main findings is understood and discussed in
detail; iv) Issues of regularization of composite operators and
renormalization of energy density have been addressed for the first
time. They enabled us to systematically check if the back reaction
can be ignored.

\bu The truncation strategy used in this series of papers is
motivated by two considerations: i) We begin with the phase space
$\ps_{\tr}$ that underlies almost all investigations of the early
universe; and, ii) The same truncation philosophy underlies other
successful applications of LQG such as the calculation of the
graviton propagator. Nonetheless, it would clearly be much better if
we could start with a full quantum gravity theory and then descend
to this truncation systematically. Could the final results be
significantly different as far as observational issues are
concerned? We believe that the answer is in the negative at least in
cases where the truncation procedure can be shown to admit
self-consistent solutions. For, in general relativity one routinely
expects first order perturbations whose back reaction is negligible
to provide an excellent approximations to the phenomenological
predictions of the exact theory, and there is no obvious reason why
the situation would be different in quantum gravity. As a simple
example to illustrate our view of the effectiveness of the truncated
theory, consider the Dirac solution of the hydrogen atom. Because it
assumes spherical symmetry prior to quantization, this truncation
excludes photons from the beginning. Therefore, at a conceptual
level, it is \emph{very} incomplete. Yet, as far as experiments are
concerned, it provides excellent approximations to answers provided
by full quantum electrodynamics till one comes to the Lamb shift. We
expect the situation to be similar for our truncated theory:
Conceptually it is surely quite incomplete vis a vis full LQG, but
we expect the full theory to provide only small corrections to
observable effects.\\

\bu However, even apart from the issue of full LQG, the current
framework can be improved in a number of ways in the near future. We
will complete our remarks by providing a few examples.

$\star$ \emph{Extensions:} First, it would be desirable to extend
the framework in several directions. Inclusion of a positive
cosmological constant would be straightforward. While it would add a
few conceptual twists along the lines of \cite{ap}, the effect on
pre-inflationary dynamics investigated here would be totally
negligible because the phenomenological value of the cosmological
constant is so small. Second, it would be worthwhile to extend the
calculations to include multi-inflatons, especially since
multi-inflatons have already been considered in LQC \cite{multi}.
Are there any significant changes in the results on power spectra? A
third extension would be to investigate cases where $\phi_{\B}$ is
much larger than 2. We gave simple physical arguments that for
$\phi_{\B} > 1.4$ the LQC results would be indistinguishable from
those of standard inflation as far as foreseeable observations are
concerned. But it is not impossible that these considerations
neglected to take into account some subtle feature of the background
dynamics for large $\phi_{\B}$. Only explicit calculations can
decide. However, since numerical simulations become prohibitively
expensive for large $\phi_{\B}$, a combination of analytical
approximations and numerics will be needed.
Finally, our preliminary results indicate that, if one were to
extend the detailed numerical calculations all the way to $\phi_{\B}
= 0.93$, the lowest value that is compatible with the 7 year WMAP
data according to \cite{as3}, one would find that the state at the
onset of inflation is so different from the BD vacuum that one has
to significantly revise the value of the inflaton mass used in this
paper. We have made a first pass at redoing the calculation using a
cyclic procedure described at the end of section \ref{s5.1} and plan
to carry out a detailed calculation. However, a more systematic and
efficient procedure extracted from the standard routines used in the
WMAP data analysis would be highly desirable.

$\star$ \emph{Quantum theory of perturbations:} To make direct
contact with calculations in the standard inflationary scenario, we
used a Fock representation for quantum states $\psi$ of
perturbations $\h{\T}, \, \h{\Q}$. This strategy is internally
consistent and yields a well-defined theory. However, from a
conceptual standpoint, it is highly desirable to use a `polymer
representation' suggested by the LQG techniques. In cases when the
energy density in the perturbations is negligible compared to that
in the background, we expect predictions of the polymer
representation to reduce to those obtained here using the Fock
representation. But if perturbations are described using the polymer
representation, quantization of the entire truncated theory would be
more firmly rooted in LQG, adding considerable conceptual coherence.
In addition, such an extension would also be valuable from
mathematical physics considerations. The correspondence between QFT
on quantum geometries $\Psi_o$ and those on the dressed effective
metrics $\t{g}_{ab}$ will continue to hold. However, LQG techniques
are likely to suggest alternate regularization and renormalization
schemes that are tailored to the `polymer representation'. The
adiabatic scheme used here has been carefully developed within QFT
on cosmological space-times over more than four decades
\cite{parker66,parker-fulling74,Anderson-Molina-Paris-Mottola,sf-book,parker-book}.
And it handles the ultraviolet divergences satisfactorily. However,
as we saw in section \ref{s2.3.2}, it does not remove infrared
divergences in the case of massless fields. Our current treatment of
infrared issues is meant to be a physically motivated but tentative
strategy. There should a better procedure that simultaneously
handles the ultraviolet \emph{and} the infrared regime. It is quite
possible that such a scheme naturally descends from techniques that
are well-adapted to LQG \cite{ttbook}. So far the literature on the
relation between LQG and regularization and renormalization in QFT
in curved space-times has remained rather general and investigations
that focus just on cosmological space-times may well provide richer
and more detailed results.

$\star$ \emph{Initial conditions:} In our choice of initial
conditions for the background state $\Psi_o$ in section \ref{s3}, we
were motivated by physical considerations. The new ingredient is the
role played by the repulsive force of origin in quantum geometry
that causes the quantum bounce. It is significant on a scale of
about $10\lp$ and effectively dilutes away the inhomogeneities.
Since regions of this size at the bounce expand out to become the
observable universe at the surface of last scattering, there is a
new mechanism to achieve the extraordinary homogeneity that is
needed in the initial state to explain the current large scale
observations. However, so far the argument is only qualitative. It
would be extremely helpful if this could be developed further via
concrete calculations, or, evidence for or against this possibility
can be gleaned from suitable numerical simulations. Another open
issue is that of further narrowing down the initial conditions,
especially for $\psi$. The three conditions we imposed  are well
motivated but still allow a very large class of $\psi$. We used the
four choices discussed in section \ref{s5.3} to verify that the
power spectra are robust and then carried out the detailed numerical
simulations using $|\psi\rangle = |0^{\rm obv}\rangle$ because there
are several reasons in favor of this choice. Are there perhaps
additional criteria that will single out this --or another state--
essentially uniquely? It would be very helpful even to significantly
narrow down the choices by using
general physical principles.\\

Finally, we would like to re-emphasize that in this series of papers
we have focused only on the quantum gravity issues. Problems related
to particle physics phenomenology and foundational issues related to
the possible quantum to classical transition remain.

\section*{Acknowledgments}
We have benefited from discussions with and comments and questions
from with a large number of colleagues, especially Francois Buchet,
David Brizuela, Stefan Hollands, Eiichiro Komatsu, Alok Ladha, Misao
Sasaki, Sarah Shandera and David Sloan. This work was supported by
the NSF grant PHY-1205388 and the Eberly research funds of Penn
state, and the Marie Curie Fellowship program of the EU.

\begin{appendix}
\section{Inclusion of an inflationary potential in the underlying framework}
\label{a1}

Since Ref. \cite{aan2} was addressed to the quantum gravity
audience, it focused on mathematical and conceptual issues
rather than phenomenological. Therefore, for simplicity of
presentation we assumed that the scalar field is massless. In
this appendix we summarize the modifications that are necessary
to incorporate an inflaton potential, such as $V(\phi) =
(1/2)m^2\phi^2$ used in the main body of this paper.

The discussion of the homogeneous phase space $\ps_o$ is the
same as that in \cite{aan2} except that the scalar (or the
Hamiltonian) constraint $\mathbb{S}_o$ now contains the
potential $V(\phi)$ (see Eq (\ref{hc})). The conceptual steps
that led to the reduced phase space $\t\ps$ of first order
perturbations are also the same as in \cite{aan2} and the
treatment of tensor perturbations in \cite{aan2} does not
require any modifications. However, for scalar perturbations,
expressions of the Hamiltonians of the final gauge invariant
Mukhanov-Sasaki variables $\Q_{\vk}$ and curvature
perturbations $\R_{\vk}$ are now more complicated. Our
discussion of the required changes will largely follow
Langlois' Hamiltonian treatment \cite{langlois}. However,
because of our primary motivation, we aim at finding
expressions of the Hamiltonian that are well suited to loop
quantization. Note also that the procedure uses only the
constraint equations; we do not assume that the background
space-time satisfies full Einstein's equations.

As discussed in \cite{aan2}, the first order constraints generate
the gauge transformations for the linear perturbations. One can
solve these constraints and isolate gauge invariant variables which
suffice to label the points of the reduced phase space $\t{\ps}_1$.
As discussed in \cite{aan2}, the Mukhanov-Sasaki variable $\Q_{\vk}$
provides a convenient choice. To arrive at them, note first that, in
the notation of \cite{aan2}, the first order metric perturbations
$\t{\q}_{ab}(\vk)$ can be expanded into scalar, vector modes and
tensor modes
\ba \tilde{\q}_{ab}(\vk) =&& S^{(1)}_{\vk}\,\qzero_{ab} +
S^{(2)}_{\vk} (\hat{k}_a\hat{k}_b - \f{1}{3} \qzero_{ab}) +
\sqrt{2}
V^{(1)}_{\vk}\, \hat{k}_{(a}\hat{x}_{b)}\nonumber\\
&+& \sqrt{2} V^{(2)}_{\vk}\, \hat{k}_{(a}\hat{y}_{b)} +
\f{1}{\sqrt{2}}\, T^{(1)}_{\vk} (\hat{x}_a\hat{x}_b -
\hat{y}_a\hat{y}_b) + \sqrt{2} T^{(2)}_{\vk}\,
(\hat{x}_{(a}\hat{y}_{b)}) \ea
where $\hat{k}$ is a unit vector in the $\vk$ direction and
$\hat{k}, \hat{x},\hat{y}$ constitutes a field of orthonormal
triads with respect to the comoving fiducial 3-metric
${\qzero}_{ab}$ in the momentum space. The Mukhanov-Sasaki
variable $\Q_{\vk}$ is then given by
\be \Q_{\vk} = \vp_{\vk} - \f{\pphi \gamma} {2 a^5 \ell^3 \b}
\, \left(S^{(1)}_{\vk} - \f{1}{3}\, S^{(2)}_{\vk}\right)\, .
\ee
where $\vp_{\vk}$ is the first order perturbation in the
inflaton.

As noted in the main text, dynamics of gauge invariant variables is
governed by the part ${\mathbb S}_2'^{(\Q)}$ of the second order
constraint obtained by (ignoring the terms that are linear in the
second order perturbation but) retaining terms that are quadratic in
the first order fields. Since this Hamiltonian is gauge invariant,
it can be expressed in terms of the Mukhanov-Sasaki variable
$\Q_{\vk}$ and its conjugate momentum ${\p}^{(\Q)}_{\vec{k}}$:
\be\label{eq:langlois1} {\mathbb S}_2'^{(\Q)} \left[ N_{\rm
hom} = a \right] = \frac{1}{2} \int \frac{d^3
k}{\left(2\pi\right)^3} \left[ \frac{1}{a^2} \left|
{\p}^{(\Q)}_{\vec{k}}\right|^2 + 2 \left( \Omega^2\right)
\left| \Q_{\vec{k}} \right|^2\right]~, \ee
with
\be \Omega^2 = -9 \frac{p_{(\phi)}^4 }{ a^4 \ell^{6} \pi_{(a)}^2} + 12
\pi G \frac{p_{(\phi)}^2}{a^2\ell^{6}}  - \frac{6 p_{(\phi)}
a^3}{\pi_{(a)}} \frac{{\rm d}V}{{\rm d} \phi} +
\frac{a^4}{2}\frac{{\rm d}^2 V}{{\rm d}\phi^2}
+\frac{a^2k^2}{2}~. \ee
Here, we have used the lapse function tailored to the conformal
time and the scale factor $a$ and its conjugate momentum
$\pi_{(a)}$ normally used in the cosmology literature
(translation to the $(\nu, \b)$ variables used in LQC can be
readily carried out using Eq.~(\ref{eq:BGvariables})). Modulo
standard factor ordering ambiguities $\Omega^2$ can be promoted
to a well-defined operator on the background Hilbert space
$\H_o$.

Another, particularly useful form of $\Omega$ arises from using
${\mathbb S}_0
=0$ to eliminate the explicit dependence on $\pi_{(a)}$,
\be \Omega^2 = \frac{1}{a^2} \left( 24 \pi G \left(\frac{E_k
E_\phi}{E_k + E_\phi}\right) - \sqrt{48\pi G}
\sqrt{\frac{E_k}{E_k + E_\phi} } \frac{{\rm d} E_\phi}{{\rm d}
\phi} + \frac{1}{2}\frac{{\rm d}^2 E_\phi}{{\rm d}\phi^2} +
\frac{a^4 k^2}{2}  \right)~, \ee
where we have defined,
\be E_k \equiv \frac{\p_{(\phi)}^2}{2}~, ~~~{\rm and} ~~~
E_\phi \equiv a^6 \ell^6 V\left(\phi\right)~, \ee
which are closely related to the kinetic and potential energy
density. Further defining the ratio,
\be \r \equiv \frac{6\kappa E_k}{E_k + E_\phi} =
\frac{3\kappa\p_{(\phi)}^2}{\frac{1}{2}p^2_{(\phi)} + a^6
\ell^6 V\left(\phi\right)}~, \ee
one finds,
\be \Omega^2 = \frac{a^4}{2} \left( \r V\left(\phi\right) -
2\sqrt{\r} V_{,\phi} + V_{,\phi\phi} +
\left(\frac{k}{a}\right)^2\right) \equiv \frac{a^2}{2}\left(\g
+ k^2\right)~. \ee
It is this form of the scalar Hamiltonian that we use in
Eq.~(\ref{pert-ham2}).

Finally, as explained in section \ref{s5.1}, comparison with
observations is simplest in terms of co-moving curvature
perturbations $\R_{\vk}$. To arrive at these gauge invariant
variables, one can perform a (background dependent) canonical
transformation within $\tilde{\Gamma}^{(1)}$, generated by
\be\label{eq:generating}
 S = \frac{z}{a} \p_{\vec{k}}^{(\Q)} \R_{\vec{k}} +
 \frac{1}{2} f \R^2_{\vec{k}}~,
\ee
where $f$ and $z$ are arbitrary function of the background
phase-space variables to begin with, but determined below.
{}From Eq.~(\ref{eq:generating}) one finds,
\be \label{eq:R} \R_{\vec{k}} = \frac{a}{z} \Q_{\vec{k}}~, \ \
{\rm and }
 \ \ \p_{\vec{k}}^{(\R)} = \frac{z}{a} \p_{\vec{k}}^{(\Q)}
+ f \R_{\vec{k}}~. \ee
The choice of $f$ and $z$ is geared to simplify the form of the
Hamiltonian in terms of $\R_{\vk}$ and $\p_{\vec{k}}^{(\R)}$.
Therefore we will now sketch the procedure and conclude with
the explicit form of $z$ that is needed in the expressions
(\ref{eq:R}) of these quantities. The Hamiltonian in terms of
$\R_{\vk}$ and $\p_{\vec{k}}^{(\R)}$ is not used in the main
text but may be useful for future investigations.

Recalling that Eq.~(\ref{eq:generating}) is a  `time' dependent
canonical transformation, with the evolution given by ${\mathbb
S}_0$, we find, %
\ba {\mathbb S}_2 '^{(\R)}\left[ N_{\rm hom} = a\right] &=&
\frac{1}{2} \int \frac{d^3k}{\left(2\pi\right)^3} \Biggl(
\frac{1}{z^2 }\left| \p_{\vec{k}}^{(\R)}\right|^2 + \left(
\frac{-2f}{z^2} + \frac{2 a}{z} \left\{ \frac{z}{a} , {\mathbb
S}_0\right\} \right) 2 Re(\p_{\vec{k}}^{(\R)} \R^{\star}_{\vec{k}}) \nonumber \\
&& +\left( 2\Omega^2\left( \frac{z}{a}\right)^2 +
\frac{f^2}{z^2} - \frac{2af}{z}\left\{\frac{z}{a} , {\mathbb
S}_{0} \right\}  + \left\{ f, {\mathbb S}_0
\right\} \right) \left| \R_{\vec{k}}\right|^2 \Biggr)~. \nonumber \\
\ea
One can check that the equations of motion for $\R$ are
independent of the choice of the background function $f$.
However, we can choose it to simplify the expression of
${\mathbb S}_2'^{(\R)}$. Specifically, for
\be
 f = a z \left\{ \frac{z}{a},{\mathbb S}_0 \right\}~,
\ee
${\mathbb S}_2'^{(\R)}$ is diagonal:
\ba\label{eq:S2dia} {\mathbb S}_2'^{(\R)}\left[ N_{\rm hom} =a
\right] &=& \frac{1}{2} \int \frac{d^3k}{\left(2\pi\right)^3}
\Biggl( \frac{1}{z^2}
\left| \p_{\vec{k}}^{(\R)}\right|^2 \nonumber \\
&& + \left( 2 \Omega^2 \left( \frac{z}{a}\right)^2 - a^2
\left\{\frac{z}{a},{\mathbb S}_0\right\}^2 + \left\{ a z\left\{
\frac{z}{a},{\mathbb S}_0 \right\}, {\mathbb S}_0 \right\}
\right) |\R_{\vec{k}}|^2\Biggr)~. \ea
Until this point the form of the function $z$ has not been
fixed. A particular simplification occurs for the specific
choice
\be\label{eq:z}
 z = \frac{-3}{4\pi G} \frac{p_{(\phi)}}{\pi_{(a)}}
\equiv \f{\gamma}{a^2\ell^3}\, \f{p_{(\phi)}}{\b} ~. \ee
(This is the relation used in section \ref{s5}). Using this
expression, one can explicitly evaluate the various Poisson
brackets, for example
\be \left\{ \frac{z}{a},{\mathbb S}_0 \right\} = \frac{3
p_{(\phi)}}{a^2} - \frac{9 p_{(\phi)}^3}{4 \pi G a^4
\pi_{(a)}^2 } - \frac{3 a^3 }{4\pi G \pi_{(a)}} \frac{{\rm
d}V}{{\rm d}\phi}~. \ee Using this in Eq.~(\ref{eq:S2dia})
gives \be {\mathbb S}_2'^{(\R)}\left[N_{\rm hom} = a\right] =
\frac{1}{2} \int \frac{{\rm d}^3k}{\left(2\pi\right)^3} \left(
\frac{1}{z^2} \left| \p_{\vec{k}}^{(\R)}\right|^2 + z^2 k^2
\left|\R_{\vec{k}} \right|^2\right)~. \ee
This form is particularly useful for cosmological calculations,
however it is clear from Eq.~(\ref{eq:z}) that the canonical
transformation between $\Q_{\vec k}$ and $\R_{\vec k}$ is ill
defined at points in the trajectory where $p_{(\phi)}=0$ or
$\pi_{(a)}=0$. In particular, $p_{(\phi)}$ can vanish during
the evolution between the bounce and the onset of slow-roll
inflation. Therefore, as discussed in \ref{s5.1}, during this
phase of evolution $\Q_{\vec{k}}$ is well defined, whilst
$\R_{\vec{k}}$ is not.\\

Finally, as we noted at the end of section \ref{s2.2.1}, the presence 
of the $m^2\phi^2$ potential also gives rise to some mathematical 
subtleties in the discussion of the quantum theory of the homogeneous 
background. These are summarized in section 4.3 of \cite{asrev} and 
will be discussed in detail in \cite{aps4}.

\end{appendix}

\end{document}